\documentclass[12pt]{article}

\catcode`\@=11
\@addtoreset{equation}{section}

\global\arraycolsep=2pt
\usepackage{mathrsfs,amsbsy,amssymb,latexsym,amsfonts,amsmath,cite}
\usepackage{graphicx,color}

\allowdisplaybreaks

\begin{document}

\begin{flushright}
\parbox{4cm}
{KUNS-2467}
\end{flushright}

\vspace*{1.5cm}

\begin{center}
{\Large\bf A deformation of quantum affine algebra \\ 
in squashed WZNW models}
\vspace*{2.5cm}\\
{\large Io Kawaguchi\footnote{E-mail:~io$\_$at$\_$gauge.scphys.kyoto-u.ac.jp} 
and Kentaroh Yoshida\footnote{E-mail:~kyoshida$\_$at$\_$gauge.scphys.kyoto-u.ac.jp}} 
\end{center}
\vspace*{0.25cm}
\begin{center}
{\it Department of Physics, Kyoto University \\ 
Kyoto 606-8502, Japan} 
\end{center}
\vspace{1cm}

\begin{abstract}
We proceed to study infinite-dimensional symmetries in two-dimensional 
squashed Wess-Zumino-Novikov-Witten (WZNW) models at the classical level.  
The target space is given by squashed S$^3$ and the isometry is $SU(2)_{\rm L}\times U(1)_{\rm R}$\,. 
It is known that $SU(2)_{\rm L}$ is enhanced to 
a couple of Yangians. We reveal here that an infinite-dimensional extension of 
$U(1)_{\rm R}$ is a deformation of quantum affine algebra, where a new deformation parameter 
is provided with the coefficient of the Wess-Zumino term. 
Then we consider the relation between the deformed quantum affine algebra and 
the pair of Yangians from the viewpoint of the left-right duality of monodromy matrices. 
The integrable structure is also discussed by computing the $r$/$s$-matrices that 
satisfy the extended classical Yang-Baxter equation. Finally two degenerate limits are discussed. 
\end{abstract}

\setcounter{footnote}{0}
\setcounter{page}{0}
\thispagestyle{empty}

\newpage

\tableofcontents 

\setcounter{footnote}{0}
\setcounter{page}{0}
\thispagestyle{empty}

\section{Introduction}

The AdS/CFT correspondence \cite{M,GKP,W} has been a fascinating topic
in the study of string theory over the decade after Maldacena's proposal. 
A tremendous amount of works have been devoted to test and generalize it.  
Nowadays, the research area covers diverse subjects. 
A great achievement in the recent progress is the discovery of the integrable structure 
behind the AdS/CFT correspondence (For a comprehensive review on this subject, see \cite{review}). 

\medskip 

In the string-theory side, the integrable structure of two-dimensional string sigma-models 
with target spacetime AdS$_5\times$S$^5$ plays an important role \cite{BPR}. It is closely related 
to the fact that AdS$_5\times$S$^5$ is described as a symmetric coset. It leads to an infinite number of 
the conserved charges constructed, for example, by following the pioneering works 
\cite{Luscher1,Luscher2,BIZZ,Bernard-Yangian,MacKay} 
(For a comprehensive textbook see \cite{AAR}). The symmetric cosets that potentially may lead to  
a holographic interpretation are classified, including spacetime fermions \cite{Zarembo-symmetric}. 

\medskip 

The next interesting issue is to consider integrable deformations of the AdS/CFT correspondence.  
There are two approaches.  
The one is an algebraic approach based on $q$-deformations of the world-sheet S-matrix 
\cite{BK,BGM,HHM,dLRT,Arutyunov}. The deformed S-matrices are explicitly constructed, 
while the target-space geometry is unclear. 
The other is a geometric approach based on deformations of target spaces of the sigma models. 
It seems likely that the deformed geometries are 
represented by non-symmetric cosets \cite{SYY} in comparison to AdS$_5\times$S$^5$ 
and hence the prescription for symmetric cosets is not available any more. 
It is necessary to develop a new method to argue the integrability. 

\medskip 
  
In the latter approach, there is a long history (For classic papers see \cite{Cherednik,FR,BFP}). 
Motivated by integrable deformations of AdS/CFT, a series of works have been done on squashed S$^3$ 
and warped AdS$_3$ \cite{KY,KOY,KYhybrid,KY-Sch,KMY-QAA,KMY-monodromy,Jordanian-KMY, ORU,OU}. 
Some specific higher-dimensional cases are discussed in \cite{BR}.  
Remarkably, the classical integrable structure of deformed sigma models was recently shown 
for arbitrary compact Lie groups and the coset cousins by Delduc, Magro and Vicedo \cite{DMV}. 
Then, they successively presented  a $q$-deformation of the AdS$_5\times$S$^5$ 
superstring \cite{DMV2}.  

\medskip 

Based on the latter approach,  we are here concerned with the classical integrable structure of 
two-dimensional Wess-Zumino-Novikov-Witten (WZNW) models with the target space squashed S$^3$\,. 
The isometry is given by $SU(2)_{\rm L} \times U(1)_{\rm R}$\,. It is partly explained in \cite{KOY} that 
there exist a couple of Yangian algebras based on $SU(2)_{\rm L}$ by explicit constructions of 
non-local conserved charges and direct computations of the Poisson brackets of the charges.   

\medskip 

In this paper we consider an infinite-dimensional extension of $U(1)_{\rm R}$\,. 
In the case without the Wess-Zumino term, it is just a classical analogue of 
a quantum affine algebra \cite{KMY-QAA}. When the Wess-Zumino term is added, 
an additional constant parameter is introduced as its coefficient. A natural question is 
what happens to the quantum affine algebra. As one may easily guess, a new kind of 
deformation is induced by the presence of the Wess-Zumino term. The resulting algebra 
is a classical analogue of the deformed quantum affine algebra. So far, it is not clear 
what is the mathematical formulation of the deformed quantum affine algebra,  
though it seems likely to be a two-parameter quantum toroidal algebra \cite{GKV}.  
In order to answer the question, it is necessary to see the first realization of the two-parameter quantum toroidal algebra. 

\medskip 

This paper is organized as follows. 
In section 2 the classical action of the squashed WZNW models is introduced.
In section 3 we consider the classical integrable structure 
based on $SU(2)_{\rm L}$\,. This is called {\it the left description}. We present 
a couple of Lax pairs and the associated monodromy matrices. 
The classical $r$/$s$-matrices are shown to satisfy the extended 
classical Yang-Baxter equation. The infinite-dimensional extensions of $SU(2)_{\rm L}$ are Yangians. 
In section 4 we argue the classical integrable structure based on $U(1)_{\rm R}$\,. 
This is called {\it the right description}. 
A Lax pair and the associated monodromy matrix are presented. 
The classical $r$/$s$-matrices 
satisfy the extended classical Yang-Baxter equation. Remarkably, 
an infinite-dimensional extension of $U(1)_{\rm R}$ is shown to be a deformation of 
quantum affine algebra, where a new deformation parameter 
is provided by the coefficient of the Wess-Zumino term.  
In section 5 the gauge equivalence between the left and right descriptions is proven. 
Under an identification between the spectral parameters, 
the {\it left} Lax pair is related to the {\it right} Lax pair via a gauge transformation.  
In section 6 we argue two degenerate limits in the right description. 
At some special points in the parameter space, the deformed quantum affine algebra 
degenerates to a Yangian, according to the enhancement of $U(1)_{\rm R}$ to $SU(2)_{\rm R}$\,. 
Section 7 is devoted to conclusion and discussion. 

\medskip 

In Appendix A, we explain the computation of the current algebra in detail. 
Appendix B provides a prescription to treat non-ultra local terms 
in computing of the Poisson brackets of the conserved charges.

\section{Preliminary}

Let us begin with the setup to fix our notation and convention. 
The metric of squashed S$^3$ is first provided in terms of an $SU(2)$ group element. 
Then the classical action of the squashed WZNW models is introduced. 
For later convenience, the equations of motion are 
explicitly written down.

\subsection{Squashed S$^3$}

The metric of round S$^3$ can be described as a $U(1)$ fibration over S$^2$\,. 
The squashing is one-parameter deformations of the $U(1)$ direction. 
The metric of squashed S$^3$ is given by 
\begin{eqnarray}
ds^2=\frac{L^2}{4}\left[d\theta^2+\sin^2\theta d\phi^2+(1+C)\left(d\psi+\cosh\theta d\phi\right)^2\right]\,. 
\label{metric-angle}
\end{eqnarray}
When $C=0$\,, the metric is reduced to that of S$^3$ with radius $L$ 
and the isometry is $SU(2)_{\rm L} \times SU(2)_{\rm R}$\,. 
When $C\neq 0$\,, the isometry is reduced to $SU(2)_{\rm L}\times U(1)_{\rm R}$\,. 

\medskip 

In order to rewrite the metric (\ref{metric-angle})\,,   
let us introduce an $SU(2)$ group element $g$ like 
\begin{eqnarray}
g &=& {\rm e}^{T^3\phi}{\rm e}^{T^2\theta}{\rm e}^{T^3\psi} \in SU(2)\,. 
\end{eqnarray}
Here the $su(2)$ generators $T^a$ $(a=1,2,3)$ satisfy the following relations
\begin{eqnarray}
\left[T^a,T^b\right]=\varepsilon^{ab}_{~~c}T^c\,, 
\label{commutation relation generator}
\end{eqnarray}
and normalized as 
\begin{eqnarray}
{\rm Tr}\left(T^aT^b\right)=-\frac{1}{2}\delta^{ab}\,. 
\label{normalization generator}
\end{eqnarray}
Note that $\varepsilon^{ab}_{~~c}$ is the totally anti-symmetric tensor normalized 
as $\varepsilon_{123}=+1$\,. 
The $su(2)$ indices are raised and lowered by $\delta^{ab}$ and its inverse, respectively. 

\medskip 

It is useful to define $T^\pm$ as 
\begin{eqnarray}
T^\pm \equiv \frac{1}{\sqrt{2}}\left(T^1\pm iT^2\right)\,. 
\end{eqnarray}
Then the commutation relations in (\ref{commutation relation generator}) are rewritten as 
\begin{eqnarray}
\left[T^\pm,T^\mp\right]=\mp iT^3\,, \qquad
\left[T^\pm,T^3\right]=\pm iT^\pm\,, 
\end{eqnarray}
and the normalization of the generators in (\ref{normalization generator}) is given by  
\begin{eqnarray}
{\rm Tr}\left(T^\pm T^\mp\right)={\rm Tr}\left(T^3T^3\right)=-\frac{1}{2}\,. 
\end{eqnarray}

\medskip 

Then the metric (\ref{metric-angle}) is rewritten in terms of the $SU(2)$ group element as 
\begin{eqnarray}
ds^2=-\frac{L^2}{2}\left[{\rm Tr}\left(J^2\right)-2C\left({\rm Tr}\left[T^3J\right]\right)^2\right]\,, 
\label{metric-group element}
\end{eqnarray}
where we have introduced the left-invariant one-form
\begin{eqnarray}
J \equiv g^{-1}dg\,. 
\end{eqnarray}
Note that $J$ can be represented by the angle variables $(\theta,\phi,\psi)$ as follows: 
\begin{eqnarray}
J&=&T^1\left(\sin\psi d\theta - \cos\psi\sin\theta d\phi\right) 
+T^2\left(\cos\psi d\theta + \sin\psi\sin\theta d\phi\right) 
+T^3\left(d\psi + \cos\theta d\phi\right) \nonumber \\
&=&T^+\frac{1}{\sqrt{2}}{\rm e}^{i\psi}\left(-id\theta - \sin\theta d\phi\right) 
+T^-\frac{1}{\sqrt{2}}{\rm e}^{-i\psi}\left(id\theta - \sin\theta d\phi\right) 
+T^3\left(d\psi + \cos\theta d\phi\right)\,. \nonumber 
\end{eqnarray}

\medskip 

With the metric (\ref{metric-group element})\,, it is easy to see the invariance 
under $SU(2)_{\rm L}\times U(1)_{\rm R}$\,. 
The $SU(2)_{\rm L}$ and $U(1)_{\rm R}$ transformations are the left- and right- multiplications,  
\begin{eqnarray}
g\to {\rm e}^{\beta_aT^a}\cdot g\cdot {\rm e}^{-\alpha T^3}\,. 
\end{eqnarray}
Here $\beta_a$ and $\alpha$ are real parameters. 

\subsection{The classical action of the squashed WZNW models}

First of all, let us introduce two-dimensional non-linear sigma models whose target space is given 
by squashed S$^3$\,. 
The classical action is 
\begin{eqnarray}
S_{\sigma M}=\frac{1}{\lambda^2}\int^{\infty}_{-\infty}\!\!\!dt\!\int^{\infty}_{-\infty}\!\!\!dx
~\eta^{\mu\nu}\left[{\rm tr}\left(J_\mu J_\nu\right)-2C{\rm tr}\left(T^3J_\mu\right){\rm tr}\left(T^3J_\nu\right)\right]\,, 
\label{action-sigma model}
\end{eqnarray}
where the parameter $\lambda$ is the coupling constant and 
the base space is a two-dimensional Minkowski spacetime with the coordinates $t$ (time) 
and $x$ (space) and 
the metric 
\begin{eqnarray}
-\eta_{tt}=\eta_{xx}=+1\,. 
\end{eqnarray}
Note that the region of the parameter $C$ is restricted $C>-1$ 
so that the positivity of the kinetic term is ensured. 

\medskip 

The next is to introduce the Wess-Zumino term on squashed S$^3$\,, 
\begin{eqnarray}
S_{WZ}=\frac{n}{12\pi}\int^1_0\!\!\!ds\int^{\infty}_{-\infty}\!\!\!dt\int^{\infty}_{-\infty}\!\!\!dx~\epsilon^{\hat{\mu}\hat{\nu}\hat{\rho}}{\rm tr}\left(\widetilde{J}_{\hat{\mu}} \widetilde{J}_{\hat{\nu}} \widetilde{J}_{\hat{\rho}} \right)\,, \qquad
\widetilde{J}\equiv \widetilde{g}^{-1}d\widetilde{g}\,,
\label{Wess-Zumino term}
\end{eqnarray}
where $n$ is an integer. 
Note that the above integral is performed on a three-dimensional base manifold spanned by $(t,x,s)$\,. 
The totally anti-symmetric tensor $\epsilon_{\hat{\mu}\hat{\nu}\hat{\rho}}$ is normalized as 
\begin{eqnarray}
\epsilon_{txs}=+1\,. 
\end{eqnarray}
The $SU(2)$ element $\widetilde{g}$ is defined on this three-dimensional manifold. 
It interporates between a constant element at $s=0$ and $g(t,x)$ at $s=1$\,: 
\begin{eqnarray}
\widetilde{g}(t,x,s=0)=g_0~:\mbox{const.}\,, \qquad
\widetilde{g}(t,x,s=1)=g(t,x)\,. 
\end{eqnarray}
Note that the Wess-Zumino term (\ref{Wess-Zumino term}) is the same as in the case of round S$^3$ 
and hence it is invariant under the $SU(2)_{\rm L} \times U(1)_{\rm R}$ symmetry 
of the sigma model action (\ref{action-sigma model})\,. 

\medskip 

Let us consider the Wess-Zumino-Novikov-Witten models defined on squashed S$^3$\,, 
which henceforth are called ``{\it squashed WZNW models}''. 
The action is given by the sum of $S_{\sigma M}$ in (\ref{action-sigma model}) 
and $S_{WZ}$ in ({\ref{Wess-Zumino term})\,: 
\begin{eqnarray}
S&=&S_{\sigma M}+S_{WZ}\,. 
\label{action}
\end{eqnarray}
The action (\ref{action}) is also $SU(2)_{\rm L}\times U(1)_{\rm R}$-invariant. 

\medskip 

From the action (\ref{action})\,, the equations of motion are obtained, 
\begin{eqnarray}
\partial^\mu J_\mu -2C{\rm tr}\left(T^3 \partial^\mu J_\mu\right)T^3 -2C{\rm tr}\left(T^3 J_\mu\right)\left[J^\mu,T^3\right] +K\epsilon^{\mu\nu}\partial_\mu J_\nu = 0\,, 
\label{eom}
\end{eqnarray}
where the new constant $K$ is defined as 
\begin{eqnarray}
K\equiv\frac{n\lambda^2}{8\pi}\,, 
\end{eqnarray}
and the totally anti-symmetric tensor $\epsilon_{\mu\nu}$ is normalized as 
\begin{eqnarray}
\epsilon_{tx}=+1\,. 
\end{eqnarray}
The $su(2)$ components of the left-invariant one-form $J^a$ are defined as 
\begin{eqnarray}
J^a\equiv-2{\rm Tr}(T^aJ)\,, 
\end{eqnarray}
or equivalently 
\begin{eqnarray}
J = T^+J^- + T^-J^+ + T^3J^3\,. 
\end{eqnarray}

\medskip 

In terms of $J^a$\,, the equations of motion are rewritten as 
\begin{eqnarray}
&&(1+C)\partial^\mu J^3_\mu +K\epsilon^{\mu\nu}\partial_\mu J^3_\nu = 0\,, \\
&&\partial^\mu J^\pm_\mu \mp iCJ^3_\mu J^{\pm,\mu} +K\epsilon^{\mu\nu}\partial_\mu J^\pm_\nu =0\,. \nonumber
\label{eom1 components}
\end{eqnarray}
By definition, the left-invariant one-form $J$ satisfies the flatness condition: 
\begin{eqnarray}
\epsilon^{\mu\nu}\left(\partial_\mu J_\nu + J_\mu J_\nu \right)=0\,. 
\end{eqnarray}
This condition can also be rewritten in terms of the components $J^a$ as 
\begin{eqnarray}
&&\epsilon^{\mu\nu}\left(\partial_\mu J^3_\nu + iJ^+_\mu J^-_\nu \right)=0\,, \\
&&\epsilon^{\mu\nu}\left(\partial_\mu J^\pm_\nu \pm i J^3_\mu J^\pm_\nu \right)=0\,. \nonumber
\label{flatness J components}
\end{eqnarray}
The flatness condition (\ref{flatness J components}) enables us to rewrite 
the equations of motion (\ref{eom1 components}) as 
\begin{eqnarray}
&&(1+C)\partial^\mu J^3_\mu -iK\epsilon^{\mu\nu}J^+_\mu J^-_\nu = 0\,, \label{eom-final} \\
&&\partial^\mu J^\pm_\mu \mp iCJ^3_\mu J^{\pm,\mu} \mp iK\epsilon^{\mu\nu}J^3_\mu J^\pm_\nu =0\,. \nonumber
\end{eqnarray}
The expressions in (\ref{eom-final}) play an important role 
in our later discussion.

\section{The left description}

In this section, we discuss the classical integrable structure of squashed WZNW models based 
on the $SU(2)_{\rm L}$ symmetry. We call it {\it left description}. This part 
contains a short review of the previous work \cite{KOY}. 

\medskip 

First of all, we construct an $SU(2)_{\rm L}$ conserved current which satisfies the flatness condition. 
With the flat and conserved current, we obtain a Lax pair and the corresponding monodromy matrix. 
Then we compute the classical $r$/$s$-matrices for the Lax pair. 
Finally, we show that the $SU(2)_{\rm L}$ symmetry is enhanced 
to the Yangian algebra ${\mathcal Y}(su(2)_{\rm L})$\,. 

\subsection{Lax pairs}

The classical action (\ref{action}) has the $SU(2)_{\rm L}$ symmetry and 
the associated conserved current is given by 
\begin{eqnarray}
j^L_\mu = gJ_\mu g^{-1} -2C{\rm tr}\left(T^3J_\mu\right) gT^3g^{-1} -K\epsilon_{\mu\nu}gJ^\nu g^{-1}\,. 
\end{eqnarray}
The conservation law of this current is equivalent to the equations of motion in (\ref{eom}) like 
\begin{eqnarray}
\partial^\mu j^L_\mu&=&g\left[\partial^\mu J_\mu -2C{\rm tr}\left(T^3 \partial^\mu J_\mu\right)T^3 -2C{\rm tr}\left(T^3 J_\mu\right)\left[J^\mu,T^3\right] +K\epsilon^{\mu\nu}\partial_\mu J_\nu\right]g^{-1} \\
&=&0\,. \nonumber
\end{eqnarray}
Note that $j_{\mu}^L$ does not satisfy the flatness condition due to the deformation. 

\medskip 

One may consider to improve $j_{\mu}^L$ so as to 
satisfy the flatness condition. This requirement leaves two improved currents \cite{KOY}, 
\begin{eqnarray}
j^{L_\pm}_\mu &=& gJ_\mu g^{-1} -2C{\rm tr}\left(T^3J_\mu\right) gT^3g^{-1} 
-K\epsilon_{\mu\nu}gJ^\nu g^{-1} \mp A\epsilon_{\mu\nu}\partial^{\nu}\left(gT^3g^{-1}\right)\,, \nonumber
\end{eqnarray}
with the coefficient $A$ represented by 
\begin{eqnarray}
&&A = \sqrt{C\left(1-\frac{K^2}{1+C}\right)}\,.
\end{eqnarray}
The subscripts $\pm$ denote the degeneracy of the improved currents. 
The improved currents satisfy the following flatness condition,
\begin{eqnarray}
\epsilon^{\mu\nu}\left(\partial_\mu j^{L_\pm}_\nu-j^{L_\pm}_\mu j^{L_\pm}_\nu\right)=0\,. 
\end{eqnarray}
When $K=0$\,, $A=\sqrt{C}$ and the improved currents constructed in \cite{KY} are reproduced.  

\medskip 

With the flat $SU(2)_{\rm L}$ currents, two Lax pairs are constructed as 
\begin{eqnarray}
&&L^{L_\pm}_t(x;\lambda_{L_\pm}) = \frac{1}{2}\left[L^{L_\pm}_+(x;\lambda_{L_\pm})
+L^{L_\pm}_-(x;\lambda_{L_\pm})\right]\,, 
\label{left-Lax pair} \\
&&L^{L_\pm}_x(x;\lambda_{L_\pm}) = \frac{1}{2}\left[L^{L_\pm}_+(x;\lambda_{L_\pm})
-L^{L_\pm}_-(x;\lambda_{L_\pm})\right]\,, \nonumber \\
&&L^{L_\pm}_+(x;\lambda_{L_\pm}) = \frac{1}{1+\lambda_{L_\pm}}j^{L_\pm}_+\,, \qquad
L^{L_\pm}_-(x;\lambda_{L_\pm}) = \frac{1}{1-\lambda_{L_\pm}}j^{L_\pm}_-\,. \nonumber
\end{eqnarray}
Here $\lambda_{L_\pm}$ are spectral parameters 
and $j^{L_\pm}_+$ and $j^{L_\pm}_-$ are defined as 
\begin{eqnarray}
j^{L_\pm}_+\equiv j^{L_\pm}_t+j^{L_\pm}_x\,, \qquad
j^{L_\pm}_-\equiv j^{L_\pm}_t-j^{L_\pm}_x\,.
\end{eqnarray}
The following commutation relation 
\begin{eqnarray}
\left[\partial_t-L^{L_\pm}_t(x;\lambda_{L_\pm}),\partial_x-L^{L_\pm}_x(x;\lambda_{L_\pm})\right]=0\,. 
\label{flatness left-Lax pair}
\end{eqnarray}
gives rise to the conservation law of the flat $SU(2)_{\rm L}$ current (equivalently, the equations of motion) 
and the flatness condition.

\medskip 

Then the monodromy matrices are defined as 
\begin{eqnarray}
U^{L_\pm}(\lambda_{L_\pm})\equiv{\rm P}\exp\left[\int^\infty_{-\infty}\!\!\!dx~
L^{L_\pm}_x(x;\lambda_{L_\pm})\right]\,. 
\end{eqnarray}
The symbol P denotes the path ordering.  
Due to the flatness of the Lax pairs in (\ref{flatness left-Lax pair})\,, the monodromy matrices are conserved,  
\begin{eqnarray}
\frac{d}{dt}U^{L_\pm}(\lambda_{L_\pm})=0\,. 
\end{eqnarray}
Thus an infinite set of conserved charges can be obtained 
by expanding the monodromy matrices with respect to $\lambda_{L_\pm}$ around appropriate points. 
For example, the monodromy matrices can be expanded 
around $\lambda_{L_{\pm}}=\infty$ as 
\begin{eqnarray}
U^{L_\pm}(\lambda_{L_\pm})=\exp\left[\sum_{n=0}^\infty\lambda_{L_\pm}^{-n-1}Q^{L_\pm}_{(n)}\right]\,. 
\end{eqnarray}
In the next subsection, we will discuss the algebra generated by $Q^{L_\pm}_{(n)}$\,. 

\medskip

Before closing this subsection, let us discuss the $r$/$s$-matrices computed 
from the Lax pairs in (\ref{left-Lax pair}) by following the Maillet formalism \cite{Maillet}. 
One can read off them from the Poisson brackets between the spatial components of the Lax pairs,  
\begin{eqnarray}
&&\Bigl\{L^{L_\pm}_x(x;\lambda_{L_\pm})\stackrel{\otimes}{,}L^{L_\pm}_x(y;\mu_{L_\pm})\Bigr\}_{\rm P} \\
&=&\Bigl[r^{L_\pm}(\lambda_{L_\pm},\mu_{L_\pm}),L^{L_\pm}_x(x;\lambda_{L_\pm})\otimes 1+1\otimes L^{L_\pm}_x(y;\mu_{L_\pm})\Bigr]\delta(x-y) \nonumber \\
&&-\Bigl[s^{L_\pm}(\lambda_{L_\pm},\mu_{L_\pm}),L^{L_\pm}_x(x;\lambda_{L_\pm})\otimes 1-1\otimes L^{L_\pm}_x(y;\mu_{L_\pm})\Bigr]\delta(x-y) \nonumber \\
&&-2s^{L_\pm}(\lambda_{L_\pm},\mu_{L_\pm})\partial_x\delta(x-y)\,. \nonumber
\end{eqnarray}
To compute the above Poisson bracket, we have to use the current algebra for $j^{L_\pm}_\mu$\,, 
\begin{eqnarray}
\Bigl\{j^{L_\pm,a}_t(x),j^{L_\pm,b}_t(y)\Bigr\}_{\rm P} &=& \varepsilon^{ab}_{~~c}j^{L_\pm,c}_t(x)\delta(x-y) -2K\delta^{ab}\partial_x\delta(x-y)\,, 
\label{left-current algebra} \\
\Bigl\{j^{L_\pm,a}_t(x),j^{L_\pm,b}_x(y)\Bigr\}_{\rm P} &=& \varepsilon^{ab}_{~~c}j^{L_\pm,c}_x(x)\delta(x-y) +\left(1+K^2+A^2\right)\delta^{ab}\partial_x\delta(x-y)\,, \nonumber \\
\Bigl\{j^{L_\pm,a}_x(x),j^{L_\pm,b}_x(y)\Bigr\}_{\rm P} &=& -\left(K^2+A^2\right)\varepsilon^{ab}_{~~c}j^{L_\pm,c}_t(x)\delta(x-y) \nonumber \\
&&-2K\varepsilon^{ab}_{~~c}j^{L_\pm,c}_x(x)\delta(x-y)-2K\delta^{ab}\partial_x\delta(x-y)\,. \nonumber
\end{eqnarray}
The explicit expressions of $r$/$s$-matrices are given by 
\begin{eqnarray}
r^{L_\pm}(\lambda_\pm,\mu_\pm) &=& \frac{h^L_{C,K}(\lambda_\pm)+h^L_{C,K}(\mu_\pm)}{2(\lambda_\pm - \mu_\pm)}\left[T^-\otimes T^+ +T^+\otimes T^- +T^3\otimes T^3\right]\,, 
\label{left r/s-matrices} \\
s^{L_\pm}(\lambda_\pm,\mu_\pm) &=& \frac{h^L_{C,K}(\lambda_\pm)-h^L_{C,K}(\mu_\pm)}{2(\lambda_\pm - \mu_\pm)}\left[T^-\otimes T^+ +T^+\otimes T^- +T^3\otimes T^3\right]\,, \nonumber 
\end{eqnarray}
where a scalar function $h^L_{C,K}(\lambda)$ is defined as 
\begin{eqnarray}
h^L_{C,K}(\lambda) \equiv \frac{A^2+(\lambda+K)^2}{1-\lambda^2}\,. \nonumber 
\end{eqnarray}
The $r$/$s$-matrices satisfy the extended classical Yang-Baxter equation\footnote{
The $r$/$s$-matrices depend on $\lambda_{L_\pm}$ and $\mu_{L_\pm}$ individually (not only $\lambda_{L_\pm}-\mu_{L_\pm}$\,) 
and they satisfy the extended classical Yang-Baxter equation. 
Thus the classification of the $r$/$s$-matrice are subtle.}, 
\begin{eqnarray}
&&\left[\left(r-s\right)^{L_\pm}_{12}(\lambda_{L_\pm},\mu_{L_\pm}),\left(r+s\right)^{L_\pm}_{13}(\lambda_{L_\pm},\nu_{L_\pm})\right] \\
&+&\left[\left(r+s\right)^{L_\pm}_{12}(\lambda_{L_\pm},\mu_{L_\pm}),\left(r+s\right)^{L_\pm}_{23}(\mu_{L_\pm},\nu_{L_\pm})\right] \nonumber \\
&+&\left[\left(r+s\right)^{L_\pm}_{13}(\lambda_{L_\pm},\nu_{L_\pm}),\left(r+s\right)^{L_\pm}_{23}(\mu_{L_\pm},\nu_{L_\pm})\right]
=0\,. \nonumber
\end{eqnarray}
It should be noted that, when $K=0$\,, the function $l^L_{C,K}(\lambda)$ is reduced to 
\begin{eqnarray}
h^L_{C,0}(\lambda) \equiv \frac{C+\lambda^2}{1-\lambda^2}\,. \nonumber 
\end{eqnarray}
Thus the $r$/$s$-matrices in (\ref{left r/s-matrices}) reproduce 
the results without the Wess-Zumino term \cite{KYhybrid}\,.

\subsection{Yangians}

So far, the monodromy matrices $U^{L_\pm}(\lambda_{L_\pm})$ have been introduced 
and an infinite number of the conserved charges $Q^{L_\pm}_{(n)}$ are obtained 
by expanding them with respect to $\lambda_{L_\pm}$\,. 

\medskip 

For the first three levels, the explicit expressions of $Q^{L_\pm}_{(n)}$ are given by 
\begin{eqnarray}
Q^{L_\pm,a}_{(0)}&=&\int^\infty_{-\infty}\!\!\!dx~j^{L_\pm,a}_t(x)\,, \\
Q^{L_\pm,a}_{(1)}&=&\frac{1}{4}\int^\infty_{-\infty}\!\!\!dx\int^\infty_{-\infty}\!\!\!dy~\epsilon(x-y)\varepsilon^a_{~bc}j^{L_\pm,b}_t(x)j^{L_\pm,c}_t(y)
-\int^\infty_{-\infty}\!\!\!~j^{L_\pm,a}_x(x)\,, \nonumber \\
Q^{L_\pm,a}_{(2)}&=&\frac{1}{12}\int^\infty_{-\infty}\!\!\!dx\int^\infty_{-\infty}\!\!\!dy\int^\infty_{-\infty}\!\!\!dz~
\epsilon(x-y)\epsilon(x-z)\delta_{bc} \nonumber \\
&&\hspace{1.5cm}\times\left[j^{L_\pm,b}_t(x)j^{L_\pm,a}_t(y)j^{L_\pm,c}_t(z)-j^{L_\pm,a}_t(x)j^{L_\pm,b}_t(y)j^{L_\pm,c}_t(z)\right] \nonumber \\
&&-\frac{1}{2}\int^\infty_{-\infty}\!\!\!dx\int^\infty_{-\infty}\!\!\!dy~
\epsilon(x-y)\varepsilon^a_{~bc}j^{L_\pm,b}_t(x)j^{L_\pm,c}_x(y)
+\int^\infty_{-\infty}\!\!\!dx~j^{L_\pm,a}_t(x)\,. \nonumber
\end{eqnarray}
Note that these charges can be directly constructed from the flat $SU(2)_{\rm L}$ 
currents $j^{L_\pm}_\mu(x)$ recursively by following the BIZZ construction \cite{BIZZ}\,. 

\medskip 

The next task is to show that the conserved charges satisfy the defining relations of Yangian 
${\mathcal Y}(su(2)_{\rm L})$\,. 
The Poisson brackets of the first two levels are given by \cite{KOY}
\begin{eqnarray}
\Bigl\{Q^{L_\pm,a}_{(0)},Q^{L_\pm,b}_{(0)}\Bigr\}_{\rm P}&=&\varepsilon^{ab}_{~~c}Q^{L_\pm,c}_{(0)}\,, \\
\Bigl\{Q^{L_\pm,a}_{(1)},Q^{L_\pm,b}_{(0)}\Bigr\}_{\rm P}&=&\varepsilon^{ab}_{~~c}Q^{L_\pm,c}_{(1)}\,, \nonumber \\
\Bigl\{Q^{L_\pm,a}_{(1)},Q^{L_\pm,b}_{(1)}\Bigr\}_{\rm P}&=&\varepsilon^{ab}_{~~c}\left[Q^{L_\pm,c}_{(2)}+\frac{1}{12}\left(Q^{L_\pm}_{(0)}\right)^2Q^{L_\pm,c}_{(0)}+2KQ^{L_\pm,c}_{(1)}\right]\,. \nonumber
\end{eqnarray}
The Serre relations are shown as 
\begin{eqnarray}
&&\Bigl\{Q^{L_\pm,3}_{(1)},\Bigl\{Q^{L_\pm,+}_{(1)},Q^{L_\pm,-}_{(1)}\Bigr\}_{\rm P}\Bigr\}_{\rm P}
=\frac{1}{4}Q^{L_\pm,3}_{(0)}\left(Q^{L_\pm,+}_{(0)}Q^{L_\pm,-}_{(1)}-Q^{L_\pm,-}_{(0)}Q^{L_\pm,+}_{(1)}\right)\,, \\
&&\Bigl\{Q^{L_\pm,\pm}_{(1)},\Bigl\{Q^{L_\pm,\pm}_{(1)},Q^{L_\pm,\mp}_{(1)}\Bigr\}_{\rm P}\Bigr\}_{\rm P}
-\Bigl\{Q^{L_\pm,3}_{(1)},\Bigl\{Q^{L_\pm,\pm}_{(1)},Q^{L_\pm,3}_{(1)}\Bigr\}_{\rm P}\Bigr\}_{\rm P} \nonumber \\
&&=\frac{1}{4}Q^{L_\pm,\pm}_{(0)}\!\left(Q^{L_\pm,\pm}_{(0)}Q^{L_\pm,\mp}_{(1)}\!-Q^{L_\pm,\mp}_{(0)}Q^{L_\pm,\pm}_{(1)}\right)
-\frac{1}{4}Q^{L_\pm,3}_{(0)}\!\left(Q^{L_\pm,\pm}_{(0)}Q^{L_\pm,3}_{(1)}\!-Q^{L_\pm,3}_{(0)}Q^{L_\pm,\pm}_{(1)}\right)\,, \nonumber \\
&&\Bigl\{Q^{L_\pm,\pm}_{(1)},\Bigl\{Q^{L_\pm,\pm}_{(1)},Q^{L_\pm,3}_{(1)}\Bigr\}_{\rm P}\Bigr\}_{\rm P}
=\frac{1}{4}Q^{L_\pm,\pm}_{(0)}\left(Q^{L_\pm,\pm}_{(0)}Q^{L_\pm,3}_{(1)}-Q^{L_\pm,3}_{(0)}Q^{L_\pm,\pm}_{(1)}\right)\,. \nonumber
\end{eqnarray}
Thus the defining relations of Yangian ${\mathcal Y}(su(2)_{\rm L})$ at the classical level are satisfied 
in the sense of Drinfeld's first realization \cite{Drinfeld1,Drinfeld2}. 

\medskip 

Here we should comment on the treatment of non-ultra local terms contained in 
the current algebra of $j^{L_\pm}_\mu(x)$\,. 
They develop ambiguities in computing the Poisson brackets of the conserved charges  
and there might be the possibility that the defining relations of ${\mathcal Y}(su(2)_{\rm L})$ are spoiled. 
In the present case, the presence of the Wess-Zumino term make the situation worse. 
It develops non-ultra local terms even in the Poisson brackets of $j^{L_\pm}_t(x)$ 
and hence cause ambiguities in computing the usual $SU(2)_{\rm L}$ Lie algebra part. 
The treatment of the non-ultra local terms is argued in Appendix B in detail.

\section{The right description}

The classical integrable structure of the squashed WZNW models can also be described  
based on $U(1)_{\rm R}$ as another description. This description is called {\it right description}. 
A Lax pair and the associated monodromy matrix are presented.  
Then an infinite-dimensional extension of $U(1)_{\rm R}$ is argued. 
The resulting algebra is a deformation of the standard quantum affine algebra. 
In the right description, the $r$/$s$-matrices are deformed by an additional term, 
in comparison to the case without the Wess-Zumino term. 

\subsection{Lax pair}

A Lax pair which respects $U(1)_{\rm R}$ is given by\footnote{The anisotropic Lax pair with $K=0$ 
is constructed originally by Cherednik \cite{Cherednik}. See also \cite{FR}.}
\begin{eqnarray}
&&L^R_t(x;\lambda_R) = \frac{1}{2}\left[L^R_+(x;\lambda_R)+L^R_-(x;\lambda_R)\right]\,, 
\label{right Lax pair}\\
&&L^R_x(x;\lambda_R) = \frac{1}{2}\left[L^R_+(x;\lambda_R)-L^R_-(x;\lambda_R)\right]\,, \nonumber \\
&&L^R_\pm(x;\lambda_R) = -\frac{\sinh\left(\alpha\pm\beta\right)}{\sinh\left[\alpha\pm(\beta+\lambda_R)\right]}
\left[T^+J^-_\pm +T^-J^+_\pm +\frac{\cosh\left(\alpha\pm\lambda_R\right)}{\cosh\alpha}T^3 J^3_\pm\right]\,. 
\nonumber
\end{eqnarray}
New constants $\alpha$ and $\beta$ are related to $C$ and $K$ like
\begin{eqnarray}
\tanh\alpha=\frac{iC}{A}\,, \qquad
\tanh\beta=\frac{iCK}{A(1+C)}\,.
\label{definition alpha beta}
\end{eqnarray}
Note that $\alpha$ and $\beta$ have the periodicities, 
\begin{eqnarray}
\alpha\sim\alpha+\pi i\,, \qquad \beta\sim\beta+\pi i\,. 
\end{eqnarray} 
In the Lax pair  (\ref{right Lax pair}), a spectral parameter $\lambda_R$ has been introduced. This is seemingly 
independent of $\lambda_{L_{\pm}}$ at this stage, 
but eventually there is a relation between them as we will see later. 
The Lax pair (\ref{right Lax pair}) is referred to as {\it the right Lax pair} hereafter. 

\medskip 

The relations given in (\ref{definition alpha beta}) imply the inequalities for $(C,K)$\,: 
\begin{eqnarray}
&&-\tanh^2\alpha=\frac{C^2}{A^2}=\frac{C(1+C)}{1+C-K^2}>-1\,, \\
&&-\tanh^2\beta=\frac{C^2K^2}{A^2(1+C)^2}=\frac{CK^2}{(1+C)(1+C-K^2)}>-1\,. \nonumber
\end{eqnarray}
With the kinematic restriction $C>-1$\,, these inequalities are equivalent to 
\begin{eqnarray}
(C-K+1)(C+K+1)(C-K^2+1)>0\,. \nonumber
\end{eqnarray}
Due to the relations in (\ref{definition alpha beta}) and the reality of $C$ and $K$\,, 
$\alpha$ and $\beta$ must be real or purely imaginary. 
When $C(C-K^2+1)<0$\,, $\alpha$ and $\beta$ are real (up to the shift of $i\pi n$ with $n\in{\mathbb Z}$)\,. 
On the other hand, when $C(C-K^2+1)>0$\,, $\alpha$ and $\beta$ are purely imaginary. 
Thus the allowed region of $C$ and $K$ can be expressed on the $(C,K)$-plane 
as depicted in Fig.\,\ref{plane:fig}. 

\begin{figure}[tbp]
\begin{center}
\includegraphics[scale=.35]{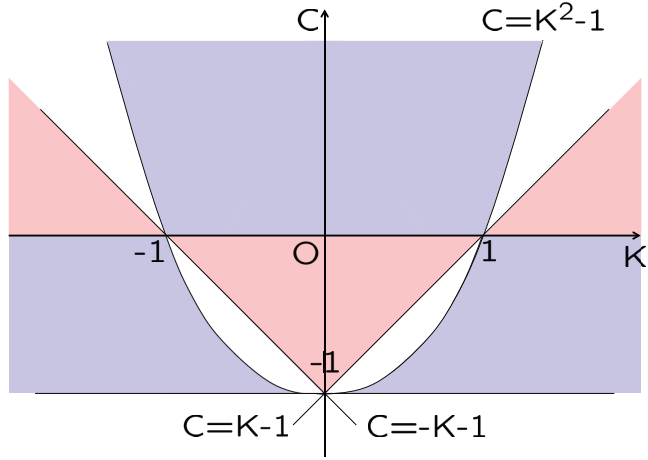}
\end{center}
\vspace*{-1cm}
\caption{\footnotesize The allowed region of $C$ and $K$\,. 
In the red region, $\alpha$ and $\beta$ are real. 
On the other hand, these are purely imaginary in the blue region. \label{plane:fig}}
\end{figure}

\medskip 

The relations given in (\ref{definition alpha beta}) can be solved for $\alpha$ and $\beta$\,, 
and hence $C$\,, $K$ (and $A$\,) are written in terms of $\alpha$ and $\beta$\,, 
\begin{eqnarray}
&&K=\frac{\sinh 2\beta}{\sinh 2\alpha}\,, \qquad
C=-\frac{\sinh(\alpha+\beta)\sinh(\alpha-\beta)}{\cosh^2\alpha}\,, 
\label{(C,K) in terms of (alpha,beta)} \\
&&A=\frac{-2i\sinh(\alpha+\beta)\sinh(\alpha-\beta)}{\sinh 2\alpha}\,. \nonumber
\end{eqnarray}
Note that $(C,K,A)$ are invariant under the shift of $\alpha$ and $\beta$ by $i\pi$\,. 

\medskip 

The equations of motion in (\ref{eom}) and the flatness condition of $J$ 
are reproduced from the commutation relation,  
\begin{eqnarray}
\left[\partial_t-L^R_t(x;\lambda_R),\partial_x-L^R_x(x;\lambda_R)\right]=0\,. 
\label{flatness right Lax pair}
\end{eqnarray}

\subsubsection*{Monodromy matrix} 

With the right Lax pair given in (\ref{right Lax pair})\,, the associated monodromy matrix is defined as 
\begin{eqnarray}
U^R(\lambda_R) \equiv {\rm P}\exp\left[\int^\infty_{-\infty}\!\!\!dx~L^R_x(x;\lambda_R)\right]\,. 
\end{eqnarray}
The flatness of the Lax pair (\ref{flatness right Lax pair}) ensures that the monodromy matrix 
is a conserved quantity,  
\begin{eqnarray}
\frac{d}{dt}U^R(\lambda_R)=0\,. 
\end{eqnarray}
Thus, by expanding $U^R(\lambda_R)$, an infinite number of conserved charges 
are constructed.

\subsection{$q$-deformation of $su(2)_{\rm R}$}

In the squashed WZNW models, the $SU(2)_{\rm R}$ symmetry, which is preserved by round S$^3$\,,  
is broken to $U(1)_{\rm R}$\,, due to the deformation term. 
The Noether current for $U(1)_{\rm R}$ is given by 
\begin{eqnarray}
j^{R,3}_\mu&=&-\left[(1+C)J^3_\mu+K\epsilon_{\mu\nu}J^{3,\nu}\right] \\
&=&-\frac{\cosh\beta}{\cosh\alpha}\left(\frac{\cosh\beta}{\cosh\alpha}
J^3_\mu+\frac{\sinh\beta}{\sinh\alpha}\epsilon_{\mu\nu}J^{3,\nu}\right)\,. \nonumber
\end{eqnarray}
For later convenience, the expressions have been given in terms of $\alpha$ and $\beta$\,, 
as well as in terms of $C$ and $K$\,. 

\medskip 

It is known that there exist non-local conserved currents which correspond to 
the broken components $T^\pm$ in the case without the Wess-Zumino term \cite{KYhybrid}. 
Let us show that this is the case even in the squashed WZNW models. 

\medskip 

For this purpose, it is helpful to introduce a non-local function $\chi(x)$\,, 
\begin{eqnarray}
\chi(x)=-\frac{1}{2}\int^\infty_{-\infty}\!\!\!dy~\epsilon(x-y)j^{R,3}_t(y)\,. 
\end{eqnarray}
This function satisfies the differential equation, 
\begin{eqnarray}
\partial_\mu\chi=-\epsilon_{\mu\nu}j^{R,3,\nu}=(1+C)\epsilon_{\mu\nu}J^{3,\nu}+KJ^3_\mu\,. 
\end{eqnarray}
This relation ensures the conservation law of the non-local currents, as we will see later. 

\medskip 

With $\chi(x)$\,, the conserved, non-local currents are constructed as 
\begin{eqnarray}
j^{R,\pm}_\mu&=&{\rm e}^{\gamma^\pm\chi}i^{R,\pm}_\mu\,, \\
i^{R,\pm}_\mu&=&-\left(J^\pm_\mu+(K\pm iA)\epsilon_{\mu\nu}J^{\pm,\nu}\right) \nonumber \\
&=&-\frac{1}{\sinh 2\alpha}\left[\sinh 2\alpha J^\pm_\mu\pm\left(\cosh 2\alpha
-{\rm e}^{\mp 2\beta}\right)\epsilon_{\mu\nu}J^{\pm,\nu}\right]\,. \nonumber
\end{eqnarray}
Here $\gamma^\pm$ are defined as 
\begin{eqnarray}
\gamma^\pm \equiv \left(\frac{1+C}{C}A\mp iK\right)^{-1}
=-i{\rm e}^{\pm\beta}\,\frac{\sinh 2\alpha}{2\cosh\beta}\,. 
\end{eqnarray}
The non-local currents $j^{R,\pm}_\mu$ give rise to the conserved charges 
\begin{eqnarray}
Q^{R,\pm}=\int^\infty_{-\infty}\!\!\!dx~j^{R,\pm}_t(x)\,, 
\end{eqnarray}
as well as the standard Noether charge of $U(1)_{\rm R}$\,, 
\begin{eqnarray}
Q^{R,3}=\int^\infty_{-\infty}\!\!\!dx~j^{R,3}_t(x)\,. 
\end{eqnarray}

\medskip 

As a next step, let us compute the algebra of $Q^{R,\pm}$ and $Q^{R,3}$\,. For this purpose, 
it is necessary to evaluate the Poisson brackets of $i^{R,\pm}_t(x)$ and $j^{R,3}_t(x)$\,,  
\begin{eqnarray}
&&\Bigl\{j^{R,3}_t(x),j^{R,3}_t(y)\Bigr\}_{\rm P}=2K\partial_x\delta(x-y)\,, \\
&&\Bigl\{i^{R,\pm}_t(x),j^{R,3}_t(y)\Bigr\}_{\rm P}=\pm ii^{R,\pm}_t(x)\delta(x-y)\,, \nonumber \\
&&\Bigl\{i^{R,\pm}_t(x),i^{R,\mp}_t(y)\Bigr\}_{\rm P}=\mp ij^{R,3}_t(x)\delta(x-y)+2K\partial_x\delta(x-y)\,. \nonumber
\end{eqnarray}
Because the Poisson brackets contain non-ultra local terms 
even for the first bracket due to non-vanishing $K$\,,  
an appropriate prescription to treat the non-ultra local terms is required 
to evaluate the Poisson brackets of $Q^{R,\pm}$ and $Q^{R,3}$\,, 
as in the case of the $su(2)_{\rm L}$ algebra in the left description. 
The prescription is argued in Appendix B in detail. 

\medskip 

The resulting algebra can be regarded as a classical analogue of $q$-deformed $su(2)_{\rm R}$ \cite{Drinfeld2,Jimbo}\,, 
\begin{eqnarray}
&&\Bigl\{Q^{R,\pm},Q^{R,3}\Bigr\}_{\rm P}=\pm iQ^{R,\pm}\,, \label{q-deformed su(2)}\\
&&\Bigl\{Q^{R,+},Q^{R,-}\Bigr\}_{\rm P}=-i\frac{(\gamma^+)^2+(\gamma^-)^2}{2\gamma^+\gamma^-}\frac{2}{(\gamma^++\gamma^-)}\sinh
\left[\frac{(\gamma^++\gamma^-)}{2}Q^{R,3}\right]\,. \nonumber
\end{eqnarray}
The $q$-parameter is given by 
\begin{eqnarray}
q={\rm e}^{(\gamma^++\gamma^-)/2}={\rm e}^\gamma\,, 
\end{eqnarray}
where $\gamma$ is a new parameter defined as  
\begin{eqnarray}
\gamma \equiv \frac{\gamma^++\gamma^-}{2}=-\frac{i}{2}\sinh 2\alpha\,. \label{gamma-expression}
\end{eqnarray}
Note that there is an unfamiliar ovarall factor 
$((\gamma^+)^2+(\gamma^-)^2)/2\gamma^+\gamma^-$
in the right-hand side of the Poisson bracket between $Q^{R,+}$ and $Q^{R,-}$\,. 
However, this factor can be absorbed by rescaling $Q^{R,\pm}$ without changing the Poisson structure. 

\medskip

Similarly, there is another set of conserved, non-local currents, 
\begin{eqnarray}
\widetilde{j}^{R,\pm}_\mu&=&{\rm e}^{-\gamma^\mp\chi}\widetilde{i}^{R,\pm}_\mu\,, \\
\widetilde{i}^{R,\pm}_\mu&=&-\left(J^\pm_\mu+(K\mp iA)\epsilon_{\mu\nu}J^{\pm,\nu}\right) \nonumber \\
&=&-\frac{1}{\sinh 2\alpha}\left[\sinh 2\alpha J^\pm_\mu\mp\left(\cosh 2\alpha-{\rm e}^{\pm 2\beta}\right)\epsilon_{\mu\nu}J^{\pm,\nu}\right]\,, \nonumber
\end{eqnarray}
and the associated charges are 
\begin{eqnarray}
\widetilde{Q}^{R,\pm}=\int^\infty_{-\infty}\!\!\!dx~\widetilde{j}^{R,\pm}_t(x)\,. 
\end{eqnarray}
Note that $\widetilde{j}^{R,\pm}$ are related to $j^{R,\pm}_\mu(x)$ by flipping the sign of $A$\,. 
The sign flip of $A$ is translated to the following transformation laws, 
\begin{eqnarray}
(\alpha,\beta)\to(-\alpha,-\beta)\,, \qquad
\gamma^\pm\to-\gamma^\mp\,. 
\end{eqnarray}
Thus, with the  sign flip of $A$\,, 
the algebra of $Q^{R,3}$ and $\widetilde{Q}^{R,\pm}$ is obtained as 
\begin{eqnarray}
&&\Bigl\{\widetilde{Q}^{R,\pm},Q^{R,3}\Bigr\}_{\rm P}=\pm i\widetilde{Q}^{R,\pm}\,, \\
&&\Bigl\{\widetilde{Q}^{R,+},\widetilde{Q}^{R,-}\Bigr\}_{\rm P}
=-i\frac{(\gamma^+)^2+(\gamma^-)^2}{2\gamma^+\gamma^-}\frac{2}{(\gamma^++\gamma^-)}\sinh
\left[\frac{(\gamma^++\gamma^-)}{2}Q^{R,3}\right]\,. \nonumber
\end{eqnarray}
The algebra is also the $q$-deformed $su(2)_{\rm R}$ with the same $q$-parameter.

\subsection{Expansions of the monodromy matrix}

The next task is to argue an infinite-dimensional extension of $q$-deformed $su(2)_{\rm R}$\,. 
For this purpose, let us consider two expansions of the monodromy matrix $U^R(\lambda_R)$ 
with respect to the spectral parameter $\lambda_R$\,. 
The expansion points are the same as in the case of squashed sigma models \cite{KMY-QAA} 
and the monodromy matrix is expanded around ${\rm Re}\lambda_R=\pm\infty$\,. 
Note that the spectral parameter $\lambda_R$ is periodically identified as 
\begin{eqnarray}
\lambda_R\sim\lambda_R+2\pi i\,, 
\end{eqnarray}
and hence the spectral parameter $\lambda_R$ is regarded as living on a cylinder. 
As a result, the point ${\rm Re}\lambda_R=+\infty$ is certainly different from the point ${\rm Re}\lambda_R=-\infty$\,. 
Thus the two expansion points give rise to two different sets of conserved charges. 
Here the expressions of the conserved charges are explicitly obtained. 

\medskip 

First of all, recall the concrete expression of $L^R_x(x;\lambda_R)\,: $
\begin{eqnarray}
&&L^R_x(x;\lambda_R) \\
&=&-\frac{1}{\sinh(\alpha+\beta+\lambda_R)\sinh(\alpha-\beta-\lambda_R)} \nonumber \\
&&\times\Bigl\{T^3\Bigl[-\frac{\sinh\alpha}{\cosh\alpha}\left(\cosh^2\beta\cosh\lambda_R\sinh\lambda_R
+\sinh\beta\cosh\beta\sinh^2\lambda_R\right)J^3_t \Bigr.\Bigr. \nonumber \\
&&\hspace{1.5cm}+\left(\sinh(\alpha+\beta)\sinh(\alpha-\beta)\cosh^2\lambda_R 
\right. \nonumber \\ &&\hspace{2cm}\Bigl.\left.
-\sinh\beta\cosh\beta\cosh\lambda_R\sinh\lambda_R
-\sinh^2\alpha\sinh^2\lambda_R\right)J^3_x\Bigr] \nonumber \\
&&\qquad+T^+\Bigl[-\sinh\alpha\cosh\alpha\sinh\lambda_RJ^-_t\Bigr. \nonumber \\
&&\hspace{2cm}\Bigl.+\left(\sinh(\alpha+\beta)\sinh(\alpha-\beta)\cosh\lambda_R 
-\sinh\beta\cosh\beta\sinh\lambda_R\right)J^-_x\Bigr] \nonumber \\
&&\qquad+T^-\Bigl[-\sinh\alpha\cosh\alpha\sinh\lambda_RJ^+_t\Bigr. \nonumber \\
&&\hspace{2cm}\Bigl.+\left(\sinh(\alpha+\beta)\sinh(\alpha-\beta)\cosh\lambda_R 
-\sinh\beta\cosh\beta\sinh\lambda_R\right)J^+_x\Bigr] \Bigr\}\,. \nonumber
\end{eqnarray}
This expression is useful to expand $U^R(\lambda_R)$\,.

\subsubsection*{Expansion around ${\rm Re}\lambda_R=-\infty$}

Let us first expand $U^R(\lambda_R)$ around ${\rm Re}\lambda_R=-\infty$\,. 
It is convenient to perform a transformation for the spectral parameter as  
\begin{eqnarray}
z_R={\rm e}^{\lambda_R}\,. 
\end{eqnarray}
The expansion around ${\rm Re}\lambda_R=-\infty$ corresponds 
to the one around $z_R=0$\,. Then the expanded monodromy matrix is expected to be the following form, 
\begin{eqnarray}
U^R(\lambda_R)={\rm e}^{q_{(0)}/2}\exp\left[\sum_{n=1}^\infty z_R^nq_{(n)}\right]{\rm e}^{q_{(0)}/2}\,. 
\label{expected}
\end{eqnarray}
Here $q_{(n)}~(n\geq 0)$ are conserved charges to be determined by a concrete expansion of $U^R(\lambda_R)$\,.

\medskip 

Note that $L^R_x(\lambda_R)$ is expanded around $z_R=0$ as 
\begin{eqnarray}
L^R_x(x;\lambda_R) &=&i\gamma^+T^3j^{R,3}_t 
-z_R{\rm e}^{2\beta}\sinh 2\alpha\left(T^-i^{R,+}_t+T^+\widetilde{i}^{R,-}_t\right) \nonumber \\
&&-z_R^2{\rm e}^{2\beta}\sinh^2 2\alpha~T^3\left(j^{R,3}_x+\frac{\cosh 2\alpha}{\sinh 2\alpha}j^{R,3}_t\right) \nonumber \\
&&-z_R^3{\rm e}^{4\beta}\sinh^2 2\alpha\left[T^-\left(i^{R,+}_x+\frac{\cosh 2\alpha}{\sinh 2\alpha}i^{R,+}_t\right)
+T^+\left(\widetilde{i}^{R,-}_x+\frac{\cosh 2\alpha}{\sinh 2\alpha}\widetilde{i}^{R,-}_t\right)\right] \nonumber \\
&&+{\mathcal O}(z_R^4)\,. \label{Lax-ex}
\end{eqnarray}
By comparing the expansion of the expected form (\ref{expected}) 
\begin{eqnarray}
U^R(\lambda_R)
&=&{\rm e}^{q_{(0)}/2}\left[
1+z_Rq_{(1)}
+z_R^2\left(q_{(2)}+\frac{1}{2}q_{(1)}^2\right) \right. \nonumber \\
&&\hspace{1.5cm}\left.+z_R^3\left(q_{(3)}+\frac{1}{2}\bigl\{q_{(1)},q_{(2)}\bigr\}+\frac{1}{6}q_{(1)}^3\right)
+{\mathcal O}(z_R^4)\right]{\rm e}^{q_{(0)}/2}\,, \nonumber
\end{eqnarray}
with the direct expansion of  $U^R(\lambda_R)$ using (\ref{Lax-ex})\,,  
the expressions of $q_{(n)}$ are fixed as follows:  
\begin{eqnarray}
&&q_{(0)}=i\gamma^+T^3Q^{R,3}_{(0)}\,, \\
&&q_{(1)}=-{\rm e}^{2\beta}\sinh 2\alpha\left(T^-Q^{R,+}_{(1)}+T^+\widetilde{Q}^{R,-}_{(1)}\right)\,, \nonumber \\
&&q_{(2)}=-\frac{i}{2}{\rm e}^{4\beta}\sinh^2 2\alpha T^3Q^{R,3}_{(2)}\,, \nonumber \\
&&q_{(3)}=-\frac{1}{4}{\rm e}^{6\beta}\sinh^3 2\alpha\left(T^-Q^{R,+}_{(3)}+T^+
\widetilde{Q}^{R,-}_{(3)}\right)\,. \nonumber
\end{eqnarray}
where the following quantities have been introduced, 
\begin{eqnarray}
Q^{R,3}_{(0)}&=&\int^\infty_{-\infty}\!\!\!dx~j^{R,3}_t(x)\quad(\,=Q^{R,3}\,)\,, \\
Q^{R,+}_{(1)}&=&\int^\infty_{-\infty}\!\!\!dx~j^{R,+}_t(x)\quad(\,=Q^{R,+}\,)\,, \quad
\widetilde{Q}^{R,-}_{(1)}=\int^\infty_{-\infty}\!\!\!dx~\widetilde{j}^{R,-}_t(x)\quad(\,=\widetilde{Q}^{R,-}\,)\,, \nonumber \\
Q^{R,3}_{(2)}&=&-\int^\infty_{-\infty}\!\!\!dx\int^\infty_{-\infty}\!\!\!dy~\epsilon(x-y)j^{R,+}_t(x)\widetilde{j}^{R,-}_t(y) \nonumber \\
&&-2i{\rm e}^{-2\beta}\int^\infty_{-\infty}\!\!\!dx~j^{R,3}_x(x)-2i{\rm e}^{-2\beta}\frac{\cosh 2\alpha}{\sinh 2\alpha}Q^{R,3}_{(0)}\,, \nonumber \\
Q^{R,+}_{(3)}&=&\frac{1}{2}\int^\infty_{-\infty}\!\!\!dx\int^\infty_{-\infty}\!\!\!dy\int^\infty_{-\infty}\!\!\!dz~\epsilon(x-y)\epsilon(x-z)\widetilde{j}^{R,-}_t(x)j^{R,+}_t(y)j^{R,+}_t(z) \nonumber \\
&&+2i{\rm e}^{-2\beta}\int^\infty_{-\infty}\!\!\!dx\int^\infty_{-\infty}\!\!\!dy~\epsilon(x-y)j^{R,+}_t(x)\left(j^{R,3}_x(y)+\frac{\cosh 2\alpha}{\sinh 2\alpha}j^{R,3}_t(y)\right) \nonumber \\
&&+\frac{4{\rm e}^{-2\beta}}{\sinh 2\alpha}\int^\infty_{-\infty}\!\!\!dxj^{R,+}_x(x)
-\frac{1}{6}\widetilde{Q}^{R,+}_{(1)}\left(Q^{R,+}_{(1)}\right)^2
+\frac{4{\rm e}^{-2\beta}\cosh 2\alpha}{\sinh^2 2\alpha}Q^{R,+}_{(1)}\,, \nonumber \\
\widetilde{Q}^{R,-}_{(3)}&=&\frac{1}{2}\int^\infty_{-\infty}\!\!\!dx\int^\infty_{-\infty}\!\!\!dy\int^\infty_{-\infty}\!\!\!dz~\epsilon(x-y)\epsilon(x-z)j^{R,+}_t(x)\widetilde{j}^{R,-}_t(y)\widetilde{j}^{R,-}_t(z) \nonumber \\
&&-2i{\rm e}^{-2\beta}\int^\infty_{-\infty}\!\!\!dx\int^\infty_{-\infty}\!\!\!dy~\epsilon(x-y)\widetilde{j}^{R,-}_t(x)\left(j^{R,3}_x(y)+\frac{\cosh 2\alpha}{\sinh 2\alpha}j^{R,3}_t(y)\right) \nonumber \\
&&+\frac{4{\rm e}^{-2\beta}}{\sinh 2\alpha}\int^\infty_{-\infty}\!\!\!dx\widetilde{j}^{R,-}_x(x)
-\frac{1}{6}Q^{R,+}_{(1)}\left(\widetilde{Q}^{R,-}_{(1)}\right)^2
+\frac{4{\rm e}^{-2\beta}\cosh 2\alpha}{\sinh^2 2\alpha}\widetilde{Q}^{R,-}_{(1)}\,. \nonumber \\
&&\hspace{5cm}\vdots \nonumber
\end{eqnarray}
These are also conserved charges and will play an important role later in studying 
the tower structure of the conserved charges that indicates an infinite-dimensional 
extension of $q$-deformed $su(2)_{\rm R}$\,. 

\subsubsection*{Expansion around ${\rm Re}\lambda_R=+\infty$}

Let us next consider another expansion around ${\rm Re}\lambda_R=+\infty$\,. 
It is helpful to introduce a new parametrization $z'_R$ defined as 
\begin{eqnarray}
z'_R = z_R^{-1}
={\rm e}^{-\lambda_R}\,. 
\end{eqnarray}
The expansion around ${\rm Re}\lambda_R=+\infty$ corresponds to the one around $z'_R=0$\,. 
Then the expected form of the expanded monodromy matrix $U^R(\lambda_R)$ is given by 
\begin{eqnarray}
U^R(\lambda_R)
={\rm e}^{\bar{q}_{(0)}/2}\exp\left[\sum_{n=1}^\infty {z'_R}^n\bar{q}_{(n)}\right]{\rm e}^{\bar{q}_{(0)}/2}\,. 
\end{eqnarray}
Here $\bar{q}_{(n)}~(n\geq 0)$ are conserved charges. 

\medskip 

The Lax pair $L^R_x(\lambda_R)$ is expanded around $z'_R=0$ as  
\begin{eqnarray}
L^R_x(x;\lambda_R) 
&=&i\gamma^-T^3j^{R,3}_t 
+z'_R{\rm e}^{-2\beta}\sinh 2\alpha\left(T^+i^{R,-}_t+T^-\widetilde{i}^{R,+}_t\right) \nonumber \\
&&-{z'_R}^2{\rm e}^{-2\beta}\sinh^2 2\alpha~T^3\left(j^{R,3}_x
-\frac{\cosh 2\alpha}{\sinh 2\alpha}j^{R,3}_t\right) \nonumber \\
&&-{z'_R}^3{\rm e}^{-4\beta}\sinh^2 2\alpha\left[T^+\left(i^{R,-}_x-\frac{\cosh 2\alpha}{\sinh 2\alpha}i^{R,-}_t\right)
+T^-\left(\widetilde{i}^{R,+}_x-\frac{\cosh 2\alpha}{\sinh 2\alpha}\widetilde{i}^{R,+}_t\right)\right] \nonumber \\
&&+{\mathcal O}({z'_R}^4)\,. \label{Lax-ex2}
\end{eqnarray}
Thus, by comparing the expansion of the expected monodromy matrix 
\begin{eqnarray}
U^R(\lambda_R)
&=&{\rm e}^{\bar{q}_{(0)}/2}\left[
1+z'_R\bar{q}_{(1)}
+{z'_R}^2\left(\bar{q}_{(2)}+\frac{1}{2}\bar{q}_{(1)}^2\right) \right. \nonumber \\
&&\hspace{1.5cm}\left.+{z'_R}^3\left(\bar{q}_{(3)}
+\frac{1}{2}\bigl\{\bar{q}_{(1)},\bar{q}_{(2)}\bigr\}+\frac{1}{6}\bar{q}_{(1)}^3\right)
+{\mathcal O}({z'_R}^4)\right]{\rm e}^{\bar{q}_{(0)}/2}\,, \nonumber
\end{eqnarray}
with the direct expansion using (\ref{Lax-ex2})\,, $\bar{q}_{(n)}$ are obtained as  
\begin{eqnarray}
&&\bar{q}_{(0)}=i\gamma^-T^3Q^{R,3}_{(0)}\,, \\
&&\bar{q}_{(1)}={\rm e}^{-2\beta}\sinh 2\alpha\left(T^+Q^{R,-}_{(1)}+T^-\widetilde{Q}^{R,+}_{(1)}\right)\,, \nonumber \\
&&\bar{q}_{(2)}=-\frac{i}{2}{\rm e}^{-4\beta}\sinh^2 2\alpha T^3\bar{Q}^{R,3}_{(2)}\,, \nonumber \\
&&\bar{q}_{(3)}=\frac{1}{4}{\rm e}^{-6\beta}\sinh^3 2\alpha\left(T^+Q^{R,-}_{(3)}+T^-
\widetilde{Q}^{R,+}_{(3)}\right)\,. \nonumber
\end{eqnarray}
Here the conserved quantities have been introduced as 
\begin{eqnarray}
Q^{R,3}_{(0)}&=&\int^\infty_{-\infty}\!\!\!dx~j^{R,3}_t(x)\quad(\,=Q^{R,3}\,)\,, \\
Q^{R,-}_{(1)}&=&\int^\infty_{-\infty}\!\!\!dx~j^{R,-}_t(x)\quad(\,=Q^{R,-}\,)\,, \quad
\widetilde{Q}^{R,+}_{(1)}=\int^\infty_{-\infty}\!\!\!dx~\widetilde{j}^{R,+}_t(x)\quad(\,=\widetilde{Q}^{R,+}\,)\,, \nonumber \\
\bar{Q}^{R,3}_{(2)}&=&\int^\infty_{-\infty}\!\!\!dx\int^\infty_{-\infty}\!\!\!dy~\epsilon(x-y)j^{R,-}_t(x)\widetilde{j}^{R,+}_t(y) \nonumber \\
&&-2i{\rm e}^{2\beta}\int^\infty_{-\infty}\!\!\!dx~j^{R,3}_x(x)+2i{\rm e}^{2\beta}\frac{\cosh 2\alpha}{\sinh 2\alpha}Q^{R,3}_{(0)}\,, \nonumber \\
Q^{R,-}_{(3)}&=&\frac{1}{2}\int^\infty_{-\infty}\!\!\!dx\int^\infty_{-\infty}\!\!\!dy\int^\infty_{-\infty}\!\!\!dz~\epsilon(x-y)\epsilon(x-z)\widetilde{j}^{R,+}_t(x)j^{R,-}_t(y)j^{R,-}_t(z) \nonumber \\
&&-2i{\rm e}^{2\beta}\int^\infty_{-\infty}\!\!\!dx\int^\infty_{-\infty}\!\!\!dy~\epsilon(x-y)j^{R,-}_t(x)\left(j^{R,3}_x(y)-\frac{\cosh 2\alpha}{\sinh 2\alpha}j^{R,3}_t(y)\right) \nonumber \\
&&-\frac{4{\rm e}^{2\beta}}{\sinh 2\alpha}\int^\infty_{-\infty}\!\!\!dxj^{R,-}_x(x)
-\frac{1}{6}\widetilde{Q}^{R,+}_{(1)}\left(Q^{R,-}_{(1)}\right)^2
+\frac{4{\rm e}^{2\beta}\cosh 2\alpha}{\sinh^2 2\alpha}Q^{R,-}_{(1)}\,, \nonumber \\
\widetilde{Q}^{R,+}_{(3)}&=&\frac{1}{2}\int^\infty_{-\infty}\!\!\!dx\int^\infty_{-\infty}\!\!\!dy\int^\infty_{-\infty}\!\!\!dz~\epsilon(x-y)\epsilon(x-z)j^{R,-}_t(x)\widetilde{j}^{R,+}_t(y)\widetilde{j}^{R,+}_t(z) \nonumber \\
&&+2i{\rm e}^{2\beta}\int^\infty_{-\infty}\!\!\!dx\int^\infty_{-\infty}\!\!\!dy~\epsilon(x-y)\widetilde{j}^{R,+}_t(x)\left(j^{R,3}_x(y)-\frac{\cosh 2\alpha}{\sinh 2\alpha}j^{R,3}_t(y)\right) \nonumber \\
&&-\frac{4{\rm e}^{2\beta}}{\sinh 2\alpha}\int^\infty_{-\infty}\!\!\!dx\widetilde{j}^{R,+}_x(x)
-\frac{1}{6}Q^{R,+}_{(1)}\left(\widetilde{Q}^{R,+}_{(1)}\right)^2
+\frac{4{\rm e}^{2\beta}\cosh 2\alpha}{\sinh^2 2\alpha}\widetilde{Q}^{R,+}_{(1)}\,. \nonumber \\
&&\hspace{5cm}\vdots \nonumber
\end{eqnarray}
These are also important to see the tower structure of the conserved charges 
in addition to the previous expansion.

\subsection{An infinite-dimensional extension of $q$-deformed $su(2)_{\rm R}$ \label{dQAA:sec}}

The next is to compute the Poisson brackets of 
$Q^R_{(n)}$ and $\widetilde{Q}^R_{(n)}$~($n\geq 0$)\,. 

\medskip 

For this purpose, the following Poisson brackets are useful, 
\begin{eqnarray}
\Bigl\{j^{R,3}_t(x),j^{R,3}_t(y)\Bigr\}_{\rm P}&=&2K\partial_x\delta(x-y)\,, \\
\Bigl\{i^{R,\pm}_t(x),j^{R,3}_t(y)\Bigr\}_{\rm P}&=&\pm ii^{R,\pm}_t(x)\delta(x-y)\,, \nonumber \\
\Bigl\{i^{R,\pm}_t(x),i^{R,\mp}_t(y)\Bigr\}_{\rm P}&=&\mp ij^{R,3}_t(x)\delta(x-y)+2K\partial_x\delta(x-y)\,. \nonumber \\
\Bigl\{\widetilde{i}^{R,\pm}_t(x),j^{R,3}_t(y)\Bigr\}_{\rm P}&=&\pm i\widetilde{i}^{R,\pm}_t(x)\delta(x-y)\,, \nonumber \\
\Bigl\{\widetilde{i}^{R,\pm}_t(x),\widetilde{i}^{R,\mp}_t(y)\Bigr\}_{\rm P}&=&\mp ij^{R,3}_t(x)\delta(x-y)+2K\partial_x\delta(x-y)\,, \nonumber \\
\Bigl\{i^{R,\pm}_t(x),\widetilde{i}^{R,\mp}_t(y)\Bigr\}_{\rm P}
&=&\mp i\Bigl[\Bigl(1\mp\frac{2\cosh^2\alpha\sinh\beta}{\cosh\beta}\Bigr)j^{R,3}_t(x) \Bigr. \nonumber \\ 
&&\hspace{2cm}\Bigl.\pm 2\sinh\alpha\cosh\alpha j^{R,3}_x(x)\Bigr]\delta(x-y) \nonumber \\
&&+2\left(K\pm iA\right)\partial_x\delta(x-y)\,, \nonumber \\
\Bigl\{j^{R,3}_x(x),i^{R,\pm}_t(y)\Bigr\}_{\rm P}
&=&\mp i\Bigl[{\rm e}^{\pm 2\beta}i^{R,\pm}_t(x) 
\pm\frac{{\rm e}^{\pm\beta}
\left(1-\cosh 2\alpha\cosh 2\beta\right)}{\sinh 2\alpha\cosh\beta}i^{R,\pm}_t(x)\Bigr]\delta(x-y)\,, 
\nonumber \\
\Bigl\{j^{R,3}_x(x),\widetilde{i}^{R,\pm}_t(y)\Bigr\}_{\rm P}
&=&\mp i\Bigl[{\rm e}^{\mp 2\beta}\widetilde{i}^{R,\pm}_t(x) 
\mp\frac{{\rm e}^{\mp\beta}
\left(1-\cosh 2\alpha\cosh 2\beta\right)}{\sinh 2\alpha\cosh\beta}
\widetilde{i}^{R,\pm}_t(x)\Bigr]\delta(x-y)\,. \nonumber
\end{eqnarray}
Then, with the above relations, the Poisson brackets are computed as  
\begin{eqnarray}
&&\Bigl\{Q^{R,\pm}_{(1)},Q^{R,3}_{(0)}\Bigr\}_{\rm P}=\pm iQ^{R,\pm}_{(1)}\,, \nonumber\\
&&\Bigl\{Q^{R,+}_{(1)},Q^{R,-}_{(1)}\Bigr\}_{\rm P}=-i\frac{(\gamma^+)^2+(\gamma^-)^2}{2\gamma^+\gamma^-}\frac{2}{(\gamma^++\gamma^-)}\sinh
\left[\frac{(\gamma^++\gamma^-)}{2}Q^{R,3}_{(0)}\right]\,, \nonumber \\
&&\Bigl\{\widetilde{Q}^{R,\pm}_{(1)},Q^{R,3}_{(0)}\Bigr\}_{\rm P}=\pm iQ^{R,\pm}_{(1)}\,, \nonumber\\
&&\Bigl\{\widetilde{Q}^{R,+}_{(1)},\widetilde{Q}^{R,-}_{(1)}\Bigr\}_{\rm P}=-i\frac{(\gamma^+)^2+(\gamma^-)^2}{2\gamma^+\gamma^-}\frac{2}{(\gamma^++\gamma^-)}\sinh
\left[\frac{(\gamma^++\gamma^-)}{2}Q^{R,3}_{(0)}\right]\,, \nonumber \\
&&\Bigl\{Q^{R,-}_{(1)},\widetilde{Q}^{R,-}_{(1)}\Bigr\}_{\rm P}=0\,, \nonumber \\
&&\Bigl\{Q^{R,+}_{(1)},\widetilde{Q}^{R,+}_{(1)}\Bigr\}_{\rm P}=0\,, \nonumber \\
&&\Bigl\{Q^{R,-}_{(1)},\widetilde{Q}^{R,+}_{(1)}\Bigr\}_{\rm P}
=i\frac{\gamma^-}{\gamma^+}\gamma\bar{Q}^{R,3}_{(2)}\,, \nonumber \\
&&\Bigl\{Q^{R,+}_{(1)},\widetilde{Q}^{R,-}_{(1)}\Bigr\}_{\rm P}
=i\frac{\gamma^+}{\gamma^-}\gamma Q^{R,3}_{(2)}\,, \nonumber \\
&&\Bigl\{\bar{Q}^{R,3}_{(2)},Q^{R,-}_{(1)}\Bigr\}_{\rm P} 
=-i\frac{\gamma^-}{\gamma^+}\gamma\left[Q^{R,-}_{(3)}
+\frac{2}{3}\widetilde{Q}^{R,+}_{(1)}\left(Q^{R,-}_{(1)}\right)^2\right]\,, \nonumber \\
&&\Bigl\{\bar{Q}^{R,3}_{(2)},\widetilde{Q}^{R,+}_{(1)}\Bigr\}_{\rm P} 
=i\frac{\gamma^-}{\gamma^+}\gamma\left[\widetilde{Q}^{R,+}_{(3)}
+\frac{2}{3}Q^{R,-}_{(1)}\left(\widetilde{Q}^{R,+}_{(1)}\right)^2\right]\,, \nonumber \\
&&\Bigl\{Q^{R,3}_{(2)},\widetilde{Q}^{R,-}_{(1)}\Bigr\}_{\rm P} 
=i\frac{\gamma^+}{\gamma^-}\gamma\left[\widetilde{Q}^{R,-}_{(3)}
+\frac{2}{3}Q^{R,+}_{(1)}\left(\widetilde{Q}^{R,-}_{(1)}\right)^2\right]\,, \nonumber \\
&&\Bigl\{Q^{R,3}_{(2)},Q^{R,+}_{(1)}\Bigr\}_{\rm P} 
=-i\frac{\gamma^+}{\gamma^-}\gamma\left[Q^{R,+}_{(3)}
+\frac{2}{3}\widetilde{Q}^{R,-}_{(1)}\left(Q^{R,+}_{(1)}\right)^2\right]\,, \nonumber \\
&&\Bigl\{Q^{R,-}_{(3)},Q^{R,-}_{(1)}\Bigr\}_{\rm P}
=-\frac{i}{3}\frac{\gamma^-}{\gamma^+}\gamma\left(Q^{R,-}_{(1)}\right)^2\bar{Q}^{R,3}_{(2)}\,, \nonumber \\
&&\Bigl\{\widetilde{Q}^{R,+}_{(3)},\widetilde{Q}^{R,+}_{(1)}\Bigr\}_{\rm P}
=\frac{i}{3}\frac{\gamma^-}{\gamma^+}\gamma\left(\widetilde{Q}^{R,+}_{(1)}\right)^2\bar{Q}^{R,3}_{(2)}\,, \nonumber \\
&&\Bigl\{\widetilde{Q}^{R,-}_{(3)},\widetilde{Q}^{R,-}_{(1)}\Bigr\}_{\rm P}
=\frac{i}{3}\frac{\gamma^+}{\gamma^-}\gamma\left(\widetilde{Q}^{R,-}_{(1)}\right)^2 Q^{R,3}_{(2)}\,, \nonumber \\
&&\Bigl\{Q^{R,+}_{(3)},Q^{R,+}_{(1)}\Bigr\}_{\rm P}
=-\frac{i}{3}\frac{\gamma^+}{\gamma^-}\gamma\left(Q^{R,+}_{(1)}\right)^2 Q^{R,3}_{(2)}\,. \nonumber
\end{eqnarray}
Again, we should be careful for non-ultra local terms. 
For the detail, see Appendix B. 

\medskip 

In addition, the Serre-like relations are evaluated as follows:  
\begin{eqnarray}
\Bigl\{Q^{R,\pm}_{(1)},\Bigl\{Q^{R,\pm}_{(1)},\Bigl\{Q^{R,\pm}_{(1)},\widetilde{Q}^{R,\mp}_{(1)}\Bigr\}_{\rm P}\Bigr\}_{\rm P}\Bigr\}_{\rm P} 
&=&-\left(\frac{\gamma^\pm}{\gamma^\mp}\gamma\right)^2\left(Q^{R,\pm}_{(1)}\right)^2
\Bigl\{Q^{R,\pm}_{(1)},\widetilde{Q}^{R,\mp}_{(1)}\Bigr\}_{\rm P}\,, 
\label{q-Serre} \\
\Bigl\{\widetilde{Q}^{R,\pm}_{(1)},\Bigl\{\widetilde{Q}^{R,\pm}_{(1)},\Bigl\{\widetilde{Q}^{R,\pm}_{(1)},
Q^{R,\mp}_{(1)}\Bigr\}_{\rm P}\Bigr\}_{\rm P}\Bigr\}_{\rm P}
&=&-\left(\frac{\gamma^\mp}{\gamma^\pm}\gamma\right)^2\left(\widetilde{Q}^{R,\pm}_{(1)}\right)^2
\Bigl\{\widetilde{Q}^{R,\pm}_{(1)},Q^{R,\mp}_{(1)}\Bigr\}_{\rm P}\,. \nonumber
\end{eqnarray}
The relations in (\ref{q-Serre}) may be interpreted as deformations of the classical analogue of $q$-Serre relations 
in the standard quantum affine algebra $U_q(\widehat{su(2)}_{\rm R})$
and ensure that the resulting algebra exhibits the similar tower structure. 
The higher-level charges are obtained from the level 1 charges 
by taking the Poisson bracket repeatedly as with $U_q(\widehat{su(2)}_{\rm R})$\,. 
The deformed algebra obtained here may be lifted up to the quantum level. 
Although we have not yet succeeded to find out its mathematical formulation,  
a two-parameter quantum toroidal algebra\cite{GKV} would be a possible candidate of it.  

\medskip

The tower structure of the conserved charges is depicted in Fig.\,\ref{tower:fig}. 
The overall coefficient of the Poisson bracket is affected by the deformation. 
When going up the tower of charges, the factor 
$\gamma^-/\gamma^+$ is multiplied to the Poisson bracket. 
On the other hand, when going down the tower, the factor 
$\gamma^+/\gamma^-$ is multiplied to the Poisson bracket. 

\begin{figure}[tbp]
\begin{center}
\includegraphics[scale=.5]{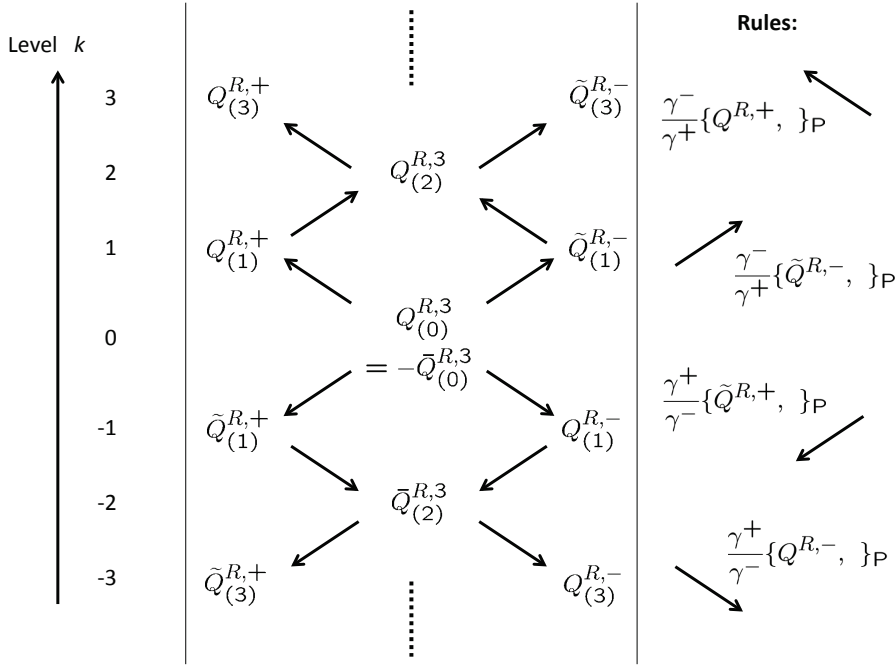}
\end{center}
\vspace*{-1.0cm}
\caption{\footnotesize The tower structure of the conserved charges. \label{tower:fig}}
\vspace*{0.5cm}
\end{figure}

\medskip 

When $\gamma^+/\gamma^-={\rm e}^{2\beta}=\pm 1$\,, 
the relations in (\ref{q-Serre}) are reduced to the classical analogue of $q$-Serre relations. 
The first condition ${\rm e}^{2\beta}=1$ means that $K=0$ and corresponds 
to squashed sigma models without the Wess-Zumino term.  
The second condition ${\rm e}^{2\beta}=-1$ means $K=0$ and $C=-1$\,. 
In the $C \to -1$ limit, $O(3)$ non-linear sigma models are reproduced 
as it is obvious from the classical action. 
However, because this limit is singular, the second condition would also be singular.

\subsection{The $r$/$s$-matrices}

The $r$/$s$-matrices associated with the right Lax pair in (\ref{right Lax pair}) are computed from the Poisson bracket 
between the spatial components of the Lax pair,
\begin{eqnarray}
\Bigl\{L^R_x(x;\lambda_R)\stackrel{\otimes}{,}L^R_x(y;\mu_R)\Bigr\}_{\rm P} 
&=&
\Bigl[r^R(\lambda_R,\mu_R),L^R_x(x;\lambda_R)\otimes 1+1\otimes L^R_x(y;\mu_R)\Bigr]\delta(x-y) \nonumber \\
&&-\Bigl[s^R(\lambda_R,\mu_R),L^R_x(x;\lambda_R)\otimes 1-1\otimes L^R_x(y;\mu_R)\Bigr]\delta(x-y) \nonumber \\
&&-2s^R(\lambda_R,\mu_R)\partial_x\delta(x-y)\,. \nonumber
\end{eqnarray}
The resulting $r$/$s$-matrices are 
\begin{eqnarray}
r^R(\lambda_R,\mu_R) &=& \frac{h^R_{\alpha,\beta}(\lambda_R)+h^R_{\alpha,\beta}(\mu_R)}{2\sinh(\lambda_R-\mu_R)} 
\left[T^-\otimes T^+ +T^+\otimes T^- +\cosh(\lambda_R-\mu_R)T^3\otimes T^3\right] 
\nonumber  \\
&&-\frac{h^R_{\alpha,\beta}(\lambda_R)-h^R_{\alpha,\beta}(\mu_R)}{2}\tanh\beta T^3\otimes T^3\,, \nonumber \\
s^R(\lambda_R,\mu_R) 
&=& \frac{h^R_{\alpha,\beta}(\lambda_R)-h^R_{\alpha,\beta}(\mu_R)}{2\sinh(\lambda_R-\mu_R)}
\left[T^-\otimes T^+ +T^+\otimes T^- +\cosh(\lambda_R-\mu_R)T^3\otimes T^3\right] \nonumber \\
&&-\frac{h^R_{\alpha,\beta}(\lambda_R)+h^R_{\alpha,\beta}(\mu_R)}{2}\tanh\beta T^3\otimes T^3\,, \label{right r/s-matrices} 
\end{eqnarray}
where a scalar function $h^R_{\alpha,\beta}(\lambda)$ is defined as 
\begin{eqnarray}
h^R_{\alpha,\beta}(\lambda) \equiv \frac{\sinh 2\alpha
\sinh^2\lambda}{2\sinh(\alpha+\beta+\lambda)\sinh(\alpha-\beta-\lambda)}\,. 
\end{eqnarray}
It is straightforward to show that the classical $r$/$s$-matrices given in (\ref{right r/s-matrices}) 
satisfy the extended classical Yang-Baxter equation, 
\begin{eqnarray}
&& \left[\left(r-s\right)^R_{12}(\lambda_R,\mu_R),\left(r+s\right)^R_{13}(\lambda_R,\nu_R)\right] 
+\left[\left(r+s\right)^R_{12}(\lambda_R,\mu_R),\left(r+s\right)^R_{23}(\mu_R,\nu_R)\right] \nonumber \\
&& \hspace*{4cm} +\left[\left(r+s\right)^R_{13}(\lambda_R,\nu_R),\left(r+s\right)^R_{23}(\mu_R,\nu_R)\right]
=0\,. \nonumber
\end{eqnarray}
Thus the classical integrability has been ensured\footnote{
The $r$/$s$-matrices depend on $\lambda_R$ and $\mu_R$ individually (not only $\lambda_R-\mu_R$\,), 
though they satisfy the extended classical Yang-Baxter equation. 
Hence it is unclear to classify the $r$/$s$-matrices in a well-known manner.}. 

\medskip 

In the previous subsection, a deformation of the quantum affine algebra has been shown explicitly. 
According to the deformation, the $r$/$s$-matrices given in (\ref{right r/s-matrices}) 
are also deformed by additional terms proportional to $\tanh\beta\,T^3\otimes T^3$\,. 

\medskip 

It seems likely that the additional terms cannot be eliminated by any gauge transformations. 
This would indicate that the deformed quantum affine algebra cannot be mapped 
to the standard quantum affine algebra. 

\medskip 

When $\beta=0$ and $\alpha\neq 0$ (i.e., $K=0$)\,, 
$h^R_{\alpha,\beta}(\lambda)$ is reduced to 
\begin{eqnarray}
h^R_{\alpha,0}(\lambda) \equiv \frac{\sinh 2\alpha
\sinh^2\lambda}{2\sinh(\alpha+\lambda)\sinh(\alpha-\lambda)}\,, 
\end{eqnarray}
and the deformation terms proportional to $\tanh\beta\, T^3\otimes T^3$ vanish. 
Thus the $r$/$s$-matrices obtained in \cite{KMY-QAA} are reproduced.

\section{The left-right duality}

Let us show the gauge-equivalence, which is referred to as {\it left-right duality}\,, 
between the right Lax pair $L^R_\mu(\lambda_R)$ and a pair of the left Lax pairs $L^{L_\pm}_\mu(\lambda_{L_\pm})$ 
under a certain relation between the spectral parameters and the rescaling of $sl(2)$ generators. 
The equivalence is shown in the case without the Wess-Zumino term \cite{KMY-monodromy}. 
The analysis here is a generalization of the result obtained in \cite{KMY-monodromy}. 

\subsection*{The fundamental domains of the spectral parameters}

It is useful to realize the fundamental domains of the spectral parameters. 

\medskip

Recall that a pair of the Lax pairs in the left description [See (\ref{left-Lax pair})] are given by 
\begin{eqnarray}
&&L^{L_\pm}_t(x;\lambda_{L_\pm}) = \frac{1}{2}\left[L^{L_\pm}_+(x;\lambda_{L_\pm})
+L^{L_\pm}_-(x;\lambda_{L_\pm})\right]\,, \nonumber \\
&&L^{L_\pm}_x(x;\lambda_{L_\pm}) = \frac{1}{2}\left[L^{L_\pm}_+(x;\lambda_{L_\pm})
-L^{L_\pm}_-(x;\lambda_{L_\pm})\right]\,, \nonumber \\
&&L^{L_\pm}_+(x;\lambda_{L_\pm}) = \frac{1}{1+\lambda_{L_\pm}}j^{L_\pm}_+\,, \qquad
L^{L_\pm}_-(x;\lambda_{L_\pm}) = \frac{1}{1-\lambda_{L_\pm}}j^{L_\pm}_-\,, \nonumber
\end{eqnarray}
where each of the spectral parameters take the values on a Riemann sphere 
with two punctures: $\lambda_{L_\pm}\in{\mathbb C}\cup\{\infty\}$\,. 
The punctures come from the fact that each Lax pair has two poles 
at $\lambda_{L_\pm}=\pm 1$\,, as depicted in Fig.\,\ref{lambdaL:fig}.

\begin{figure}[tbp]
\begin{center}
\includegraphics[scale=.4]{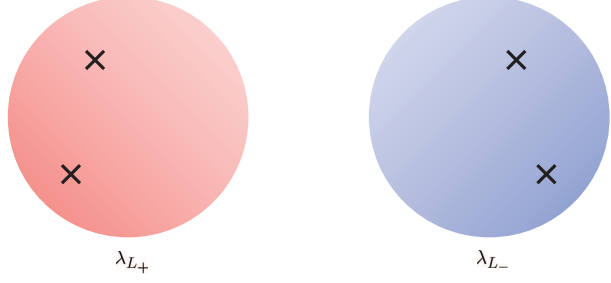}
\end{center}
\vspace*{-3.5cm}
\caption{\footnotesize Two Riemann spheres with two punctures. \label{lambdaL:fig}}
\vspace*{0.5cm}
\end{figure}

\medskip

On the other hand, the Lax pair in the right description [See (\ref{right Lax pair})] is given by  
\begin{eqnarray}
&&L^R_t(x;\lambda_R) = \frac{1}{2}\left[L^R_+(x;\lambda_R)+L^R_-(x;\lambda_R)\right]\,, \nonumber \\
&&L^R_x(x;\lambda_R) = \frac{1}{2}\left[L^R_+(x;\lambda_R)-L^R_-(x;\lambda_R)\right]\,, \nonumber \\
&&L^R_\pm(x;\lambda_R) = -\frac{\sinh\left(\alpha\pm\beta\right)}{\sinh\left[\alpha\pm(\beta+\lambda_R)\right]}
\left[T^+J^-_\pm +T^-J^+_\pm +\frac{\cosh\left(\alpha\pm\lambda_R\right)}{\cosh\alpha}T^3 J^3_\pm\right]\,. 
\nonumber
\end{eqnarray}
The spectral parameter $\lambda_R$ is periodically identified as 
\begin{eqnarray}
\lambda_R\sim\lambda_R+2\pi i\,, 
\end{eqnarray}
and hence it takes the values on a cylinder. 
Because the right Lax pair is regular in the ${\rm Re}\lambda_R\to\pm\infty$ limit, 
the fundamental domain of spectral parameter can be regarded as a Riemann sphere under the map 
\begin{eqnarray}
\lambda_R\to z_R={\rm e}^{\lambda_R}\,. 
\end{eqnarray}
It is obvious that the right Lax pair has four poles at 
\begin{eqnarray}
z_R={\rm e}^{\pm\alpha-\beta}\,,~ -{\rm e}^{\pm\alpha-\beta}\,. 
\end{eqnarray}
Thus the domain of the spectral parameter $z_R$ is regarded as a Riemann sphere 
with four punctures as depicted in Fig.\,\ref{z-sphere:fig}\footnote{
When $\alpha$ and $\beta$ are real, the four poles are on the real axis of the $z_R$-plane. 
When $\alpha$ and $\beta$ are purely imaginary, the four poles are on the unit circle with the center at the origin. }. 

\begin{figure}[tbp]
\vspace*{-1.0cm}
\begin{center}
\includegraphics[scale=.4]{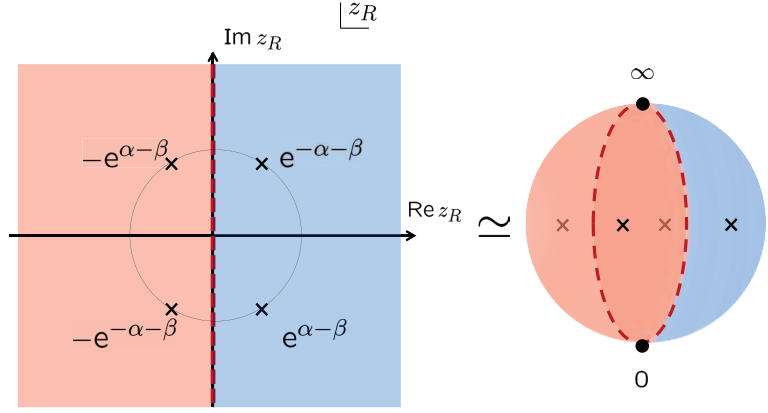}
\end{center}
\vspace*{-2.0cm}
\caption{\footnotesize The Riemann sphere with four punctures 
for $\alpha$ and $\beta$ are purely imaginary. \label{z-sphere:fig}}
\vspace*{0.5cm}
\end{figure}

\medskip 

In the following, we show that the Riemann sphere of $z_R$ is related to 
a pair of the Riemann spheres of  $\lambda_{L_\pm}$\,. 

\subsection*{The reduced right descriptions}

As a next step, we introduce the {\it reduced right descriptions}. 
Lax pairs in the reduced right descriptions are obtained from the original right Lax pair 
through the following isomorphisms of the $sl(2)$ algebra, 
\begin{eqnarray}
&&T^\pm ~~\to~~{\rm e}^{\pm\lambda_R}T^\pm={\rm e}^{-i\lambda_RT^3}T^\pm{\rm e}^{i\lambda_RT^3} \label{reduced right Lax pair +} \\
 {\rm or}\quad &&T^\pm ~~\to~~{\rm e}^{\mp\lambda_R}T^\pm={\rm e}^{i\lambda_RT^3}T^\pm{\rm e}^{-i\lambda_RT^3}\,. \label{reduced right Lax pair -}
\end{eqnarray}
Because $T^3$ is invariant under the isomorphisms, the generators can be rewritten into 
\begin{eqnarray}
T^a \to{\rm e}^{\mp i\lambda_RT^3}T^a{\rm e}^{\pm i\lambda_RT^3}\,. 
\end{eqnarray}

\medskip 

One of the resulting Lax pairs is 
\begin{eqnarray}
L^{R_+}_\pm(x;\lambda_{R_+})
&=& -\frac{\sinh\left(\alpha\pm\beta\right)}{\sinh\left[\alpha\pm\left(\beta+\lambda_{R_+}\right)\right]} \\
&&\quad\times\left[{\rm e}^{+\lambda_{R_+}}T^+J^-_\pm +{\rm e}^{-\lambda_{R_+}}T^-J^+_\pm +\frac{\cosh\left(\alpha\pm\lambda_{R_+}\right)}{\cosh\alpha}T^3 J^3_\pm\right]\,. \nonumber
\end{eqnarray}
For later convenience, we change the subscript of the spectral parameter from $R$ to $R_+$\,. 
This reduced right Lax pair corresponds to the former isomorphism (\ref{reduced right Lax pair +})\,. 
Similarly,  the other reduced right Lax pair is defined as 
\begin{eqnarray}
L^{R_-}_\pm(x;\lambda_{R_-})
&=& -\frac{\sinh\left(\alpha\pm\beta\right)}{\sinh\left[\alpha\pm\left(\beta+\lambda_{R_-}\right)\right]} \\
&&\quad\times\left[{\rm e}^{-\lambda_{R_-}}T^+J^-_\pm +{\rm e}^{+\lambda_{R_-}}T^-J^+_\pm +\frac{\cosh\left(\alpha\pm\lambda_{R_-}\right)}{\cosh\alpha}T^3 J^3_\pm\right]\,. \nonumber
\end{eqnarray}
For this reduced right Lax pair, its spectral parameter is denoted as $\lambda_{R_-}$\,. 
This reduced right Lax pair corresponds to the later isomorphism (\ref{reduced right Lax pair -})\,. 

\medskip 

Note that the domains of the spectral parameters $\lambda_{R_\pm}$ have the periodicities,  
\begin{eqnarray}
\lambda_{R_\pm}\sim\lambda_{R_\pm}+\pi i\,, 
\end{eqnarray}
and hence the squares of $z_{R_{\pm}}$
\begin{eqnarray}
z_{R_\pm}^2={\rm e}^{2\lambda_{R_\pm}}
\end{eqnarray}
live on the Riemann spheres\footnote{
Note that $z_{R_\pm}$ and $-z_{R_\pm}$ cannot be distinguished and 
hence $z_{R_\pm}^2$ takes the value on a Riemann sphere.}. 
Each of the reduced right Lax pairs $L^{R_\pm}_\mu(\lambda_{R_\pm})$ 
has two poles in its fundamental domain,  
\begin{eqnarray}
z_{R_\pm}^2={\rm e}^{2\alpha-2\beta} \quad {\rm and} \quad {\rm e}^{-2\alpha-2\beta}\,. 
\end{eqnarray}
As a result, the fundamental domains of $z_{R_\pm}^2$ are a pair of Riemann spheres with two punctures. 

\medskip 

So far, we have shown that the right description is decomposed to a couple of 
the reduced right descriptions. The statement we want to show is that 
each of the Lax pair in the reduced right description is equivalent to 
each of the Lax pairs in the left description through a gauge transformation with 
a certain identification of the spectral parameters. 

\subsubsection*{A relation of the spectral parameters}

Then the next task is to find out a relation of the spectral parameters in the left and right 
descriptions.  
Because both $\lambda_{L_\pm}$ and $z_{R_\pm}^2$ live on the two-punctured Riemann spheres, 
they should be related through a M\"obius transformation, 
\begin{eqnarray}
z_{R_\pm}^2=\frac{a\lambda_{L_\pm}+b}{c\lambda_{L_\pm}+d}\,. 
\end{eqnarray}
The constant parameter should be fixed by the pole structure and 
the correspondence of the expansion points for Yangians. 

\medskip 

The first requirement is that the position of the poles should be mapped each other. 
This condition leads to 
\begin{eqnarray}
{\rm e}^{\pm 2\alpha-2\beta}=\frac{a\pm b}{c\pm d}\,. 
\end{eqnarray}
Now we have two possibilities: 
the pole at $z_{R_\pm}^2={\rm e}^{2\alpha-2\beta}$ corresponds to 1) the pole at $\lambda_{L_\pm}=1$\,, 
or 2) the pole at $\lambda_{L_\pm}=-1$\,. However, we should here take the possibility 1) so that 
the resulting M\"obius transformation reproduces the result 
in \cite{KMY-monodromy} when $K=0$\,.

\medskip 

Then, recall that the Yangian charges $Q^{L_\pm}_{(n)}$ are obtained 
by expanding the (reduced) right Lax pairs around $z_{R_\pm}^2=1$\,. 
From this information, the following relation is obtained,  
\begin{eqnarray}
1=\frac{a}{c}\,. 
\end{eqnarray}
Thus the final result is given by 
\begin{eqnarray}
z_{R_\pm}^2=\frac{\sinh 2\alpha\lambda_{L_\pm}+\cosh 2\alpha-{\rm e}^{-2\beta}}{\sinh 2\alpha\lambda_{L_\pm}
-\cosh 2\alpha+{\rm e}^{2\beta}}=\frac{\lambda_{L_\pm}+K+iA}{\lambda_{L_\pm}+K-iA}\,, 
\label{Mobius map between z_Rpm and lambda_Lpm}
\end{eqnarray}
or equivalently, 
\begin{eqnarray}
\frac{\lambda_{L_\pm}+K}{iA}=\frac{1}{\tanh\lambda_{R_\pm}}\,. 
\end{eqnarray}

\medskip 

Here we comment on the connection between the two-punctured $\lambda_L$-Riemann spheres 
and the four-punctured $z_R$-Riemann sphere. 
The relation between $z_R$ and $\lambda_L$ is 
also given by (\ref{Mobius map between z_Rpm and lambda_Lpm})\,:
\begin{eqnarray}
z_R^2=\frac{\lambda_{L_\pm}+K+iA}{\lambda_{L_\pm}+K-iA}\,. 
\label{Mobius map between z_R and lambda_Lpm}
\end{eqnarray}
However the interpretation is more involved because $-z_R$ is certainly a different from the point $z_R$\,. 
As a relation between $z_R$ and $\lambda_{L_\pm}$ (not $z_R^2$ and $\lambda_{L_\pm}$\,)\,, 
the relation in (\ref{Mobius map between z_R and lambda_Lpm}) indicates that 
there exists a cut between $\lambda_{L_\pm}=-K+iA$ and $\lambda_{L_\pm}=-K-iA$\,. 
Two-punctured $\lambda_L$-Riemann spheres are connected along the cut and combined into a single Riemann sphere with four punctures. 
The resulting four-punctured Riemann sphere is depicted in Fig.\ref{lambdaL0:fig}. 

\begin{figure}[tbp]
\begin{center}
\includegraphics[scale=.4]{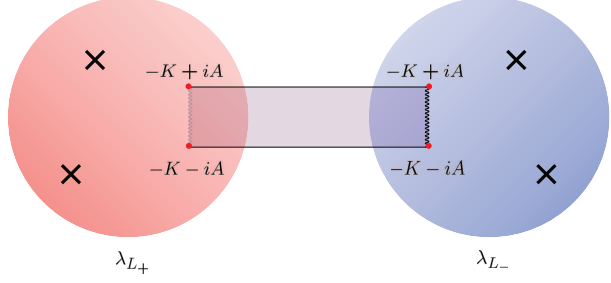}
\end{center}
\vspace*{-3.5cm}
\caption{\footnotesize 
Two-punctured $\lambda_L$-Riemann spheres are connected along the cut. \label{lambdaL0:fig}}
\vspace*{0.5cm}
\end{figure}

\subsection*{The gauge equivalence}

We concentrate on showing the gauge equivalence between $L^{R_+}_\mu(x;\lambda_{R_+})$ and 
$L^{L_+}_\mu(x;\lambda_{L_+})$\,. 
The analysis for the case with (-)-subscript is also similar. 
The gauge transformation is generated by $g$ and the relation between the spectral parameters is given by 
\begin{eqnarray}
\frac{\lambda_{L_+}+K}{iA}=\frac{1}{\tanh\lambda_{R_+}}\,. 
\label{relation between spectral parameters+}
\end{eqnarray}

\medskip 

First of all, the left Lax pair $L^{L_+}_\mu(x;\lambda_{L_+})$ can be rewritten as 
\begin{eqnarray}
&&L^{L_+}_\pm(x;\lambda_{L_+}) \\
&=& \frac{1}{1\pm \lambda_{L_+}} g \left( J_\pm - 2C{\rm tr}\left(T^3J_\pm\right)T^3 \mp KJ_\pm \mp A\left[J_\pm,T^3\right] \right)g^{-1} \nonumber \\
&=& \frac{1}{1\pm \lambda_{L_+}} g \left[ T^+\left(1 \mp K \mp iA\right)J^-_\pm +T^-\left(1 \mp K \pm iA\right)J^+_\pm \right. \nonumber \\
&&\hspace{5cm}\left.+T^3\left(1+C \mp K\right)J^3_\pm \right]g^{-1}\,. \nonumber 
\end{eqnarray}
Then a gauge-transformation of the left Lax pair is given by  
\begin{eqnarray}
&&\Bigl[L^{L_+}_\pm(x;\lambda_{L_+})\Bigr]^g \\
&\equiv& g^{-1}L^{L_+}_\pm(x;\lambda_{L_+})g -g^{-1}\partial_\pm g   \nonumber \\
&=& -J_\pm+\frac{1}{1\pm \lambda_{L_+}} \left[ T^+\left(1 \mp K \mp iA\right)J^-_\pm +T^-\left(1 \mp K \pm iA\right)J^+_\pm \right. \nonumber \\
&&\hspace{5cm}\left.+T^3\left(1+C \mp K\right)J^3_\pm \right] \nonumber \\
&=&-\left[\frac{\lambda_{L_+} +K +iA}{\lambda_{L_+}\pm 1}T^+J^-_\pm +\frac{\lambda_{L_+} 
+K -iA}{\lambda_{L_+}\pm 1}T^-J^+_\pm +\frac{\lambda_{L_+} +K \mp C}{\lambda_{L_+}\pm 1}T^3J^3_\pm \right]\,. \nonumber
\end{eqnarray}
Finally, with the relation between the spectral parameters (\ref{relation between spectral parameters+})\,, 
it becomes the reduced right Lax pair: 
\begin{eqnarray}
&&\Bigl[L^{L_+}_\pm(x;\lambda_{L_+})\Bigr]^g \\
&=& -\frac{\sinh\left(\alpha\pm\beta\right)}{\sinh\left[\alpha\pm\left(\beta+\lambda_{R_+}\right)\right]} 
\left[{\rm e}^{+\lambda_{R_+}}T^+J^-_\pm +{\rm e}^{-\lambda_{R_-}}T^-J^+_\pm +\frac{\cosh\left(\alpha\pm\lambda_{R_+}\right)}{\cosh\alpha}T^3J^3_\pm \right] \nonumber \\
&=& L^{R_+}_\pm(x;\lambda_{R_+})\,. \nonumber
\end{eqnarray}

\medskip 

Similarly, with the spectral-parameter relation 
\begin{eqnarray}
\frac{\lambda_{L_-}+K}{iA}=\frac{1}{\tanh\lambda_{R_-}}\,, 
\label{relation between spectral parameters-}
\end{eqnarray}
we can show the gauge equivalence between the other left Lax pair $L^{L_-}_\mu(x;\lambda_{L_-})$ and 
the other reduced right Lax pair $L^{R\-}_\mu(x;\lambda_{R_-})$\,: 
\begin{eqnarray}
\Bigl[L^{L_-}_\pm(x;\lambda_{L_-})\Bigr]^g =L^{R_-}_\pm(x;\lambda_{R_-})\,. 
\end{eqnarray}

\subsubsection*{Summary}

Let us summarize the results obtained so far.  
The two reduced right Lax pairs $L^{R_\pm}_\mu(x;\lambda_{R_\pm})$ have been introduced. 
Then they are obtained from the right Lax pair through the  $sl(2)$ isomorphisms, 
\begin{eqnarray}
T^a\to{\rm e}^{\mp i\lambda_RT^3}T^a{\rm e}^{\pm i\lambda_RT^3}\,. 
\label{sl(2) isomorphism}
\end{eqnarray}
The reduced right Lax pairs $L^{R_\pm}_\mu(x;\lambda_{R_\pm})$ are related to the left Lax pairs $L^{L_\pm}_\mu(x;\lambda_{L_\pm})$ 
through the gauge transformation generated by $g$\,, 
\begin{eqnarray}
\Bigl[L^{L_\pm}_\mu(x;\lambda_{L_\pm})\Bigr]^g =L^{R_\pm}_\mu(x;\lambda_{R_\pm})\,, 
\end{eqnarray}
under the relations between the spectral parameters 
\begin{eqnarray}
\frac{\lambda_{L_\pm}+K}{iA}=\frac{1}{\tanh\lambda_{R_\pm}}\,. 
\label{relation between spectral parameters}
\end{eqnarray}

\medskip

So far we have shown that $U^{L_\pm}(\lambda_{L_\pm})$ are gauge-equivalent to $U^{R_\pm}(\lambda_{R_\pm})$ 
and hence the charges for the deformed $U_q(\widehat{su(2)}_{\rm R})$ 
can also be obtained from $U^{L_\pm}(\lambda_{L_\pm})$\,. One can read off 
the expansion points from the relation (\ref{relation between spectral parameters}) and 
the fact that the expansion points for $U^{R_\pm}(\lambda_{R_\pm})$ are $z_{R_\pm}=0$ and $z_{R_\pm}=\infty$\,. 
Thus the charges for the deformed $U_q(\widehat{su(2)}_{\rm R})$ are obtained 
by expanding $U^{L_\pm}(\lambda_{L_\pm})$ with respect to 
$\lambda_{L_\pm}$ around $\lambda_{L_\pm}=-K-iA$ and $\lambda_{L_\pm}=-K+iA$\,. 
The locations of poles and the expansion points for the deformed $U_q(\widehat{su(2)}_{\rm R})$ 
and ${\mathcal Y}(su(2)_{\rm L})$ are summarized in Tab.\,\ref{list:tab}\,.

\begin{table}[htbp]
\vspace*{0.5cm}
\begin{center}
\begin{tabular}{c||c|c}
\hline 
Charges $\setminus$ Monodromies  & $\qquad U^{R_\pm}(\lambda_{R_\pm}) \qquad$ & $\qquad U^{L_\pm}(\lambda_{L_\pm}) \qquad$ \\
\hline\hline 
$Q^{R,3}_{(0)}\,,Q^{R,+}_{(1)}\,,\widetilde{Q}^{R,-}_{(1)}$ & $0$ & $-K-iA$ \\
\hline
$Q^{R,3}_{(0)}\,,Q^{R,-}_{(1)}\,,\widetilde{Q}^{R,+}_{(1)}$ & $\infty$ & $-K+iA$ \\ 
\hline
$Q^{L,a}_{(0)}\,,Q^{L,a}_{(1)}$ & $1$ & $\infty$ \\
\hline\hline 
local charges & ${\rm e}^{\pm 2\alpha-2\beta}$ & $\pm 1$ \\
\hline 
\end{tabular}
\caption{\footnotesize The conserved charges and the expansion points of monodromy matrices are listed. 
The expansion points of $U^{R_\pm}(\lambda_{R_\pm})$ are described in terms of $z_{R_\pm}^2$\,. 
\label{list:tab}}
\end{center}
\end{table}

\subsubsection*{The gauge transformation of the $r/s$-matrices}

The $r$/$s$-matrices for the reduced right Lax pairs can be obtained 
from the $r$/$s$-matrices for the left Lax pairs. 
Recall the gauge-transformation laws of the $r$/$s$-matrices, 
\begin{eqnarray}
&&(r+s)^{R_\pm}(\lambda_{R_\pm},\mu_{R_\pm})\delta(x-y) 
\label{(r+s) gauge transformation}\\
&=& g^{-1}(x)\otimes g^{-1}(y) \nonumber \\
&&\times\Biggl[(r+s)^{L_\pm}(\lambda_{L_\pm},\mu_{L_\pm})\delta(x-y)-\Bigl\{L^{L_\pm}_x(x;\lambda_{L_\pm})\stackrel{\otimes}{,}g(y)\Bigr\}_{\rm P}1\otimes g^{-1}(y)\Biggr] \nonumber \\
&&\times g(x)\otimes g(y)\,, \nonumber \\
&&(r-s)^{R_\pm}(\lambda_{R_\pm},\mu_{R_\pm})\delta(x-y) \\
&=& g^{-1}(x)\otimes g^{-1}(y) \nonumber 
\label{(r-s) gauge transformation}\\
&&\times\Biggl[(r-s)^{L_\pm}(\lambda_{L_\pm},\mu_{L_\pm})\delta(x-y)-\Bigl\{g(x)\stackrel{\otimes}{,}L^{L_\pm}_x(y;\mu_{L_\pm})\Bigr\}_{\rm P}1\otimes g^{-1}(y)\Biggr] \nonumber \\
&&\times g(x)\otimes g(y)\,. \nonumber
\end{eqnarray}
Thus the Poisson brackets 
\begin{eqnarray}
&&\Bigl\{L^{L_\pm}_x(x;\lambda_L)\otimes,g(y)\Bigr\}_{\rm P}1\otimes g^{-1}(y) 
\label{Lg Poisson}\\
&=&\frac{-1}{1-\lambda_L^2}\left\{\left(-\lambda_L-\frac{K}{1+C}\right)\left(T^-\otimes T^+ +T^+\otimes T^- +T^3\otimes T^3\right) \right. \nonumber \\
&&\qquad\qquad \left.\pm A\left[T^-\otimes T^+ +T^+\otimes T^- +T^3\otimes T^3,1\otimes gT^3g^{-1}(y)\right] \right. \nonumber \\
&&\qquad\qquad \left. +\frac{CK}{1+C}\left[\left[T^-\otimes T^+ +T^+\otimes T^- +T^3\otimes T^3,1\otimes gT^3g^{-1}(y)\right],1\otimes gT^3g^{-1}(y)\right]\right\} \nonumber \\
&&\qquad\times\delta(x-y)\,, \nonumber \\
&&\Bigl\{g(x)\otimes,L^{L_\pm}_x(y;\mu_L)\Bigr\}_{\rm P}g^{-1}(x)\otimes 1 
\label{gL Poisson}\\
&=&\frac{1}{1-\mu_L^2}\left\{\left(-\mu_L-\frac{K}{1+C}\right)\left(T^-\otimes T^+ +T^+\otimes T^- +T^3\otimes T^3\right) \right. \nonumber \\
&&\qquad\qquad \left.\pm A\left[T^-\otimes T^+ +T^+\otimes T^- +T^3\otimes T^3,gT^3g^{-1}(x)\otimes 1\right] \right. \nonumber \\
&&\qquad\qquad \left. +\frac{CK}{1+C}\left[\left[T^-\otimes T^+ +T^+\otimes T^- +T^3\otimes T^3,gT^3g^{-1}(x)\otimes 1\right],gT^3g^{-1}(x)\otimes 1\right]\right\} \nonumber \\
&&\qquad\times\delta(x-y)\,, \nonumber
\end{eqnarray}
are needed to compute the $r$/$s$-matrices for the reduced right Lax pair. 

\medskip 

With the explicit expression of the $r$/$s$-matrices for the left Lax pairs (\ref{left r/s-matrices})\,, 
the relation (\ref{Lg Poisson}) and the maps between spectral parameters (\ref{relation between spectral parameters})\,, 
we can evaluate the right hand side of (\ref{(r+s) gauge transformation}) as 
\begin{eqnarray}
&&(r+s)^{R_\pm}(\lambda_{R_\pm},\mu_{R_\pm})\delta(x-y) \\
&=&\frac{1}{\sinh(\lambda_{R_\pm}-\mu_{R_\pm})}\frac{\sinh 2\alpha\sinh^2\lambda_{R_\pm}}{2\sinh(\alpha+\beta+\lambda_{R_\pm})\sinh(\alpha-\beta-\lambda_{R_\pm})} \nonumber \\
&&\times\Biggl\{\left[\cosh(\lambda_{R_\pm}-\mu_{R_\pm}) -\tanh\beta\sinh(\lambda_{R_\pm}-\mu_{R_\pm})\right]T^3\otimes T^3 \nonumber \\
&&\qquad +{\rm e}^{\pm(\lambda_{R_\pm}-\mu_{R_\pm})}T^-\otimes T^+ +{\rm e}^{\mp(\lambda_{R_\pm}-\mu_{R_\pm})}T^+\otimes T^-\Biggr\}\delta(x-y)\,. \nonumber
\end{eqnarray}
Similarly, the following can be obtained 
\begin{eqnarray}
&&(r-s)^{R_\pm}(\lambda_{R_\pm},\mu_{R_\pm})\delta(x-y) \\
&=&\frac{1}{\sinh(\lambda_{R_\pm}-\mu_{R_\pm})}\frac{\sinh 2\alpha\sinh^2\mu_{R_\pm}}{2\sinh(\alpha+\beta+\mu_{R_\pm})\sinh(\alpha-\beta-\mu_{R_\pm})} \nonumber \\
&&\times\Biggl\{\left[\cosh(\lambda_{R_\pm}-\mu_{R_\pm}) +\tanh\beta\sinh(\lambda_{R_\pm}-\mu_{R_\pm})\right]T^3\otimes T^3 \nonumber \\
&&\qquad +{\rm e}^{\pm(\lambda_{R_\pm}-\mu_{R_\pm})}T^-\otimes T^+ +{\rm e}^{\mp(\lambda_{R_\pm}-\mu_{R_\pm})}T^+\otimes T^-\Biggr\}\delta(x-y)\,. \nonumber
\end{eqnarray}
Thus the resulting $r$/$s$-matrices for the reduced right Lax pairs are 
\begin{eqnarray}
r^{R_\pm}(\lambda_{R_\pm},\mu_{R_\pm}) &=& \frac{h^R_{\alpha,\beta}(\lambda_{R_\pm})+h^R_{\alpha,\beta}(\mu_{R_\pm})}{2\sinh(\lambda_{R_\pm}-\mu_{R_\pm})} \nonumber \\
&&\times\left[{\rm e}^{\pm(\lambda_{R_\pm}-\mu_{R_\pm})}T^-\otimes T^+ +{\rm e}^{\mp(\lambda_{R_\pm}-\mu_{R_\pm})}T^+\otimes T^- \right. \nonumber \\
&&\qquad \left.+\cosh(\lambda_{R_\pm}-\mu_{R_\pm})T^3\otimes T^3\right] \nonumber \\
&&-\frac{\left(h^R_{\alpha,\beta}(\lambda_{R_\pm})-h^R_{\alpha,\beta}(\mu_{R_\pm})\right)}{2}\tanh\beta T^3\otimes T^3\,, \nonumber \\
s^{R_\pm}(\lambda_{R_\pm},\mu_{R_\pm}) &=& \frac{h^R_{\alpha,\beta}(\lambda_{R_\pm})-h^R_{\alpha,\beta}(\mu_{R_\pm})}{2\sinh(\lambda_{R_\pm}-\mu_{R_\pm})} \nonumber \\
&&\times\left[{\rm e}^{\pm(\lambda_{R_\pm}-\mu_{R_\pm})}T^-\otimes T^+ +{\rm e}^{\mp(\lambda_{R_\pm}-\mu_{R_\pm})}T^+\otimes T^- \right. \nonumber \\
&&\qquad\left.+\cosh(\lambda_{R_\pm}-\mu_{R_\pm})T^3\otimes T^3\right] \nonumber \\
&&-\frac{\left(h^R_{\alpha,\beta}(\lambda_{R_\pm})+h^R_{\alpha,\beta}(\mu_{R_\pm})\right)}{2}\tanh\beta T^3\otimes T^3\,. \nonumber
\end{eqnarray}
The $r$/$s$-matrices for the right Lax pair (\ref{right r/s-matrices}) are obtained from these $r$/$s$-matrices for the reduced right Lax pairs 
by taking the isomorphism of $sl(2)$ (\ref{sl(2) isomorphism}) into account.

\section{The degenerate limits}

So far, it has been shown that $U(1)_{\rm R}$ is enhanced to $q$-deformed $su(2)_{\rm R}$\,. 
Then let us argue the $q\to 1$ limit in which $q$-deformed $su(2)_{\rm R}$ 
degenerates to the original $su(2)_{\rm R}$\,. 

\medskip 

In fact, there are two kinds of the $q \to 1$ limit, which are referred to as  {\it the degenerate limits}. 
Since the $q$-parameter is written as $q={\rm e}^\gamma$\,, the limits are specified by the condition, $\gamma=0$\,. 
From (\ref{gamma-expression})\,, this condition is equivalent to 
\begin{eqnarray}
\alpha=0 \qquad {\rm or} \qquad \alpha=\frac{\pi i}{2}\,. 
\end{eqnarray}
In the following, we will argue each of the limits. 

\subsection{$\alpha=0$}

We begin with the $\alpha\to 0$ limit. Recall the relation between the original parameters $(C,K)$ 
and the parameters $(\alpha,\beta)$\,, (\ref{(C,K) in terms of (alpha,beta)})\,. 
Due to the relation and the finiteness of the original parameters, 
the condition $\alpha=0$ requires the condition $\beta=0$\,. 

\medskip 

Let us rescale $\alpha$ and $\beta$ as 
\begin{eqnarray}
\alpha\to\epsilon\,\alpha\,, \qquad
\beta\to\epsilon\,\beta\,, 
\end{eqnarray}
and take the $\epsilon\to 0$ limit. 
Then the relation (\ref{(C,K) in terms of (alpha,beta)}) is reduced to the following: 
\begin{eqnarray}
K=\frac{\beta}{\alpha}\,, \qquad 
C=0\,, \qquad
A=0\,. 
\end{eqnarray}
Thus this limit reproduces the undeformed $SU(2)$ WZNW models. 

\medskip

Next we consider the limit for the right Lax pair. 
Let us perform the redefinition of spectral parameter,  
\begin{eqnarray}
\lambda_R\to\epsilon\alpha\lambda_R-\epsilon\beta\,, 
\end{eqnarray}
and take the $\epsilon\to 0$ limit. 
Then the Lax pair is evaluated as 
\begin{eqnarray}
L^R_\pm(x;\lambda_R)= -\frac{1\pm K}{1\pm\lambda_R}\left[T^+J^-_\pm +T^-J^+_\pm +T^3 J^3_\pm\right]\,. 
\end{eqnarray}

\medskip 

The same procedure gives rise to the $r$/$s$-matrices,  
\begin{eqnarray}
r^R(\lambda_R,\mu_R)
&=& \frac{h^R_0(\lambda_R)+h^R_0(\mu_R)}{2(\lambda_R-\mu_R)} 
\left[T^-\otimes T^+ +T^+\otimes T^- +T^3\otimes T^3\right]\,, \\
s^R(\lambda_R,\mu_R)
&=& \frac{h^R_0(\lambda_R)-h^R_0(\mu_R)}{2(\lambda_R-\mu_R)}
\left[T^-\otimes T^+ +T^+\otimes T^- +T^3\otimes T^3\right]\,. \nonumber 
\end{eqnarray}
Here the function $h^R_0(\lambda)$ is given by 
\begin{eqnarray}
h^R_0(\lambda)=\frac{\left(\lambda-K\right)^2}{1-\lambda^2}\,. 
\end{eqnarray}

\medskip

Since $\gamma^\pm=0$ and $A=0$\,, 
two sets of non-local currents $j^{R,\pm}_\mu$ and $\widetilde{j}^{R,\pm}_\mu$ degenerate into 
\begin{eqnarray}
j^{R,\pm}_\mu=\widetilde{j}^{R,\pm}_\mu=-\left(J^\pm_\mu+K\epsilon_{\mu\nu}J^{\pm,\nu}\right)\equiv {\mathcal J}^{R,\pm}_\mu\,. 
\end{eqnarray}
For later convenience, let us introduce the following matrix valued conserved current: 
\begin{eqnarray}
{\mathcal J}^R_\mu=T^+{\mathcal J}^{R,-}_\mu+T^-{\mathcal J}^{R,+}_\mu+T^3{\mathcal J}^{R,3}_\mu\,, 
\end{eqnarray}
where the $T^3$ component ${\mathcal J}^{R,3}_\mu$ is defined as 
\begin{eqnarray}
{\mathcal J}^{R,3}_\mu \equiv j^{R,3}_\mu =-\left(J^3_\mu+K\epsilon_{\mu\nu}J^{3,\nu}\right)\,. 
\end{eqnarray}
It is an easy task to show that this conserved current satisfies the flatness condition: 
\begin{eqnarray}
\epsilon^{\mu\nu}\left(\partial_\mu {\mathcal J}^R_\nu -{\mathcal J}^R_\mu {\mathcal J}^R_\nu\right)=0\,,
\end{eqnarray}
and the right Lax pair can be written in terms of the conserved current ${\mathcal J}^R_\mu$ as 
\begin{eqnarray}
L^R_\pm(x;\lambda_R)=\frac{{\mathcal J}^R_\pm}{1\pm\lambda_R}\,. 
\end{eqnarray}

\medskip 

Then it is possible to construct an infinite number of conserved charges from ${\mathcal J}^R_\mu$\,, by following the BIZZ construction \cite{BIZZ}: 
\begin{eqnarray}
{\mathcal Y}^{R,a}_{(0)}&=&\int^\infty_{-\infty}\!\!\!dx~{\mathcal J}^{R,a}_t(x)\,, \\
{\mathcal Y}^{R,a}_{(1)}&=&\frac{1}{4}\int^\infty_{-\infty}\!\!\!dx\int^\infty_{-\infty}\!\!\!dy~\varepsilon^a_{~bc}
\epsilon(x-y){\mathcal J}^{R.b}_t(x){\mathcal J}^{R.c}_t(y)
-\int^\infty_{-\infty}\!\!\!dx~{\mathcal J}^{R.a}_x(x)\,, \nonumber \\
&&\hspace{3cm}\vdots \nonumber
\end{eqnarray}
The current algebra for ${\mathcal J}^{R,a}_\mu$ is given by 
\begin{eqnarray}
\Bigl\{{\mathcal J}^{R,a}_t(x),{\mathcal J}^{R,b}_t(y)\Bigr\}_{\rm P}
&=&\varepsilon^{ab}_{~~c}{\mathcal J}^{R,c}_t(x)+2K\delta^{ab}\partial_x\delta(x-y)\,, \\
\Bigl\{{\mathcal J}^{R,a}_t(x),{\mathcal J}^{R,b}_x(y)\Bigr\}_{\rm P}
&=&\varepsilon^{ab}_{~~c}{\mathcal J}^{R,c}_x(x)+\left(1+K^2\right)\delta^{ab}\partial_x\delta(x-y)\,, \nonumber \\
\Bigl\{{\mathcal J}^{R,a}_x(x),{\mathcal J}^{R,b}_x(y)\Bigr\}_{\rm P}
&=&-K^2\varepsilon^{ab}_{~~c}{\mathcal J}^{R,c}_t(x)+2K\varepsilon^{ab}_{~~c}{\mathcal J}^{R,c}_x(x)\delta(x-y) \nonumber \\
&&+2K\delta^{ab}\partial_x\delta(x-y)\,. \nonumber
\end{eqnarray}
With this current algebra, one can show that the algebra formed by a set of ${\mathcal Y}^R_{(n)}$ is 
Yangian ${\mathcal Y}(su(2)_{\rm L})$ with an appropriate regularization of non-ultra local terms.

\subsection{$\alpha=\pi i/2$}

The other degenerate limit is next considered. Again, due to the relation (\ref{(C,K) 
in terms of (alpha,beta)}) and the finiteness of $(C,K)$\,, 
the condition $\alpha=\pi i/2$ implies that $\beta=\pi i/2$\,. 
So it is useful to redefine $\alpha$ and $\beta$ as 
\begin{eqnarray}
\alpha \quad \to \quad \frac{\pi i}{2} + \epsilon\,\alpha\,, \qquad
\beta \quad \to \quad \frac{\pi i}{2} + \epsilon\,\beta\,. 
\end{eqnarray}
The $\epsilon\to 0$ limit with the redefinition leads to the following relations,  
\begin{eqnarray}
K=\frac{\beta}{\alpha}\,, \qquad 
C=\frac{\beta^2}{\alpha^2}-1\,, \qquad
A=0\,. 
\end{eqnarray}
This limit describes the points specified by $C=K^2-1$\,. 
As in the case with $\alpha=0$\,, the coefficient of the improvement term vanishes, $A=0$\,. 
At the points, the effect of the squashing parameter $C$ is canceled by the Wess-Zumino term. 
As a result, the right description becomes isotropic even though the metric of target space is deformed. 

\medskip

To see the right Lax pair at the points,  
the spectral parameter should be redefined as  
\begin{eqnarray}
\lambda_R \quad \to \quad  \epsilon\,\alpha\,\lambda_R - \epsilon\,\beta\,. 
\end{eqnarray}
Then the $\epsilon\to 0$ limit leads to the following expression of the Lax pair,  
\begin{eqnarray}
L^R_\pm(x;\lambda_R)= -\frac{1\pm K}{1\pm\lambda_R}\left[T^+J^-_\pm +T^-J^+_\pm +\left(1\mp K\pm\lambda_R\right)T^3 J^3_\pm\right]\,. 
\label{right Lax pair alpha=pii/2}
\end{eqnarray}

\medskip 

Similarly, the $r$/$s$-matrices are obtained as 
\begin{eqnarray}
r^R(\lambda_R,\mu_R)
&=& \frac{h^R_{\pi i/2}(\lambda_R)+h^R_{\pi i/2}(\mu_R)}{2(\lambda_R-\mu_R)} 
\left[T^-\otimes T^+ +T^+\otimes T^- +T^3\otimes T^3\right] \\
&&-\frac{h^R_{\pi i/2}(\lambda_R)-h^R_{\pi i/2}(\mu_R)}{2K}T^3\otimes T^3\,, \nonumber \\
s^R(\lambda_R,\mu_R)
&=& \frac{h^R_{\pi i/2}(\lambda_R)-h^R_{\pi i/2}(\mu_R)}{2(\lambda_R-\mu_R)}
\left[T^-\otimes T^+ +T^+\otimes T^- +T^3\otimes T^3\right] \nonumber \\
&&-\frac{h^R_{\pi i/2}(\lambda_R)+h^R_{\pi i/2}(\mu_R)}{2K}T^3\otimes T^3\,, \nonumber
\end{eqnarray}
where the function $h^R_{\pi i/2}(\lambda)$ is defined as 
\begin{eqnarray}
h^R_{\pi i/2}(\lambda) \equiv \frac{\left(\lambda-K\right)^2}{1-\lambda^2}\,. 
\end{eqnarray}

\medskip

In this limit, the following relations are satisfied, 
\[
\gamma^\pm=-\gamma^\mp=\pm i/K\,, \qquad A=0\,.
\] 
Then $\widetilde{j}^{R,\pm}_\mu$ coincide with $j^{R,\pm}_\mu$ as 
\begin{eqnarray}
j_{\mu}^{R,\pm}=\widetilde{j}^{R,\pm}_\mu=-{\rm e}^{\pm i\chi/K}
\left(J^\pm+K\epsilon_{\mu\nu}J^{\pm,\nu}\right) \equiv {\mathcal J}^{R,\pm}_\mu\,.
\end{eqnarray}
Note that ${\mathcal J}^{R,\pm}_\mu$ remain non-local currents, 
in contrast to the $\alpha=0$ case. It is useful to define a matrix valued conserved current as 
\begin{eqnarray}
{\mathcal J}^R_\mu=T^+{\mathcal J}^{R,-}_\mu+T^-{\mathcal J}^{R,+}_\mu+T^3{\mathcal J}^{R,3}_\mu\,, 
\end{eqnarray}
where the $T^3$ component ${\mathcal J}^{R,3}_\mu$ is defined as 
\begin{eqnarray}
{\mathcal J}^{R,3}_\mu \equiv -j^{R,3}_\mu = \left(K^2J^3_\mu+K\epsilon_{\mu\nu}J^{3,\nu}\right)\,. 
\end{eqnarray}
Here the sign of $j^{R,3}_\mu$ is flipped for later convenience. 
Although ${\mathcal J}^R_\mu$ is non-local, it satisfies the flatness condition,
\begin{eqnarray}
\epsilon^{\mu\nu}\left(\partial_\mu {\mathcal J}^R_\nu-{\mathcal J}^R_\mu {\mathcal J}^R_\nu\right)=0\,. 
\end{eqnarray}
Thus an infinite number of conserved charges can be constructed from ${\mathcal J}^R_\mu$ like 
\begin{eqnarray}
{\mathcal Y}^{R,a}_{(0)}&=&\int^\infty_{-\infty}\!\!\!dx~{\mathcal J}^{R,a}_t(x)\,, \\
{\mathcal Y}^{R,a}_{(1)}&=&\frac{1}{4}\int^\infty_{-\infty}\!\!\!dx\int^\infty_{-\infty}\!\!\!dy~\varepsilon^a_{~bc}
\epsilon(x-y){\mathcal J}^{R.b}_t(x){\mathcal J}^{R.c}_t(y)
-\int^\infty_{-\infty}\!\!\!dx~{\mathcal J}^{R.a}_x(x)\,, \nonumber \\
&&\hspace{3cm}\vdots \nonumber
\end{eqnarray}
Again, the higher-level charges ${\mathcal Y}^R_{(n\geq 2)}$ 
are obtained recursively by the BIZZ procedure \cite{BIZZ}. 

\medskip 

In addition, another Lax pair can be obtained from the flat current ${\mathcal J}^R_\mu$ as 
\begin{eqnarray}
{\mathcal L}^R_\pm(\lambda_R)=\frac{{\mathcal J}^R_\pm}{1\pm\lambda_R}\,, 
\label{non local right Lax pair}
\end{eqnarray}
where we use the same spectral parameter $\lambda_R$\,, for later convenience. 
This Lax pair is non-local and hence it is different from the Lax pair (\ref{right Lax pair alpha=pii/2})\,. 
However there exists a gauge transformation which connect the non-local Lax pair (\ref{non local right Lax pair}) 
to the right Lax pair (\ref{right Lax pair alpha=pii/2})\,. 

\medskip 

Let us find an appropriate gauge function\footnote{ 
The strategy is similar to the one to consider the Jordanian twists 
in the Schr\"odinger sigma models \cite{Jordanian-KMY}.}. An important observation is that,   
at $\lambda_R=\infty$\,, the right Lax pair (\ref{right Lax pair alpha=pii/2}) does not vanish 
but approaches to a finite quantity as follows: 
\begin{eqnarray}
L^R_\pm(\lambda_R=\infty)=-T^3(1\pm K)J^3_\pm=-T^3\partial_\pm\left[\frac{\chi}{K}\right]\,. 
\end{eqnarray}
The gauge transformation must be chosen to eliminate this finite part 
because the non-local Lax pair (\ref{non local right Lax pair}) vanishes at $\lambda_R=\infty$\,. 
Thus the gauge transformation is generated by  
\begin{eqnarray}
F(x)={\rm e}^{-T^3\frac{\chi(x)}{K}}\,. \label{true}
\end{eqnarray}
Note that a constant $SU(2)$ element may be multiplied from the right and thus 
the function $F(x)$ is not determined uniquely. However, the choice in (\ref{true}) is appropriate 
to relate the right Lax pair (\ref{right Lax pair alpha=pii/2}) 
to the non-local Lax pair (\ref{non local right Lax pair})\,, as we will see just below. 

\medskip 

By performing a non-local gauge transformation generated by $F(x)$ in (\ref{true})\,,  
the right Lax pair (\ref{right Lax pair alpha=pii/2}) 
can be rewritten as 
\begin{eqnarray}
L^R_\pm(\lambda_R) \to \left[L_\pm^R(\lambda_R)\right]^{F}&=&F^{-1}L^R_\pm(\lambda_R)F-F^{-1}\partial_\pm F \\
&=&\frac{{\mathcal J}^R_\pm}{1\pm\lambda_R}={\mathcal L}^R_\pm(\lambda_R)\,. \nonumber
\end{eqnarray}
According to the gauge transformation, the monodromy matrix is also transformed,  
\begin{eqnarray}
{\mathcal U}^R(\lambda_R)=\left[U^R(\lambda_R)\right]^{F} 
&=&F^{-1}(x=\infty)U^R(\lambda_R)F(x=-\infty) \\
&=&{\rm e}^{-\frac{T^3Q^{R,3}}{2K}}U^R(\lambda_R){\rm e}^{-\frac{T^3Q^{R,3}}{2K}}\,. \nonumber
\end{eqnarray}
This relation may be interpreted as a classical analogue of the Reshetikhin twist \cite{R}. 
It is interesting to reveal the relation to the quantum twist from the viewpoint of the mathematical 
formulation.  

\medskip 

In addition, it is worth to note that the flat and conserved current ${\mathcal J}^R_\mu$ can be written as 
\begin{eqnarray}
{\mathcal J}^R_\mu=-\left(G^{-1}\partial_\mu G+K\epsilon_{\mu\nu}G^{-1}\partial^\nu G\right)\,, \qquad
G=gF\,. 
\end{eqnarray}
Then the current algebra of ${\mathcal J}^{R,a}_\mu$ is given by 
\begin{eqnarray}
\Bigl\{{\mathcal J}^{R,a}_t(x),{\mathcal J}^{R,b}_t(y)\Bigr\}_{\rm P}&=&\varepsilon^{ab}_{~~c}{\mathcal J}^{R,c}_t(x)
+2K\delta^{ab}\partial_x\delta(x-y)\,, \\
\Bigl\{{\mathcal J}^{R,a}_t(x),{\mathcal J}^{R,b}_x(y)\Bigr\}_{\rm P}&=&\varepsilon^{ab}_{~~c}{\mathcal J}^{R,c}_x(x)
+\left(1+K^2\right)\delta^{ab}\partial_x\delta(x-y)\,, \nonumber \\
\Bigl\{{\mathcal J}^{R,a}_x(x),{\mathcal J}^{R,b}_x(y)\Bigr\}_{\rm P}
&=&-K^2\varepsilon^{ab}_{~~c}{\mathcal J}^{R,c}_t(x)+2K\varepsilon^{ab}_{~~c}{\mathcal J}^{R,c}_x(x)
\delta(x-y) \nonumber \\
&&+2K\delta^{ab}\partial_x\delta(x-y)\,. \nonumber
\end{eqnarray}
Again, the infinite-dimensional algebra of ${\mathcal Y}^R_{(n)}$ is ${\mathcal Y}(su(2)_{\rm R})$\,.

\section{Conclusion and Discussion}

We have considered the classical integrable structure of the squashed WZNW models 
based on an infinite-dimensional extension of $U(1)_{\rm R}$\,. 
The system contains two deformation parameters. The one is provided by the coefficient of the Wess-Zumino term.  
The other is the squashing parameter of target space. 
We have constructed an anisotropic Lax pair and computed the associated $r$/$s$-matrices, 
which may be regarded as one-parameter deformations 
of the results without the Wess-Zumino term \cite{KYhybrid,KMY-QAA}. 

\medskip

Due to the presence of the Wess-Zumino term, an infinite-dimensional extension of $su(2)_{\rm R}$ 
is given by a deformation of the standard quantum affine algebra, which contain two parameters. 
The deformed algebra contains two sets of $q$-deformed $su(2)_{\rm R}$ 
and the tower structure of the conserved charges is quite similar to 
that of the standard quantum affine algebra. 
The only modification appears in the coefficient of the Poisson bracket. 
It also changes the classical $q$-Serre relations. 
The resulting algebra seems likely to be the classical analogue of a two-parameter quantum toroidal algebra\cite{GKV}. 
However, it is not sure so far for this point because it seems that the first realization of the algebra has not been constructed yet.

\medskip

The left-right duality has also been revealed by showing that the left Lax pairs are gauge-equivalent to the right Lax pair 
with the relation between the spectral parameters and the isomorphism of the $sl(2)$ algebra. 
This is a generalization of the duality in the case without the Wess-Zumino term. 
The right Lax pair can also be decomposed to a pair of the reduced right Lax pairs, each of which  
is equivalent to the corresponding left Lax pair. 

\medskip

In addition, two degenerate limits have been considered. They are realized at the points $C=0$ 
(corresponding to $\alpha=\beta=0$\,) and the points $C=K^2-1$ (corresponding to $\alpha=\beta=\pi i/2$\,)\,. 
In the former case, the target space is undeformed and the right description becomes the isotropic. 
In the latter case, while the metric of the target space is still squashed, 
the effect of the squashing is compensated by taking an appropriate value of the coefficient of the Wess-Zumino term, 
so that the right description becomes the isotropic. 

\medskip 

In each of the degenerate limits, we have constructed an $SU(2)_{\rm R}$ 
conserved current satisfying the flatness condition. With this current, 
another Lax pair can be constructed. For the later case, the construction is more involved and 
a non-local gauge-transformation is needed. 
In addition, Yangian generators have also been constructed. 

\medskip 

There are some open issues. It would be interesting to study the fast-moving limit \cite{Kruczenski} of 
the squashed WZNW models 
(For squashed S$^3$ and warped AdS$_3$ see \cite{Wen} and \cite{KameYoshi}, respectively). 
Some applications of the results may be considered in the context of string theory. In this direction, 
for a deformation of AdS$_3$/CFT$_2$\,, see an earlier paper \cite{null3}. 
For warped AdS$_3$ and squashed S$^3$ geometries in string theory, 
for example, see \cite{OU0,SS,DLR,DG,Eoin}.

\medskip 

We hope that some applications of the deformed quantum affine algebra would be found and be 
a key ingredient in studying the integrability of the AdS/CFT correspondence.

\subsection*{Acknowledgment}

We would like to thank Takuya Matsumoto for useful discussions. 
The work of IK was supported by the Japan Society for the Promotion of Science (JSPS)\,. 

\appendix

\section*{Appendix}

\section{The Poisson brackets of $J^a_\mu$ and $j^{L_\pm}_\mu$}

The Poisson brackets of $J^a_\mu$ play an important role in computing the 
current algebra of $j^{L_\pm}_\mu$\,. The computation 
is straightforward but very messy. In order to use the canonical Poisson brackets of the dynamical variables, 
the computation is described in terms of angle variables. Then, by using the Poisson brackets, the current algebra 
of  $j^{L_\pm}_\mu$ is computed.  

\medskip 

\subsection{The Poisson brackets of $J^a_\mu$}

The classical action (\ref{action}) is composed of the two parts $S=S_{\sigma M}+S_{WZ}$\,. 
Each part is expressed in terms of the angle variables $(\theta,\phi,\psi)$ as follows:   
\begin{eqnarray}
&&S_{\sigma M}=\frac{1}{2\lambda^2}\int^{\infty}_{-\infty}\!\!\!dt\int^{\infty}_{-\infty}\!\!\!dx 
\left[\dot{\theta}^2+\sin^2\theta\dot{\phi}^2+(1+C)\left(\dot{\psi}+\cos\theta\dot{\phi}\right)^2\right. 
\nonumber \\ &&\qquad\qquad\qquad\qquad\qquad\qquad\left.-\theta'^2-\sin^2\theta\phi'^2
-(1+C)\left(\psi'+\cos\theta\phi'\right)^2\right]\,, \nonumber \\
&&S_{WZ}=-\frac{n}{8\pi}\int^{\infty}_{-\infty}\!\!\!dt\int^{\infty}_{-\infty}\!\!\!dx 
\left(\cos\theta\dot{\phi}\psi'-\cos\theta\dot{\psi}\phi'\right)\,. 
\end{eqnarray}
The symbols ``dot'' and ``prime'' denote the derivatives with respect to $t$ and $x$\,, respectively. 

\medskip 

The conjugate momenta $(\Pi_\theta,\Pi_\phi,\Pi_\psi)$ for $(\theta,\phi,\psi)$ are given by 
\begin{eqnarray}
\Pi_\theta&=&\dot{\theta}\,, \nonumber \\
\Pi_\phi&=&\sin^2\theta\dot{\phi}+(1+C)\cos\theta\left(\dot{\psi}
+\cos\theta\dot{\phi}\right)-K\cos\theta\psi'\,, \nonumber \\
\Pi_\psi&=&(1+C)\left(\dot{\psi}+\cos\theta\dot{\phi}\right)+K\cos\theta\phi'\,. 
\end{eqnarray}
The ``velocity'' variables are 
\begin{eqnarray}
\dot{\theta}&=&\widehat{\Pi}_\theta\,, \nonumber \\
\dot{\phi}&=&\frac{1}{\sin^2\theta}\widehat{\Pi}_\phi
-\frac{\cos\theta}{\sin^2\theta}\widehat{\Pi}_\psi\,, \nonumber \\
\dot{\psi}&=&\left(\frac{1}{1+C}-\frac{\cos^2\theta}{\sin^2\theta}\right)\widehat{\Pi}_\psi
-\frac{\cos\theta}{\sin^2\theta}\widehat{\Pi}_\phi\,, 
\end{eqnarray}
where $(\widehat{\Pi}_\theta,\widehat{\Pi}_\phi,\widehat{\Pi}_\psi)$ are defined as 
\begin{eqnarray}
\widehat{\Pi}_\theta \equiv \Pi_\theta\,,\quad 
\widehat{\Pi}_\phi \equiv \Pi_\phi+K\cos\theta\psi'\,, \quad 
\widehat{\Pi}_\psi \equiv \Pi_\psi-K\cos\theta\phi'\,. 
\end{eqnarray}
Then the Poisson brackets of the canonical variables are summarized 
in Tab.\,\ref{canonical variables Poisson brackets}. 

\begin{table}[htbp]
\vspace*{0.5cm}
\begin{center}
\begin{tabular}{c||c|c|c|c|c|c}
 & $\theta$ & $\phi$ & $\psi$ & $\widehat{\Pi}_\theta$ & $\widehat{\Pi}_\phi$ & $\widehat{\Pi}_\psi$ \\
\hline \hline
$\theta$ & $0$ & $0$ & $0$ & $1$ & $0$ & $0$ \\
\hline
$\phi$ & $0$ & $0$ & $0$ & $0$ & $1$ & $0$ \\
\hline
$\psi$ & $0$ & $0$ & $0$ & $0$ & $0$ & $1$ \\
\hline
$\widehat{\Pi}_\theta$ & $-1$ & $0$ & $0$ & $0$ & $K\sin\theta\psi'$ & $-K\sin\theta\phi'$ \\
\hline
$\widehat{\Pi}_\phi$ & $0$ & $-1$ & $0$ & $-K\sin\theta\psi'$ & $0$ & $K\sin\theta\theta'$ \\
\hline
$\widehat{\Pi}_\psi$ & $0$ & $0$ & $-1$ & $K\sin\theta\phi'$ & $-K\sin\theta\theta'$ & $0$ \\
\hline
\end{tabular}
\caption{The Poisson brackets of the angle variables and the canonical momenta.}
\label{canonical variables Poisson brackets}
\end{center}
\end{table}

With the canonical variables, the components of $J$ are expressed as 
\begin{eqnarray}
&&J^\pm_t=\frac{1}{\sqrt{2}}{\rm e}^{\mp i\psi}\left(\pm i\widehat{\Pi}_\theta 
-\frac{1}{\sin\theta}\widehat{\Pi}_\phi +\frac{\cos\theta}{\sin\theta}\widehat{\Pi}_\psi\right)\,, \nonumber \\
&&J^3_t=\frac{1}{1+C}\widehat{\Pi}_\psi\,, \nonumber \\
&&J^\pm_x=\frac{1}{\sqrt{2}}{\rm e}^{\mp i\psi}\left(\pm i\theta' -\sin\theta \phi'\right)\,, \nonumber \\
&&J^3_x=\psi' +\cos\theta \phi'\,. 
\label{left-invariant one-form angle variables}
\end{eqnarray}
By using the Poisson brackets in Tab.\,\ref{canonical variables Poisson brackets} 
and the expressions given in (\ref{left-invariant one-form angle variables}), 
the Poisson brackets of $J^a_\mu$ are computed like 
\begin{eqnarray}
\left\{J^{\pm}_t(x),J^{\mp}_t(y)\right\}_{\rm P}&=&\mp i\left[-(1+C)J^3_t +KJ^3_x\right](x)\delta(x-y)\,, \\
\left\{J^{\pm}_t(x),J^3_t(y)\right\}_{\rm P}&=&\pm i\left[-\frac{1}{1+C}J^{\pm}_t 
+\frac{K}{1+C}J^\pm_x\right](x)\delta(x-y)\,, \nonumber \\ 
\left\{J^{\pm}_t(x),J^{\mp}_x(y)\right\}_{\rm P}&=&\mp i\left[-J^3_x\right](x)\delta(x-y)
+\partial_x\delta(x-y)\,, \nonumber \\ 
\left\{J^{\pm}_t(x),J^3_x(y)\right\}_{\rm P}&=&\pm i\left[-J^{\pm}_x\right](x)\delta(x-y)\,, \nonumber \\ 
\left\{J^3_t(x),J^{\pm}_x(y)\right\}_{\rm P}&=&\mp i\left[-\frac{1}{1+C}J^{\pm}_x\right](x)\delta(x-y)\,, \nonumber \\ 
\left\{J^3_t(x),J^3_x(y)\right\}_{\rm P}&=&\frac{1}{1+C}\partial_x\delta(x-y)\,, \nonumber \\
\left\{J^{\pm}_x(x),J^{\mp}_t(y)\right\}_{\rm P}&=&\mp i\left[-J^3_x\right](x)\delta(x-y)
+\partial_x\delta(x-y)\,, \nonumber \\ 
\left\{J^{\pm}_x(x),J^3_t(y)\right\}_{\rm P}&=&\pm i\left[-\frac{1}{1+C}J^{\pm}_x\right](x)\delta(x-y)\,, \nonumber \\ 
\left\{J^3_x(x),J^{\pm}_t(y)\right\}_{\rm P}&=&\mp i\left[-J^{\pm}_x\right](x)\delta(x-y)\,, \nonumber \\ 
\left\{J^3_x(x),J^3_t(y)\right\}_{\rm P}&=&\frac{1}{1+C}\partial_x\delta(x-y)\,.  
\label{Poisson J}
\end{eqnarray}
Similarly, the Poisson brackets between $g$ and $J^a_\mu$ are evaluated as 
\begin{eqnarray}
\left\{g(x),J^\pm_t(y)\right\}_{\rm P}&=& g(x)T^\pm\delta(x-y)\,, \nonumber \\
\left\{g(x),J^3_t(y)\right\}_{\rm P}&=& \frac{1}{1+C}g(x)T^3\delta(x-y)\,. 
\label{Poisson g J}
\end{eqnarray}

\subsection{The current algebra of $j^{L_\pm}_\mu$}

To compute the current algebra of $j^{L_\pm}_\mu$\,, it is convenient to 
rewrite $j^{L_\pm}_\mu$ in terms of $J$ as 
\begin{eqnarray}
j^{L_\pm}_\mu &=& g\left\{J_\mu -2C{\rm tr}\left(T^3J_\mu\right)T^3 
-K\epsilon_{\mu\nu}J^\nu \mp A\epsilon_{\mu\nu}\left[J^\nu,T^3\right]\right\}g^{-1} \nonumber \\
&=& g\left\{T^+\left[J^-_\mu -\left(K \pm iA\right)\epsilon_{\mu\nu}J^{-,\nu}\right] 
+T^-\left[J^+_\mu -\left(K \mp iA\right)\epsilon_{\mu\nu}J^{+,\nu}\right] \right. \nonumber \\
&&\left. +T^3\left[(1+C)J^3_\mu -K\epsilon_{\mu\nu}J^{3,\nu}\right]\right\}g^{-1}\,. 
\end{eqnarray}
With the relations in (\ref{Poisson J}) and  (\ref{Poisson g J})\,, 
the current algebra is computed as  
\begin{eqnarray}
\Bigl\{j^{L_\pm,a}_t(x),j^{L_\pm,b}_t(y)\Bigr\}_{\rm P} &=& \varepsilon^{ab}_{~~c}j^{L_\pm,c}_t(x)\delta(x-y) 
-2K\delta^{ab}\partial_x\delta(x-y)\,, \nonumber \\
\Bigl\{j^{L_\pm,a}_t(x),j^{L_\pm,b}_x(y)\Bigr\}_{\rm P} &=& \varepsilon^{ab}_{~~c}j^{L_\pm,c}_x(x)\delta(x-y) 
+\left(1+K^2+A^2\right)\delta^{ab}\partial_x\delta(x-y)\,, \nonumber \\
\Bigl\{j^{L_\pm,a}_x(x),j^{L_\pm,b}_x(y)\Bigr\}_{\rm P} &=& -\left(K^2+A^2\right)
\varepsilon^{ab}_{~~c}j^{L_\pm,c}_t(x)\delta(x-y) \nonumber \\
&&-2K\varepsilon^{ab}_{~~c}j^{L_\pm,c}_x(x)\delta(x-y)-2K\delta^{ab}\partial_x\delta(x-y)\,. \nonumber
\end{eqnarray}
In addition, the Poisson brackets between $g$ and $j^{L_+}_\mu$ are also obtained as 
\begin{eqnarray}
&&\Bigl\{g(x),j^{L_+,a}_t(y)\Bigr\}_{\rm P}= T^ag(x)\delta(x-y)\,, \\
&&\Bigl\{g(x),j^{L_+,a}_x(y)\Bigr\}_{\rm P}= \left\{-\frac{K}{1+C}T^a+A\left[T^a,gT^3g^{-1}\right] \right. \nonumber \\
&&\hspace{4cm}\left.+\frac{CK}{1+C}\left[\left[T^a,gT^3g^{-1}\right],gT^3g^{-1}\right]\right\}g(x)
\delta(x-y)\,. \nonumber
\end{eqnarray}

\section{A prescription to treat non-ultra local terms}

We present the computations of 1) Yangian, 2) $q$-deformed $su(2)_{\rm R}$\,, and 3) a deformed quantum affine 
algebra. In particular, a prescription to treat non-ultra local terms is carefully described in each case. 

\subsection{Yangian ${\mathcal Y}(su(2)_{\rm L})$}

Let us first compute Yangian ${\mathcal Y}(su(2)_{\rm L})$\,. 
Non-ultra local terms appear in the computations of the Poisson brackets of $Q^{L_\pm}_{(n)}$ 
and hence we should be careful of the order of limits. 
The subscripts of $L$\,, ($\pm$) are omitted for simplicity henceforth.

\subsubsection*{The Poisson brackets at level 0}

The first is the Poisson brackets of the level 0 charges. 
The charges are regularized as 
\begin{eqnarray}
Q^{L,a}_{(0)}(X_0,X_1)=\int^{X_1}_{-X_0}\!\!\!dx~j^{L,a}_t(x)\,. 
\end{eqnarray}
Then the Poisson brackets of the regularized charges are given by 
\begin{eqnarray}
&&\Bigl\{Q^{L,a}_{(0)}(X_0,X_1),Q^{L,b}_{(0)}(Y_0,Y_1)\Bigr\}_{\rm P} 
=\int^{X_1}_{-X_0}\!\!\!dx\int^{Y_1}_{-Y_0}\!\!\!dy\left[-2K\delta^{ab}\partial_x\delta(x-y)
+\varepsilon^{ab}_{~~c}\,j^{L,c}_t(x)\delta(x-y)\right]\,. \nonumber
\end{eqnarray}
The first term contains a derivative of the delta function, called a non-ultra local term. 
This term develops an ambiguity depending on the order of limits as follows: 
\begin{eqnarray}
&&\Bigl\{Q^{L,a}_{(0)}(X_0,X_1),Q^{L,b}_{(0)}(Y_0,Y_1)\Bigr\}_{\rm P} \\
&=&-2K\delta^{ab}\left[\theta(X_0-Y_0)-\theta(X_1-Y_1)\right] 
+\varepsilon^{ab}_{~~c}Q^{L,c}_{(0)}(\min(X_0,Y_0),\min(X_1,Y_1))\,. \nonumber
\end{eqnarray}
In the following, we will not write down unambiguous terms explicitly as follows:
\begin{eqnarray}
&&\Bigl\{Q^{L,a}_{(0)}(X_0,X_1),Q^{L,b}_{(0)}(Y_0,Y_1)\Bigr\}_{\rm P} 
\\ &=&
-2K\delta^{ab}\left[\theta(X_0-Y_0)-\theta(X_1-Y_1)\right]
+({\rm no~ambiguity})\,. \nonumber
\end{eqnarray}
Note that this ambiguity arises due to the presence of the Wess-Zumino term. 
When $K=0$\,, there is no ambiguity at the level 0 as usual. 

\medskip 

Now, one may follow a prescription proposed in \cite{MacKay}\,, 
\begin{eqnarray}
X_0=X_1\equiv X\,, \qquad
Y_0=Y_1\equiv Y\,, \qquad
\end{eqnarray}
so as to make the ambiguous term vanish. Then, by taking the limits $X\to \infty$ and $Y\to \infty$\,,  
the following is obtained,  
\begin{eqnarray}
\Bigl\{Q^{L,a}_{(0)},Q^{L,b}_{(0)}\Bigr\}_{\rm P}
=\varepsilon^{ab}_{~~c}Q^{L,c}_{(0)}\,. 
\end{eqnarray}

\subsubsection*{The Poisson brackets at level 1}

The next is to regularize the level 1 charges. 
The level 1 charges are regularized as 
\begin{eqnarray}
Q^{L,a}_{(1)}(X,X',X'')
&=&\frac{1}{4}\int^X_{-X}\!\!\!dx\int^{X'}_{-X'}\!\!\!dx'~
\epsilon(x-x')\,\varepsilon^a_{~bc}\,j^{L,b}_t(x)j^{L,c}_t(x') 
-\int^{X''}_{-X''}\!\!\!dx''j^{L,a}_x(x'')\,. \nonumber
\end{eqnarray}
The Poisson brackets of the regularized level 1 and level 0 charges are 
\begin{eqnarray}
&&\Bigl\{Q^{L,a}_{(1)}(X,X',X''),Q^{L,b}_{(0)}(Y)\Bigr\}_{\rm P} \nonumber \\
&=&2K\,\varepsilon^{ab}_{~~c}\left[\theta(X-Y)Q^{L,c}_{(0)}(\min(X',Y))+\theta(X'-Y)Q^{L,c}_{(0)}(\min(X,Y))\right] 
\nonumber \\ &&
+({\rm no~ambiguity})\,. 
\end{eqnarray}
Now the unambiguous terms are given by
\begin{eqnarray}
\varepsilon^{ab}_{~~c}\,Q^{L,c}_{(1)}\,, 
\end{eqnarray}
and hence the ambiguous terms have to vanish. 
Thus the $Y\to\infty$ limit should be taken before the $X,X',X''\to\infty$ limits.    
As a result, the following is obtained,  
\begin{eqnarray}
\Bigl\{Q^{L,a}_{(1)},Q^{L,b}_{(0)}\Bigr\}_{\rm P}=\varepsilon^{ab}_{~~c}\,Q^{L,a}_{(1)}\,. 
\end{eqnarray}

\subsubsection*{The Poisson brackets at level 2}

The Poisson brackets of the regularized level 1 charges are computed here. 
Those give rise to the brackets at level 2, 
\begin{eqnarray}
&&\Bigl\{Q^{L,a}_{(1)}(X,X',X''),Q^{L,b}_{(1)}(Y,Y',Y'')\Bigr\}_{\rm P} \\
&=&-\frac{K}{8}\theta(X-Y)\int^{X'}_{-X'}\!\!\!dx'\int^{Y'}_{-Y'}\!\!\!dy' \nonumber \\
&&\hspace{3cm}\times\left[\epsilon(x'+Y)\epsilon(y'+Y)-\epsilon(x'-Y)\epsilon(y'-Y)\right] \nonumber \\
&&\hspace{3cm}\times\left[\delta^{ab}\delta_{cd}j^{L,c}_t(x')j^{L,d}_t(y')-j^{L,b}_t(x')j^{L,a}_t(y')\right] \nonumber \\
&&-\frac{K}{8}\theta(X'-Y)\int^{X}_{-X}\!\!\!dx\int^{Y'}_{-Y'}\!\!\!dy' \nonumber \\
&&\hspace{3cm}\times\left[\epsilon(x+Y)\epsilon(y'+Y)-\epsilon(x-Y)\epsilon(y'-Y)\right] \nonumber \\
&&\hspace{3cm}\times\left[\delta^{ab}\delta_{cd}j^{L,c}_t(x)j^{L,d}_t(y')-j^{L,b}_t(x)j^{L,a}_t(y')\right] \nonumber \\
&&-\frac{K}{8}\theta(X-Y')\int^{X'}_{-X'}\!\!\!dx'\int^{Y}_{-Y}\!\!\!dy \nonumber \\
&&\hspace{3cm}\times\left[\epsilon(x'+Y')\epsilon(y+Y')-\epsilon(x'-Y')\epsilon(y-Y')\right] \nonumber \\
&&\hspace{3cm}\times\left[\delta^{ab}\delta_{cd}j^{L,c}_t(x')j^{L,d}_t(y)-j^{L,b}_t(x')j^{L,a}_t(y)\right] \nonumber \\
&&-\frac{K}{8}\theta(X'-Y')\int^{X}_{-X}\!\!\!dx\int^{Y}_{-Y}\!\!\!dy \nonumber \\
&&\hspace{3cm}\times\left[\epsilon(x+Y')\epsilon(y+Y')-\epsilon(x-Y')\epsilon(y-Y')\right] \nonumber \\
&&\hspace{3cm}\times\left[\delta^{ab}\delta_{cd}j^{L,c}_t(x)j^{L,d}_t(y)-j^{L,b}_t(x)j^{L,a}_t(y)\right] \nonumber \\
&&+\frac{(1+C)^2+K^2}{4(1+C)}\theta(X-Y'')\varepsilon^{ab}_{~~c}\int^{X'}_{-X'}\!\!\!dx'
\left[\epsilon(x'+Y'')-\epsilon(x'-Y'')\right]j^{L,c}_t(x') \nonumber \\
&&+\frac{(1+C)^2+K^2}{4(1+C)}\theta(X'-Y'')\varepsilon^{ab}_{~~c}\int^{X}_{-X}\!\!\!dx
\left[\epsilon(x+Y'')-\epsilon(x-Y'')\right]j^{L,c}_t(x) \nonumber \\
&&+\frac{(1+C)^2+K^2}{4(1+C)}\theta(Y-X'')\varepsilon^{ab}_{~~c}\int^{Y'}_{-Y'}\!\!\!dy'
\left[\epsilon(y'+X'')-\epsilon(y'-X'')\right]j^{L,c}_t(y') \nonumber \\
&&+\frac{(1+C)^2+K^2}{4(1+C)}\theta(Y'-X'')\varepsilon^{ab}_{~~c}\int^{Y}_{-Y}\!\!\!dy
\left[\epsilon(y+X'')-\epsilon(y-X'')\right]j^{L,c}_t(y) \nonumber \\
&&+({\rm no~ambiguity})\,. \nonumber
\end{eqnarray}
By taking the $Y,Y',Y'''\to\infty$ limits before sending $X,X',X''$ to infinity, the following expression is obtained,
\begin{eqnarray}
&&\Bigl\{Q^{L,a}_{(1)}(X,X',X''),Q^{L,b}_{(1)}\Bigr\}_{\rm P} \\
&=&\min_{Y,Y',Y''\to\infty}\Bigl\{Q^{L,a}_{(1)}(X,X',X''),Q^{L,b}_{(1)}(Y,Y',Y'')\Bigr\}_{\rm P} \nonumber \\
&=&\frac{(1+C)^2+K^2}{2(1+C)}\varepsilon^{ab}_{~~c}\int^{\infty}_{-\infty}\!\!\!dy
\left[\epsilon(y+X'')-\epsilon(y-X'')\right]j^{L,c}_t(y) 
+({\rm no~ambiguity})\,. \nonumber
\end{eqnarray}
Then the ambiguous terms are given by 
\begin{eqnarray}
\frac{(1+C)^2+K^2}{1+C}\varepsilon^{ab}_{~~c}\,Q^{L,c}_{(0)}\,, 
\end{eqnarray}
by taking the $X,X',X''\to\infty$ limits. 

\medskip 

On the other hand, since the unambiguous terms are represented by 
\begin{eqnarray}
\varepsilon^{ab}_{~~c}\left[Q^{L,c}_{(2)}+\frac{1}{12}\left(Q^L_{(0)}\right)^2Q^{L,c}_{(0)}+2KQ^{L,c}_{(1)}
-\frac{(1+C)^2+K^2}{1+C}Q^{L,c}_{(0)}\right]\,,
\end{eqnarray}
the net result is 
\begin{eqnarray}
\Bigl\{Q^{L,a}_{(1)},Q^{L,b}_{(1)}\Bigr\}_{\rm P}
&=&\varepsilon^{ab}_{~~c}\left[Q^{L,c}_{(2)}+\frac{1}{12}\left(Q^L_{(0)}\right)^2Q^{L,c}_{(0)}+2KQ^{L,c}_{(1)}\right]\,.
\end{eqnarray}

\subsubsection*{Serre relations}

The Serre-relations are finally considered. One of them is represented by  
\begin{eqnarray}
\Bigl\{Q^{L,3}_{(1)},\Bigl\{Q^{L,+}_{(1)},Q^{L,-}_{(1)}\Bigr\}_{\rm P}\Bigr\}_{\rm P}
&=&\frac{1}{4}Q^{L,3}_{(0)}\left[Q^{L,+}_{(0)}Q^{L,-}_{(1)}-Q^{L,-}_{(0)}Q^{L,+}_{(1)}\right]\,. 
\label{Serre1}
\end{eqnarray}
Let us first introduce the following quantity: 
\begin{eqnarray}
&&\Bigl\{Q^{L,+}_{(1)},Q^{L,-}_{(1)}\Bigr\}_{\rm P}(X,X',X'',X''',X'''',X''''') \label{reguralized Q+Q-}\\
&=&-\frac{i}{4}\int^{X}_{-X}\!\!\!dx\int^{X'}_{-X'}\!\!\!dx'\int^{X''}_{-X''}\!\!\!dx''~
\epsilon(x-x')\epsilon(x-x'') \nonumber \\
&&\hspace{2cm}\times\left[j^{L,+}_t(x)j^{L,3}_t(x')j^{L,-}_t(x'')+j^{L,-}_t(x)j^{L,3}_t(x')j^{L,+}_t(x'')\right. \nonumber \\
&&\hspace{5cm}\left.+j^{L,3}_t(x)j^{L,3}_t(x')j^{L,3}_t(x'')\right] \nonumber \\
&&-\frac{1}{2}\int^{X'''}_{-X'''}\!\!\!dx'''\int^{X''''}_{-X''''}\!\!\!dx''''~\epsilon(x'''-x'''') \nonumber \\
&&\hspace{2cm}\times\left[j^{L,+}_t(x''')j^{L,-}_x(x'''')-j^{L,-}_t(x''')j^{L,+}_x(x'''')\right] \nonumber \\
&&-i\int^{X'''''}_{-X'''''}\!\!\!dx'''''j^{L,3}_t(x''''') \nonumber \\
&&+K\int^{X'''}_{-X'''}\!\!\!dx'''\int^{X''''}_{-X''''}\!\!\!dx''''~\epsilon(x'''-x'''')
j^{L,+}_t(x''')j^{L,-}_t(x'''') \nonumber \\
&&+2iK\int^{X'''''}_{-X'''''}\!\!\!dx'''''j^{L,3}_x(x''''')\,. \nonumber 
\end{eqnarray}
Note that the regularized $Q^{L,3}_{(1)}$ is given by 
\begin{eqnarray}
Q^{L,3}_{(1)}(Y,Y',Y'') \label{reguralized Q3} 
=\frac{i}{2}\int^{Y}_{-Y}\!\!\!dy\int^{Y'}_{-Y'}\!\!\!dy'~\epsilon(y-y')
j^{L,+}_t(y)j^{L,-}_t(y')
-\int^{Y''}_{-Y''}\!\!\!dy''~j^{L,3}_x(y'')\,. 
\end{eqnarray}
By using the quantities in (\ref{reguralized Q+Q-}) and (\ref{reguralized Q3})\,, the following relation is obtained, 
\begin{eqnarray}
&&\Bigl\{Q^{L,3}_{(1)}(Y,Y',Y''),\Bigl\{Q^{L,+}_{(1)},Q^{L,-}_{(1)}\Bigr\}_{\rm P}(X,X',X'',X''',X'''',X''''')\Bigr\}_{\rm P} \\
&=&\frac{K}{2}\!\!\int^{\min(X,X'',Y)}_{-\min(X,X'',Y)}\!\!\!\!\!dx\!\int^{X'}_{-X'}\!\!\!\!dx'\!\!\int^{Y'}_{-Y'}\!\!\!\!dy'
\epsilon(x-x')\epsilon(x-y')j^{L,+}_t(x)j^{L,3}_t(x')j^{L,-}_t(y') \nonumber \\
&&-\frac{K}{2}\!\!\int^{\min(X,X'',Y')}_{-\min(X,X'',Y')}\!\!\!\!\!dx\!\int^{X'}_{-X'}\!\!\!\!dx'\!\!\int^{Y}_{-Y}\!\!\!\!dy~
\epsilon(x-x')\epsilon(x-y)j^{L,-}_t(x)j^{L,3}_t(x')j^{L,+}_t(y) \nonumber \\
&&+\frac{K}{2}\!\!\int^{\min(X,X',Y')}_{-\min(X,X',Y')}\!\!\!\!\!dx'\!\!\int^{X''}_{-X''}\!\!\!\!dx''\!\!\!\int^{Y}_{-Y}\!\!\!\!dy~
\epsilon(x'\!-y)\epsilon(x'\!-x'')j^{L,3}_t(x')j^{L,-}_t(x'')j^{L,+}_t(y) \nonumber \\
&&+\frac{K}{2}\!\!\int^{X'}_{-X'}\!\!\!\!dx'\!\!\int^{\min(X,X'',Y')}_{-\min(X,X'',Y')}\!\!\!\!\!dx''\!\!\!\int^{Y}_{-Y}\!\!\!\!dy~
\epsilon(x''\!\!-y)\epsilon(x''\!\!-x')j^{L,-}_t(x'')j^{L,3}_t(x')j^{L,+}_t(y) \nonumber \\
&&-\frac{K}{2}\!\!\int^{\min(X,X',Y)}_{-\min(X,X',Y)}\!\!\!\!\!dx'\!\!\int^{X''}_{-X''}\!\!\!\!dx''\!\!\!\int^{Y'}_{-Y'}\!\!\!\!dy'
\epsilon(x'\!-y')\epsilon(x'\!-x'')j^{L,3}_t(x')j^{L,+}_t(x'')j^{L,-}_t(y') \nonumber \\
&&-\frac{K}{2}\!\!\int^{X'}_{-X'}\!\!\!\!dx'\!\!\int^{\min(X,X'',Y)}_{-\min(X,X'',Y)}\!\!\!\!\!dx''\!\!\!\int^{Y'}_{-Y'}\!\!\!\!dy'
\epsilon(x''\!\!-y')\epsilon(x''\!\!-x')j^{L,+}_t(x'')j^{L,3}_t(x')j^{L,-}_t(y') \nonumber \\
&&+\frac{K}{4}\theta(Y-X'')\!\!\int^{X}_{-X}\!\!\!dx\int^{X'}_{-X'}\!\!\!dx'\int^{Y'}_{-Y'}\!\!\!dy'
\epsilon(x-x')j^{L,+}_t(x)j^{L,3}_t(x')j^{L,-}_t(y') \nonumber \\
&&\hspace{3cm}\times\left[\epsilon(y'+X'')\epsilon(x+X'')-\epsilon(y'-X'')\epsilon(x-X'')\right] \nonumber \\
&&-\frac{K}{4}\theta(Y'\!-X'')\!\!\int^{X}_{-X}\!\!\!dx\int^{X'}_{-X'}\!\!\!dx'\int^{Y}_{-Y}\!\!\!dy~
\epsilon(x-x')j^{L,-}_t(x)j^{L,3}_t(x')j^{L,+}_t(y) \nonumber \\
&&\hspace{3cm}\times\left[\epsilon(y+X'')\epsilon(x+X'')-\epsilon(y-X'')\epsilon(x-X'')\right] \nonumber \\
&&-\frac{K}{4}\theta(Y'\!-X)\!\!\int^{X'}_{-X'}\!\!\!dx'\int^{X''}_{-X''}\!\!\!dx''\int^{Y}_{-Y}\!\!\!dy~
j^{L,3}_t(x')j^{L,-}_t(x'')j^{L,+}_t(y) \nonumber \\
&&\hspace{2.5cm}\times\left[\epsilon(y+\!X)\epsilon(x'\!+\!X)\epsilon(x''\!\!+\!X)
\!-\!\epsilon(y-\!X)\epsilon(x'\!-\!X)\epsilon(x''\!\!-\!X)\right] \nonumber \\
&&-\frac{K}{4}\theta(Y-X)\!\!\int^{X'}_{-X'}\!\!\!dx'\int^{X''}_{-X''}\!\!\!dx''\int^{Y'}_{-Y'}\!\!\!dy'
j^{L,3}_t(x')j^{L,+}_t(x'')j^{L,-}_t(y') \nonumber \\
&&\hspace{2.5cm}\times\left[\epsilon(y'\!+\!X)\epsilon(x'\!+\!X)\epsilon(x''\!\!+\!X)
\!-\!\epsilon(y'\!-\!X)\epsilon(x'\!-\!X)\epsilon(x''\!\!-\!X)\right] \nonumber \\
&&+\frac{i}{2}\frac{(1+C)^2+K^2}{1+C}\int^{\min(X''',X'''',Y)}_{-\min(X''',X'''',Y)}\!\!\!dx'''\int^{Y'}_{-Y'}\!\!\!dy'
\epsilon(x'''-y')j^{L,+}_t(x''')j^{L,-}_t(y') \nonumber \\
&&-\frac{i}{2}\frac{(1+C)^2+K^2}{1+C}\int^{\min(X''',X'''',Y')}_{-\min(X''',X'''',Y')}\!\!\!dx'''\int^{Y}_{-Y}\!\!\!dy~
\epsilon(x'''-y)j^{L,+}_t(y)j^{L,-}_t(x''') \nonumber \\
&&+iK\int^{\min(X''',X'''',Y')}_{-\min(X''',X'''',Y')}\!\!\!dx''''\int^{Y}_{-Y}\!\!\!dy~
\epsilon(y-x'''')j^{L,+}_t(y)j^{L,-}_x(x'''') \nonumber \\
&&-iK\int^{\min(X''',X'''',Y)}_{-\min(X''',X'''',Y)}\!\!\!dx''''\int^{Y'}_{-Y'}\!\!\!dy'
\epsilon(x''''-y')j^{L,+}_x(x'''')j^{L,-}_t(y') \nonumber \\
&&-\frac{i}{2}\frac{(1+C)^2+K^2}{1+C}\int^{\min(X,X',Y'')}_{-\min(X,X',Y'')}\!\!\!dx\int^{X''}_{-X''}\!\!\!dx''
\epsilon(x-x'')j^{L,+}_t(x)j^{L,-}_t(x'') \nonumber \\
&&-\frac{i}{2}\frac{(1+C)^2+K^2}{1+C}\int^{\min(X,X',Y'')}_{-\min(X,X',Y'')}\!\!\!dx\int^{X''}_{-X''}\!\!\!dx''
\epsilon(x-x'')j^{L,-}_t(x)j^{L,+}_t(x'') \nonumber \\
&&+\frac{i}{2}\frac{(1+C)^2+K^2}{1+C}\int^{\min(X,X',Y'')}_{-\min(X,X',Y'')}\!\!\!dx'\int^{X''}_{-X''}\!\!\!dx''
\epsilon(x'-x'')j^{L,3}_t(x')j^{L,3}_t(x'') \nonumber \\
&&+\frac{i}{2}\frac{(1+C)^2+K^2}{1+C}\int^{X'}_{-X'}\!\!\!dx'\int^{\min(X,X'',Y'')}_{-\min(X,X'',Y'')}\!\!\!dx''
\epsilon(x''-x')j^{L,3}_t(x'')j^{L,3}_t(x') \nonumber \\
&&-\frac{i}{2}\frac{(1+C)^2+K^2}{1+C}\int^{\min(X,X',Y'')}_{-\min(X,X',Y'')}\!\!\!dx\int^{X''}_{-X''}\!\!\!dx''
\epsilon(x-x'')j^{L,3}_t(x)j^{L,3}_t(x'') \nonumber \\
&&-\frac{i}{2}\frac{(1+C)^2+K^2}{1+C}\int^{\min(X,X'',Y'')}_{-\min(X,X'',Y'')}\!\!\!dx\int^{X'}_{-X'}\!\!\!dx'
\epsilon(x-x')j^{L,3}_t(x)j^{L,3}_t(x') \nonumber \\
&&+\frac{i}{4}\frac{(1+C)^2+K^2}{1+C}\theta(Y-X'''')
\int^{X'''}_{-X'''}dx'''\int^{Y'}_{-Y'}\!\!\!dy'
j^{L,+}_t(x''')j^{L,-}_t(y') \nonumber \\
&&\hspace{3cm}\times\left[\epsilon(x'''+X'''')\epsilon(y'+X'''')-\epsilon(x'''-X'''')\epsilon(y'-X'''')\right] \nonumber \\
&&+\frac{i}{4}\frac{(1+C)^2+K^2}{1+C}\theta(Y'-X'''')
\int^{X'''}_{-X'''}dx'''\int^{Y}_{-Y}\!\!\!dy~
j^{L,-}_t(x''')j^{L,+}_t(y) \nonumber \\
&&\hspace{3cm}\times\left[\epsilon(x'''+X'''')\epsilon(y+X'''')-\epsilon(x'''-X'''')\epsilon(y-X'''')\right] \nonumber \\
&&-\frac{i}{2}K\theta(Y'-X''')
\int^{X''''}_{-X''''}dx''''\int^{Y}_{-Y}\!\!\!dy~
j^{L,-}_x(x'''')j^{L,+}_t(y) \nonumber \\
&&\hspace{3cm}\times\left[\epsilon(x''''+X''')\epsilon(y+X''')-\epsilon(x''''-X''')\epsilon(y-X''')\right] \nonumber \\
&&-\frac{i}{2}K\theta(Y-X''')
\int^{X''''}_{-X''''}dx''''\int^{Y'}_{-Y'}\!\!\!dy'
j^{L,+}_x(x'''')j^{L,-}_t(y') \nonumber \\
&&\hspace{3cm}\times\left[\epsilon(x''''+X''')\epsilon(y'+X''')-\epsilon(x''''-X''')\epsilon(y'-X''')\right] \nonumber \\
&&+\frac{i}{4}\frac{(1+C)^2+K^2}{1+C}\theta(Y''-X')
\int^{X}_{-X}dx\int^{X''}_{-X''}\!\!\!dx''
j^{L,+}_t(x)j^{L,-}_t(x'') \nonumber \\
&&\hspace{6cm}\times\epsilon(x-x'')\left[\epsilon(x+X')-\epsilon(x-X')\right] \nonumber \\
&&+\frac{i}{4}\frac{(1+C)^2+K^2}{1+C}\theta(Y''-X')
\int^{X}_{-X}dx\int^{X''}_{-X''}\!\!\!dx''
j^{L,-}_t(x)j^{L,+}_t(x'') \nonumber \\
&&\hspace{6cm}\times\epsilon(x-x'')\left[\epsilon(x+X')-\epsilon(x-X')\right] \nonumber \\
&&+\frac{i}{4}\frac{(1+C)^2+K^2}{1+C}\theta(Y''-X)
\int^{X'}_{-X'}dx'\int^{X''}_{-X''}\!\!\!dx''
j^{L,3}_t(x')j^{L,3}_t(x'') \nonumber \\
&&\hspace{4.5cm}\times\left[\epsilon(x'+X)\epsilon(x''+X)-\epsilon(x'-X)\epsilon(x''-X)\right] \nonumber \\
&&+\frac{i}{4}\frac{(1+C)^2+K^2}{1+C}\theta(Y''-X')
\int^{X}_{-X}dx\int^{X''}_{-X''}\!\!\!dx''
j^{L,3}_t(x)j^{L,3}_t(x'') \nonumber \\
&&\hspace{6cm}\times\epsilon(x-x'')\left[\epsilon(x+X')-\epsilon(x-X')\right] \nonumber \\
&&+\frac{i}{4}\frac{(1+C)^2+K^2}{1+C}\theta(Y''-X'')
\int^{X}_{-X}dx\int^{X'}_{-X'}\!\!\!dx'
j^{L,3}_t(x)j^{L,3}_t(x') \nonumber \\
&&\hspace{6cm}\times\epsilon(x-x')\left[\epsilon(x+X'')-\epsilon(x-X'')\right] \nonumber \\
&&+({\rm no~ambiguity})\,. \nonumber
\end{eqnarray}
Then, by taking the limits in which $X$\,, $X'$\,, $X''$\,, $X'''$\,, $X''''$ and $X'''''$ are sent to infinity,  
the expression is simplified as  
\begin{eqnarray}
&&\Bigl\{Q^{L,3}_{(1)}(Y,Y',Y''),\Bigl\{Q^{L,+}_{(1)},Q^{L,-}_{(1)}\Bigr\}_{\rm P}\Bigr\}_{\rm P} \\
&=&\frac{K}{2}\int^{Y}_{-Y}\!\!\!dx\int^{\infty}_{-\infty}\!\!\!dx'\int^{Y'}_{-Y'}\!\!\!dy'~
\epsilon(x-x')\epsilon(x-y')
j^{L,+}_t(x)j^{L,3}_t(x')j^{L,-}_t(y') \nonumber \\
&&-\frac{K}{2}\int^{Y'}_{-Y'}\!\!\!dx\int^{\infty}_{-\infty}\!\!\!dx'\int^{Y}_{-Y}\!\!\!dy~
\epsilon(x-x')\epsilon(x-y)
j^{L,-}_t(x)j^{L,3}_t(x')j^{L,+}_t(y) \nonumber \\
&&+\frac{K}{2}\int^{Y'}_{-Y'}\!\!\!dx'\int^{\infty}_{-\infty}\!\!\!dx''\int^{Y}_{-Y}\!\!\!dy~
\epsilon(x'-y)\epsilon(x'-x'')
j^{L,3}_t(x')j^{L,-}_t(x'')j^{L,+}_t(y) \nonumber \\
&&+\frac{K}{2}\int^{\infty}_{-\infty}\!\!\!dx'\int^{Y'}_{-Y'}\!\!\!dx''\int^{Y}_{-Y}\!\!\!dy~
\epsilon(x''-y)\epsilon(x''-x')
j^{L,-}_t(x'')j^{L,3}_t(x')j^{L,+}_t(y) \nonumber \\
&&-\frac{K}{2}\int^{Y}_{-Y}\!\!\!dx'\int^{\infty}_{-\infty}\!\!\!dx''\int^{Y'}_{-Y'}\!\!\!dy'~
\epsilon(x'-y')\epsilon(x'-x'')
j^{L,3}_t(x')j^{L,+}_t(x'')j^{L,-}_t(y') \nonumber \\
&&-\frac{K}{2}\int^{\infty}_{-\infty}\!\!\!dx'\int^{Y}_{-Y}\!\!\!dx''\int^{Y'}_{-Y'}\!\!\!dy'~
\epsilon(x''-y')\epsilon(x''-x')
j^{L,+}_t(x'')j^{L,3}_t(x')j^{L,-}_t(y') \nonumber \\
&&+\frac{i}{2}\frac{(1+C)^2+K^2}{1+C}\int^{Y}_{-Y}\!\!\!dx'''\int^{Y'}_{-Y'}\!\!\!dy'~
\epsilon(x'''-y')j^{L,+}_t(x''')j^{L,-}_t(y') \nonumber \\
&&-\frac{i}{2}\frac{(1+C)^2+K^2}{1+C}\int^{Y'}_{-Y'}\!\!\!dx'''\int^{Y}_{-Y}\!\!\!dy~
\epsilon(x'''-y)j^{L,+}_t(y)j^{L,-}_t(x''') \nonumber \\
&&+iK\int^{Y'}_{-Y'}\!\!\!dx''''\int^{Y}_{-Y}\!\!\!dy~
\epsilon(y-x'''')j^{L,+}_t(y)j^{L,-}_x(x'''') \nonumber \\
&&-iK\int^{Y}_{-Y}\!\!\!dx''''\int^{Y'}_{-Y'}\!\!\!dy'~
\epsilon(x''''-y')j^{L,+}_x(x'''')j^{L,-}_t(y') \nonumber \\
&&-\frac{i}{2}\frac{(1+C)^2+K^2}{1+C}\int^{Y''}_{-Y''}\!\!\!dx\int^{\infty}_{-\infty}\!\!\!dx''~
\epsilon(x-x'')j^{L,+}_t(x)j^{L,-}_t(x'') \nonumber \\
&&-\frac{i}{2}\frac{(1+C)^2+K^2}{1+C}\int^{Y''}_{-Y''}\!\!\!dx\int^{\infty}_{-\infty}\!\!\!dx''~
\epsilon(x-x'')j^{L,-}_t(x)j^{L,+}_t(x'') \nonumber \\
&&+\frac{i}{2}\frac{(1+C)^2+K^2}{1+C}\int^{Y''}_{-Y''}\!\!\!dx'\int^{\infty}_{-\infty}\!\!\!dx''~
\epsilon(x'-x'')j^{L,3}_t(x')j^{L,3}_t(x'') \nonumber \\
&&+\frac{i}{2}\frac{(1+C)^2+K^2}{1+C}\int^{\infty}_{-\infty}\!\!\!dx'\int^{Y''}_{-Y''}\!\!\!dx''~
\epsilon(x''-x')j^{L,3}_t(x'')j^{L,3}_t(x') \nonumber \\
&&-\frac{i}{2}\frac{(1+C)^2+K^2}{1+C}\int^{Y''}_{-Y''}\!\!\!dx\int^{\infty}_{-\infty}\!\!\!dx''~
\epsilon(x-x'')j^{L,3}_t(x)j^{L,3}_t(x'') \nonumber \\
&&-\frac{i}{2}\frac{(1+C)^2+K^2}{1+C}\int^{Y''}_{-Y''}\!\!\!dx\int^{\infty}_{-\infty}\!\!\!dx'~
\epsilon(x-x')j^{L,3}_t(x)j^{L,3}_t(x') \nonumber \\
&&+({\rm no~ambiguity})\,. \nonumber
\end{eqnarray}
Finally, by sending $Y$\,, $Y'$ and $Y''$ to infinity, 
the ambiguous terms are written as 
\begin{eqnarray}
&& \Bigl\{Q^{L,3}_{(1)},\Bigl\{Q^{L,+}_{(1)},Q^{L,-}_{(1)}\Bigr\}_{\rm P}\Bigr\}_{\rm P} \\
&=&iK\int^\infty_{-\infty}\!\!\!dx\int^\infty_{-\infty}\!\!\!dy~
\epsilon(x-y)\left[j^{L,+}_t(x)j^{L,-}x(y)-j^{L,+}_x(x)j^{L,-}_t(y)\right] 
+({\rm no~ambiguity})\,. \nonumber
\end{eqnarray}
The unambiguous terms are evaluated as 
\begin{eqnarray}
&&\frac{1}{4}Q^{L,3}_{(0)}\left[Q^{L,+}_{(0)}Q^{L,-}_{(1)}-Q^{L,-}_{(0)}Q^{L,+}_{(1)}\right] \\ &&
-iK\int^\infty_{-\infty}\!\!\!dx\int^\infty_{-\infty}\!\!\!dy~
\epsilon(x-y)\left[j^{L,+}_t(x)j^{L,-}x(y)-j^{L,+}_x(x)j^{L,-}_t(y)\right]\,. \nonumber
\end{eqnarray}
Thus the relation (\ref{Serre1}) has been shown, though the expression was quite messy 
in the middle of the computation. 
The other Serre relations can also be shown in a similar way.

\subsection{$q$-deformed $su(2)_{\rm R}$}

The Poisson brackets of $Q^{R,3}$ and $Q^{R,\pm}$ are next computed. 

\medskip 

The level 0 charge $Q^{R,3}$ and the non-local field $\chi(x)$ are regularized as, respectively,  
\begin{eqnarray}
&&Q^{R,3}(X_0,X_1)=\int^{X_1}_{-X_0}\!\!\!dx~j^{R,3}_t(x)\,, \\
&&\chi(x;X_0,X_1)=-\frac{1}{2}\int^{X_1}_{-X_0}\!\!\!dx'~\epsilon(x-x')j^{R,3}_t(x')\,, 
\quad x\in[-X_0,X_1]\,. \nonumber
\end{eqnarray}
Note that for $\chi(x;X_0,X_1)$ the domain of $x$ is restricted. 
The restriction prevents us from taking the $x\to\pm\infty$ limit 
before $-X_0$ and $X_1$ are sent to infinity. Hence the relation
\begin{eqnarray}
\lim_{x\to\pm\infty}\chi(x;X_0,X_1)=\mp\frac{1}{2}Q^{R,3}(X_0,X_1) \nonumber
\end{eqnarray}
is not valid any more. The Poisson brackets of the regularized charges $Q^{R,3}$ and $\chi(x)$ are 
\begin{eqnarray}
\Bigl\{Q^{R,3}(X_0,X_1),Q^{R,3}(Y_0,Y_1)\Bigr\}_{\rm P}
&=&2K\Bigl[\theta(X_0-Y_0)-\theta(X_1-Y_1)\Bigr]\,, \label{Poisson Q3Q3} \\
\Bigl\{\chi(x;X_0,X_1),Q^{R,3}(Y_0,Y_1)\Bigr\}_{\rm P} 
&=& -K\Bigl[2\theta(x-Y_1)-2\theta(x+Y_0) \Bigr. \nonumber \\ 
&& \Bigl. \hspace*{3cm} +\theta(Y_0-X_0)+\theta(Y_1-X_1)\Bigr]\,, \nonumber \\
\Bigl\{\chi(x;X_0,X_1),\chi(y;Y_0,Y_1)\Bigr\}_{\rm P} &=&K\Bigl[-\epsilon(x-y)+\theta(x-Y_1)+\theta(x+Y_0) 
 -\theta(y-X_1) \Bigr. \nonumber \\ 
&& \Bigl. \qquad -\theta(y+X_0) 
+\frac{1}{2}\theta(X_0-Y_0)-\frac{1}{2}\theta(X_1-Y_1)\Bigr]\,. \nonumber
\end{eqnarray}

\medskip 

Next we define the regularized level 1 charges as 
\begin{eqnarray}
Q^{R,\pm}(X_0,X_1;X'_0,X'_1)&=&\int^{X_1}_{-X_0}\!\!\!dx~{\rm e}^{\gamma^\pm\chi(x;X'_0,X'_1)}i^{R,\pm}_t(x)\,, 
\quad 
[-X_0,X_1]\subset [-X'_0,X'_1]\,. \nonumber
\end{eqnarray}
Note that the interval $[-X_0,X_1]$ is included in the interval $[-X'_0,X'_1]$\,. 
Then the Poisson brackets of the regularized charges $Q^{R,3}$ and $Q^{R,\pm}$ are given by 
\begin{eqnarray}
&&\Bigl\{Q^{R,\pm}(X_0,X_1;X'_0,X'_1),Q^{R,3}(Y_0,Y_1)\Bigr\}_{\rm P} \\
&=&\pm iQ^{R,\pm}\bigl(\min(X_0,Y_0),\min(X_1,Y_1);X'_0,X'_1\bigr) \nonumber \\
&&\pm\frac{i}{2}\left(1-\frac{\gamma^\pm}{\gamma^\mp}\right)\left[2Q^{R,\pm}
\bigl(\min(X_0,Y_0),\min(X_1,Y_1);X'_0,X'_1\bigr)\right. \nonumber \\
&&\hspace{3cm}\left.-[\theta(Y_0-X'_0)+\theta(Y_1-X'_1)]Q^{R,\pm}(X_0,X_1;X'_0,X'_1)\right]\,, \nonumber \\
&&\Bigl\{Q^{R,+}(X_0,X_1;X'_0,X'_1),Q^{R,-}(Y_0,Y_1;Y'_0,Y'_1)\Bigr\}_{\rm P} \nonumber \\
&=&i\frac{\gamma^-}{\gamma^+}\frac{1}{\gamma^++\gamma^-}
\left[{\rm e}^{\gamma^+\chi(\min(X_1,Y_1);X'_0,X'_1)+\gamma^-\chi(\min(X_1,Y_1);Y'_0,Y'_1)} \right. \nonumber \\
&&\hspace{3cm}\left.-{\rm e}^{\gamma^+\chi(-\min(X_0,Y_0);X'_0,X'_1)+\gamma^-\chi(-\min(X_0,Y_0);Y'_0,Y'_1)}\right] 
\nonumber \\
&&-i\frac{\gamma^+-\gamma^-}{\gamma^+\gamma^-}\left[\theta(X_0-Y_0)
{\rm e}^{\gamma^+\chi(-Y_0;X'_0,X'_1)+\gamma^-\chi(-Y_0;Y'_0,Y'_1)}\right. \nonumber \\
&&\hspace{2.5cm}\left.-\theta(X_1-Y_1){\rm e}^{\gamma^+\chi(Y_1;X'_0,X'_1)+\gamma^-\chi(Y_1;Y'_0,Y'_1)}\right] \nonumber \\
&&-\frac{1}{2}\left(\gamma^+-\gamma^-\right)Q^{R,-}(Y_0,Y_1;Y'_0,Y'_1) \nonumber \\
&&~\times\int^{X_1}_{-X_0}\!\!\!dx\left[\theta(x-Y'_1)+\theta(x+Y'_0)\right]{\rm e}^{\gamma^+\chi(x;X'_0,X'_1)}i^{R,+}_t(x) 
\nonumber \\
&&+\frac{1}{2}\left(\gamma^+-\gamma^-\right)Q^{R,+}(X_0,X_1;X'_0,X'_1) \nonumber \\
&&~\times\int^{Y_1}_{-Y_0}\!\!\!dy\left[\theta(y-X'_1)+\theta(y+X'_0)\right]{\rm e}^{\gamma^-\chi(y;Y'_0,Y'_1)}i^{R,-}_t(y) 
\nonumber \\
&&-\frac{1}{4}\left(\gamma^+-\gamma^-\right)[\theta(X'_0-Y'_0)-\theta(X'_1-Y'_1)] \nonumber \\
&&~\times Q^{R,+}(X_0,X_1;X'_0,X'_1)Q^{R,-}(Y_0,Y_1;Y'_0,Y'_1)\,. \nonumber
\end{eqnarray}

\subsubsection*{The order of limits}

Let us consider the order of limits in the Poisson brackets computed so far. 

\medskip 

The first is the Poisson brackets in (\ref{Poisson Q3Q3})\,, 
which depend on the order of limits. Note that the bracket $\{Q^{R,3},Q^{R,3}\}_{\rm P}$ 
should vanish. 
A prescription is to take a specific regularization as follows: 
\begin{eqnarray}
Q^{R,3}(X)=Q^{R,3}(X,X)\,. 
\end{eqnarray}
Then the Poisson bracket vanishes properly,  
\begin{eqnarray}
\Bigl\{Q^{R,3}(X),Q^{R,3}(Y)\Bigr\}_{\rm P}=0\,. 
\end{eqnarray}
Thus, after taking the limits, the desired result is obtained as 
\begin{eqnarray}
\Bigl\{Q^{R,3},Q^{R,3}\Bigr\}_{\rm P}=0\,. 
\end{eqnarray}

\medskip 

Similarly, the level 1 charges can be regularized as  
\begin{eqnarray}
Q^{R,\pm}(X;X')=Q^{R,\pm}(X,X;X',X')\,. 
\end{eqnarray}
The Poisson brackets of $Q^{R,\pm}(X,X')$ and $Q^{R,3}(Y)$ are 
\begin{eqnarray}
&& \Bigl\{Q^{R,\pm}(X;X'),Q^{R,3}(Y)\Bigr\}_{\rm P}  = \pm iQ^{R,\pm}\bigl(\min(X,Y);X'\bigr) 
\label{Poisson QpmQ3} \\ && \hspace*{4cm}
\pm i\left(1-\frac{\gamma^\pm}{\gamma^\mp}\right)\left[Q^{R,\pm}\bigl(\min(X,Y);X'\bigr)
-\theta(Y-X')Q^{R,\pm}(X;X')\right]\,, \nonumber
\end{eqnarray}
and the Poisson bracket between $Q^{R,\pm}(X,X')$ is 
\begin{eqnarray}
&&\Bigl\{Q^{R,+}(X;X'),Q^{R,-}(Y;Y')\Bigr\}_{\rm P} \label{Poisson QpQm}\\
&=&i\frac{\gamma^-}{\gamma^+}\frac{1}{\gamma^++\gamma^-}
\left[{\rm e}^{\gamma^+\chi(\min(X,Y);X')+\gamma^-\chi(\min(X,Y);Y')} 
-{\rm e}^{\gamma^+\chi(-\min(X,Y);X')+\gamma^-\chi(-\min(X,Y);Y')}\right] \nonumber \\
&&-i\frac{\gamma^+-\gamma^-}{\gamma^+\gamma^-}\theta(X-Y)
\left[{\rm e}^{\gamma^+\chi(-Y;X')+\gamma^-\chi(-Y;Y')}
-{\rm e}^{\gamma^+\chi(Y;X')+\gamma^-\chi(Y;Y')}\right] \nonumber \\
&&-\frac{1}{2}\left(\gamma^+-\gamma^-\right)Q^{R,-}\!(Y;Y') 
\int^{X}_{-X}\!\!\!dx\left[\theta(x-Y')+\theta(x+Y')\right]{\rm e}^{\gamma^+\chi(x;X')}i^{R,+}_t(x) \nonumber \\
&&+\frac{1}{2}\left(\gamma^+-\gamma^-\right)\!Q^{R,+}(X;X') 
\int^{Y}_{-Y}\!\!\!dy\left[\theta(y-X')+\theta(y+X')\right]{\rm e}^{\gamma^-\chi(y;Y')}i^{R,-}_t(y)\,, \nonumber
\end{eqnarray}
where $\chi(x;X)$ is defined as 
\begin{eqnarray}
\chi(x;X) \equiv \chi(x;X,X)\,. 
\end{eqnarray}
Thus the Poisson brackets in (\ref{Poisson QpmQ3}) and (\ref{Poisson QpQm}) 
depend on the order of limits. The appearance of the order of limits is due to the presence 
of the Wess-Zumino term. When $K=0$\,, 
$\gamma_+=\gamma_-$ 
and no ambiguity exists,  as discussed in \cite{KYhybrid}. 

\medskip 

Let us consider the order of limits for the Poisson bracket (\ref{Poisson QpmQ3})\,. 
A prescription is to take the limit $Y\to\infty$ before the limit $X'\to\infty$\,: 
\begin{eqnarray}
\Bigl\{Q^{R,\pm}(X;X'),Q^{R,3}\Bigr\}_{\rm P}=\pm iQ^{R,\pm}\bigl(X;X'\bigr)\,. 
\end{eqnarray}
Then one can obtain the desired expression, 
\begin{eqnarray}
\Bigl\{Q^{R,\pm},Q^{R,3}\Bigr\}_{\rm P}=\pm iQ^{R,\pm}\,, 
\end{eqnarray}
by sending $X$ and $X'$ to infinity. 

\medskip 

The next is the order of limits in (\ref{Poisson QpQm})\,. 
A prescription is to take the $X'\to\infty$ limit before the $X\to\infty$ limit. 
Thus the regularized charge is introduced as  
\begin{eqnarray}
Q^{R,\pm}(X)=\lim_{X'\to\infty}Q^{R,\pm}(X;X')\,. 
\end{eqnarray}
The Poisson bracket between $Q^{R,\pm}$ is 
\begin{eqnarray}
\Bigl\{Q^{R,+}(X),Q^{R,-}(Y)\Bigr\}_{\rm P} 
&=&i\frac{\gamma^-}{\gamma^+}\frac{1}{\gamma^++\gamma^-}\left[{\rm e}^{(\gamma^++\gamma^-)\chi(\min(X,Y))}
-{\rm e}^{(\gamma^++\gamma^-)\chi(-\min(X,Y))}\right] \nonumber \\
&&-i\frac{\gamma^+-\gamma^-}{\gamma^+\gamma^-}\theta(X-Y)
\left[{\rm e}^{(\gamma^++\gamma^-)\chi(-Y)}-{\rm e}^{(\gamma^++\gamma^-)\chi(Y)}\right]\,. \nonumber
\end{eqnarray}
The second term depends on the order of limits. 
This ambiguity appears as only the overall factor  
and hence it can be absorbed by rescaling the charges. 
Here the order of limits is fixed in the following way: 
\begin{eqnarray}
\Bigl\{Q^{R,+},Q^{R,-}\Bigr\}_{\rm P}
&\equiv&\lim_{X\to\infty}\Bigl\{Q^{R,+}(X),Q^{R,-}(X)\Bigr\}_{\rm P} \nonumber \\
&=&\frac{i}{2}\frac{(\gamma^+)^2+(\gamma^-)^2}{\gamma^+\gamma^-(\gamma^++\gamma^-)}
\lim_{X\to\infty}\left[{\rm e}^{(\gamma^++\gamma^-)
\chi(X)}-{\rm e}^{(\gamma^++\gamma^-)\chi(-X)}\right] \nonumber \\
&=&-i\frac{(\gamma^+)^2+(\gamma^-)^2}{2\gamma^+\gamma^-}\frac{2}{(\gamma^++\gamma^-)}\sinh
\left[\frac{(\gamma^++\gamma^-)}{2}Q^{R,3}\right]\,. 
\end{eqnarray}
Thus the $q$-deformed $su(2)$ algebra has been shown with some prescriptions for 
the order of limits, where the $q$-parameter is defined as 
\[
q \equiv \exp\left[\frac{\gamma_+ + \gamma_-}{2}\right]\,. 
\]

\subsection{A deformation of quantum affine algebra}

Let us here present how to compute the deformed quantum affine algebra presented in Sec.\,\ref{dQAA:sec}. 
In the middle of the computation, again, it is necessary to introduce some prescriptions to treat 
the ambiguities coming from the order of limits. 

\medskip 

The Poisson bracket between $Q^{R,-}_{(1)}$ and $\widetilde{Q}^{R,+}_{(1)}$ is first computed. 
For the regularized charges, it is given by  
\begin{eqnarray}
&&\Bigl\{Q^{R,-}_{(1)}(X),\widetilde{Q}^{R,+}_{(1)}(Y)\Bigr\}_{\rm P} \label{no-amb} \\
&=&\frac{1}{2}{\rm e}^{-2\beta}\!\sinh 2\alpha\!
\left[\int^X_{-X}\!\!\!dx\int^Y_{-Y}\!\!\!dy~\epsilon(x-y)j^{R,-}_t(x)\widetilde{j}^{R,+}_t(y) \right. \nonumber \\
&&\hspace{2cm}\left.-2i{\rm e}^{2\beta}\!\int^{\min(X,Y)}_{-\min(X,Y)}\!\!\!dx~j^{R,3}_x(x)
+2i{\rm e}^{2\beta}\frac{\cosh 2\alpha}{\sinh 2\alpha}Q^{R,3}_{(0)}(\min(X,Y))\right] \nonumber \\
&=&\frac{1}{2}{\rm e}^{-2\beta}\!\sinh 2\alpha \bar{Q}^{R,3}_{(2)}(X,Y,\min(X,Y))\,. \nonumber
\end{eqnarray}
Here the level 2 charge $\bar{Q}^{R,3}_{(2)}$ is regularized as 
\begin{eqnarray}
\bar{Q}^{R,3}_{(2)}(X,X',X'')
&=&\int^X_{-X}\!\!\!dx\int^{X'}_{-X'}\!\!\!dy~\epsilon(x-y)j^{R,-}_t(x)\widetilde{j}^{R,+}_t(y) \\
&&-2i{\rm e}^{2\beta}\!\int^{X''}_{-X''}\!\!\!dx~j^{R,3}_x(x)
+2i{\rm e}^{2\beta}\frac{\cosh 2\alpha}{\sinh 2\alpha}Q^{R,3}_{(0)}(X'')\,. \nonumber
\end{eqnarray}
Note that no ambiguous term is contained in this Poisson bracket (\ref{no-amb})\,. 
Thus, by taking the $X,Y\to\infty$ limits, we obtain the following: 
\begin{eqnarray}
\Bigl\{Q^{R,-}_{(1)},\widetilde{Q}^{R,+}_{(1)}\Bigr\}_{\rm P}
=\frac{1}{2}{\rm e}^{-2\beta}\!\sinh 2\alpha \bar{Q}^{R,3}_{(2)}\,.
\end{eqnarray}

\medskip 

As a next step, the Poisson bracket between $\bar{Q}^{R,3}_{(2)}$ and $Q^{R,-}_{(1)}$ is computed. 
For the regularized charges, it is evaluated as 
\begin{eqnarray}
&&\Bigl\{\bar{Q}^{R,3}_{(2)}(X,X',X''),Q^{R,-}_{(1)}(Y)\Bigr\}_{\rm P} \\
&=&-\frac{1}{2}{\rm e}^{-2\beta}\!\sinh 2\alpha \nonumber \\
&&\times\Biggl[\int^X_{-X}\!\!\!dx\!\int^{X'}_{-X'}\!\!\!\!dx'\!\!\int^Y_{-Y}\!\!\!dy~
\epsilon(x-x')\!\left[\epsilon(x-y)-\epsilon(x'-y)\right]j^{R,-}_t(x)\widetilde{j}^{R,+}_t(x')j^{R,-}_t(y) \Biggr. \nonumber \\
&&\hspace{1cm}-2i{\rm e}^{2\beta}\!\int^X_{-X}\!\!\!\!dx\!\!\int^{\min(X',Y)}_{-\min(X',Y)}\!\!\!\!\!dx'
\epsilon(x-x')j^{R,-}_t(x)\!\left(j^{R,3}_x(x')\!-\!\frac{\cosh 2\alpha}{\sinh 2\alpha}j^{R,3}_t(x')\!\right) \nonumber \\
&&\hspace{1cm}-\frac{4{\rm e}^{2\beta}}{\sinh 2\alpha}\int^{\min(X'',Y)}_{-\min(X'',Y)}\!\!\!dx''j^{R,-}_x(x'')
+\frac{4{\rm e}^{2\beta}\cosh 2\alpha}{\sinh^2 2\alpha}
Q^{R,-}_{(1)}(\min(X'',Y))\Biggr] \nonumber \\
&&+4(K-iA)\left\{Q^{R,-}_{(1)}(\min(X,Y))-\theta(Y-X')Q^{R,-}_{(1)}(X)\right\} \nonumber \\
&=&-\frac{1}{2}{\rm e}^{-2\beta}\sinh 2\alpha \nonumber \\
&&\times\!\Biggl[\!Q^{R,-}_{(3)}(X,X',Y,X,\min(X',Y),\min(X'',Y))
\!+\!\frac{2}{3}Q^{R,-}_{(1)}(X)\widetilde{Q}^{R,+}_{(1)}(X')Q^{R,-}_{(1)}(Y)\Biggr. \nonumber \\
&&\Biggl.+\frac{1}{2}\int^X_{-X}\!\!\!dx\!\int^{X'}_{-X'}\!\!\!\!dx'\!\!\int^Y_{-Y}\!\!\!dy~
\epsilon(x-y)\!\left[\epsilon(x-x')\!+\!\epsilon(y-x')\right] 
j^{R,-}_t(x)\widetilde{j}^{R,+}_t(x')j^{R,-}_t(x'')\Biggr] \nonumber \\
&&+4(K-iA)\left\{Q^{R,-}_{(1)}(\min(X,Y))-\theta(Y-X')Q^{R,-}_{(1)}(X)\right\}\,. \nonumber
\end{eqnarray}
Here the last term 
\begin{eqnarray}
4(K-iA)\left\{Q^{R,-}_{(1)}(\min(X,Y))-\theta(Y-X')Q^{R,-}_{(1)}(X)\right\}
\end{eqnarray}
depends on the order of limits. 
Note that the level 3 charge $Q^{R,-}_{(3)}$ is regularized as 
\begin{eqnarray}
&&Q^{R,-}_{(3)}(X,X',X'',X''',X'''',X''''') 
\\&=&\frac{1}{2}\int^X_{-X}\!\!\!dx\int^{X'}_{-X'}\!\!\!dx'\int^{X''}_{-X''}\!\!\!dx''~
\epsilon(x'-x)\epsilon(x'-x'')j^{R,-}_t(x)\widetilde{j}^{R,+}_t(x')j^{R,-}_t(x'') \nonumber \\
&&-2i{\rm e}^{2\beta}\int^{X'''}_{-X'''}\!\!\!dx'''\int^{X''''}_{-X''''}\!\!\!dx''''~\epsilon(x'''-x'''') 
j^{R,-}_t(x''')\left(j^{R,3}_x(x'''')-\frac{\cosh 2\alpha}{\sinh 2\alpha}j^{R,3}_t(x'''')\right) \nonumber \\
&&-\frac{4{\rm e}^{2\beta}}{\sinh 2\alpha}\int^{X'''''}_{-X'''''}\!\!\!dx'''''~j^{R,-}_x(x''''')
+\frac{4{\rm e}^{2\beta}\cosh 2\alpha}{\sinh^2 2\alpha}Q^{R,-}_{(1)}(X''''') \nonumber \\
&&-\frac{1}{6}Q^{R,-}_{(1)}(X)\widetilde{Q}^{R,+}_{(1)}(X')Q^{R,-}_{(1)}(X'')\,. \nonumber
\end{eqnarray}
A prescription is to take the $Y\to\infty$ limit before $X $ is sent to infinity 
so that the ambiguous term vanishes:  
\begin{eqnarray}
&& \Bigl\{\bar{Q}^{R,3}_{(2)}(X,X',X''),Q^{R,-}_{(1)}\Bigr\}_{\rm P} \nonumber \\
&=&\lim_{Y\to\infty}\Bigl\{\bar{Q}^{R,3}_{(2)}(X,X',X''),Q^{R,-}(Y)\Bigr\}_{\rm P} \nonumber \\
&=&-\frac{1}{2}{\rm e}^{-2\beta}\!\sinh 2\alpha\Bigl[Q^{R,-}_{(3)}(X,X',\infty,X,X',X'')
+\frac{2}{3}Q^{R,-}_{(1)}(X)\widetilde{Q}^{R,+}_{(1)}(X')Q^{R,-}_{(1)}\Bigr] \nonumber \\
&&-\frac{1}{4}{\rm e}^{-2\beta}\!\sinh 2\alpha\int^X_{-X}\!\!\!dx\int^{X'}_{-X'}\!\!\!dx'
\int^\infty_{-\infty}\!\!\!dy~\epsilon(x-y)\left[\epsilon(x-x')+\epsilon(y-x')\right] \nonumber \\
&&\hspace{7cm}\times j^{R,-}_t(x)\widetilde{j}^{R,+}_t(x')j^{R,-}_t(y)\,. \nonumber 
\end{eqnarray}
By taking the remaining limits, we obtain the following result: 
\begin{eqnarray}
\Bigl\{\bar{Q}^{R,3}_{(2)},Q^{R,-}_{(1)}\Bigr\}_{\rm P} 
=-\frac{1}{2}{\rm e}^{-2\beta}\!\sinh 2\alpha\left[Q^{R,-}_{(3)}
+\frac{2}{3}\widetilde{Q}^{R,+}_{(1)}\left(Q^{R,-}_{(1)}\right)^2\right]\,.
\end{eqnarray}

\medskip 

It is interesting to see a higher-level Poisson bracket, for example, 
the Poisson bracket between $Q^{R,-}_{(3)}$ and $Q^{R,-}_{(1)}$\,. 
For the regularized charges, it is given by 
\begin{eqnarray}
&&\Bigl\{Q^{R,-}_{(3)}(X,X',X'',X''',X'''',X'''''),Q^{R,-}_{(1)}(Y)\Bigr\}_{\rm P} \\
&=&-\frac{1}{4}{\rm e}^{-2\beta}\sinh 2\alpha \nonumber \\
&&\times\Biggl[
-\frac{1}{2}Q^{R,-}_{(1)}(X)Q^{R,-}_{(1)}(Y)\int^{X'}_{-X'}\!\!\!dx'\int^{X''}_{-X''}\!\!\!dx''~\epsilon(x'-x'')\widetilde{j}^{R,+}_t(x')j^{R,-}_t(x'') \Biggr. \nonumber \\
&&\qquad+\frac{1}{2}Q^{R,-}_{(1)}(X'')Q^{R,-}_{(1)}(Y)\int^X_{-X}\!\!\!dx\int^{X'}_{-X'}\!\!\!dx'~\epsilon(x-x')j^{R,-}_t(x)\widetilde{j}^{R,+}_t(x') \nonumber \\
&&\qquad+\frac{1}{2}\int^X_{-X}\!\!\!dx\int^{X'}_{-X'}\!\!\!dx'\int^{X''}_{-X''}\!\!\!dx''\int^Y_{-Y}\!\!\!dy \nonumber \\
&&\hspace{2cm}\times\Bigl\{\epsilon(x'-x'')\epsilon(y-x)\left[\epsilon(y-x')+\epsilon(x-x')\right]\Bigr. \nonumber \\
&&\hspace{4cm}\Bigl.+\epsilon(x'-x)\epsilon(y-x'')\left[\epsilon(y-x')+\epsilon(x''-x')\right]\Bigr\} \nonumber \\
&&\hspace{2cm}\times j^{R,-}_t(x)\widetilde{j}^{R,+}_t(x')j^{R,-}_t(x'')j^{R,-}_t(y) \nonumber \\
&&\qquad-2i{\rm e}^{2\beta}Q^{R,-}_{(1)}(X)Q^{R,-}_{(1)}(X'')\int^{\min(X',Y)}_{-\min(X',Y)}\!\!\!dx'
\left[j^{R,3}_x(x')-\frac{\cosh 2\alpha}{\sinh 2\alpha}j^{R,3}_t(x')\right] \nonumber \\
&&\qquad+4i{\rm e}^{2\beta}\int^X_{-X}\!\!\!dx\int^{\min(X',Y)}_{-\min(X',Y)}\!\!\!dx'\int^{X''}_{-X''}\!\!\!dx''\epsilon(x-x')\epsilon(x-x'') \nonumber \\
&&\hspace{3cm}\times j^{R,-}_t(x)\left[j^{R,3}_x(x')-\frac{\cosh 2\alpha}{\sinh 2\alpha}j^{R,3}_t(x')\right]j^{R,-}_t(x'') \nonumber \\
&&\qquad-4i{\rm e}^{2\beta}\int^{X'''}_{-X'''}\!\!\!dx'''\int^{X''''}_{-X''''}\!\!\!dx''''\int^Y_{-Y}\!\!\!dy\epsilon(x'''-x'''')\epsilon(x'''-y) \nonumber \\
&&\hspace{3cm}\times j^{R,-}_t(x''')\left[j^{R,3}_x(x'''')-\frac{\cosh 2\alpha}{\sinh 2\alpha}j^{R,3}_t(x'''')\right]j^{R,-}_t(y) \nonumber \\
&&\qquad-\frac{8{\rm e}^{2\beta}}{\sinh 2\alpha}\int^{X'''}_{-X'''}\!\!\!dx'''\int^{\min(X'''',Y)}_{-\min(X'''',Y)}\!\!\!dx''''~
\epsilon(x'''-x'''')j^{R,-}_t(x''')j^{R,-}_x(x'''') \nonumber \\
&&\qquad-\frac{8{\rm e}^{2\beta}}{\sinh 2\alpha}\int^{X'''''}_{-X'''''}\!\!\!dx'''''\int^Y_{-Y}\!\!\!dy~
\epsilon(x'''''-y)j^{R,-}_x(x''''')j^{R,-}_t(y) \nonumber \\
&&\qquad+\frac{8{\rm e}^{2\beta}\cosh 2\alpha}{\sinh^2 2\alpha}\!
\int^{X'''}_{-X'''}\!\!\!dx'''\int^{\min(X'''',Y)}_{-\min(X'''',Y)}\!\!\!dx''''
\epsilon(x'''-x'''') \nonumber \\
&&\hspace{6cm}\times j^{R,-}_t(x''')j^{R,-}_t(x'''') \nonumber \\
&&\qquad\Biggl.-\frac{1}{3}Q^{R,-}_{(1)}(X)Q^{R,-}_{(1)}(X'')\bar{Q}^{R,3}_{(2)}(Y,X',\min(X',Y))\Biggr] \nonumber \\
&&+(K-iA)\theta(X'-Y) \nonumber \\
&&\qquad\times\int^X_{-X}\!\!\!dx\int^{X''}_{-X''}\!\!\!dx''
\left[\epsilon(x+Y)\epsilon(x''+Y)-\epsilon(x-Y)\epsilon(x''-Y)\right] 
j^{R,-}_t(x)j^{R,-}_t(x'')\,. \nonumber
\end{eqnarray}
Note that the last term
\begin{eqnarray}
&&(K-iA)\theta(X'-Y)\int^X_{-X}\!\!\!dx\int^{X''}_{-X''}\!\!\!dx'' \\
&&\hspace{2.5cm}\times\left[\epsilon(x+Y)\epsilon(x''+Y)-\epsilon(x-Y)\epsilon(x''-Y)\right]
j^{R,-}_t(x)j^{R,-}_t(x'')\,. \nonumber
\end{eqnarray}
is proportional to the step function and hence seems likely to be an ambiguous term. 
However, this is not the case. 
This term does not depend on the order of limits and it is reduced to zero after taking all of the limits. 
Thus the resulting Poisson bracket is given by 
\begin{eqnarray}
\Bigl\{Q^{R,-}_{(3)},Q^{R,-}_{(1)}\Bigr\}_{\rm P}
=-\frac{1}{6}{\rm e}^{-2\beta}\sinh 2\alpha\left(Q^{R,-}_{(1)}\right)^2\bar{Q}^{R,3}_{(2)}\,.
\end{eqnarray}
Similarly, the other Poisson brackets are computed, though those are not touched here.

\end{document}